\documentclass[12pt,letterpaper]{article}
\usepackage[T1]{fontenc}
\usepackage[utf8]{inputenc}
\usepackage{newtxtext}
\usepackage{amsmath,amsthm,mathtools}
\usepackage{newtxmath,bm}
\usepackage[left=.85in,right=.85in,top=.70in,bottom=.70in]{geometry}
\usepackage{microtype,multicol}
\usepackage{xcolor,graphicx,booktabs,array}
\usepackage{tikz,pgfplots}
\pgfplotsset{compat=1.18}
\usetikzlibrary{arrows.meta,calc,positioning,fit,backgrounds,decorations.pathreplacing,shapes.geometric}
\usepackage[numbers,sort&compress]{natbib}
\usepackage[font=small,labelfont=bf]{caption}
\usepackage{placeins,needspace,etoolbox}
\usepackage{hyperref}
\newcommand{\doi}[1]{\href{https://doi.org/#1}{\nolinkurl{#1}}}
\hypersetup{colorlinks=true,linkcolor=blue!40!black,citecolor=blue!40!black,urlcolor=blue!40!black,
 bookmarksdepth=2,pdfauthor={Mark M. Wilde},
 pdftitle={Convex splitting by binary quantum hypothesis testing}}
\numberwithin{equation}{section}
\newtheorem{theorem}{Theorem}[section]
\newtheorem{lemma}[theorem]{Lemma}

\newtheorem{corollary}[theorem]{Corollary}
\DeclareMathOperator{\Tr}{Tr}
\newcommand{\E}{\mathcal E}
\newcommand{\norm}[1]{\left\|#1\right\|}
\definecolor{inkblue}{RGB}{36,69,101}
\definecolor{softblue}{RGB}{232,240,247}
\definecolor{inkgreen}{RGB}{28,103,87}
\definecolor{softgreen}{RGB}{233,244,239}
\makeatletter
\g@addto@macro\normalsize{%
  \setlength{\abovedisplayskip}{4.6pt plus 1.5pt minus 1.4pt}%
  \setlength{\belowdisplayskip}{4.6pt plus 1.5pt minus 1.4pt}%
  \setlength{\abovedisplayshortskip}{3pt plus 2pt}%
  \setlength{\belowdisplayshortskip}{4pt plus 2pt minus 2pt}}
\makeatother
\makeatletter
\renewcommand{\section}{\@startsection{section}{1}{\z@}{-2.8ex plus -.7ex minus -.2ex}{1.5ex plus .2ex}{\normalfont\Large\bfseries}}
\renewcommand{\paragraph}{\@startsection{paragraph}{4}{\z@}{1.8ex plus .5ex minus .2ex}{-1em}{\normalfont\normalsize\bfseries}}
\renewcommand{\subsection}{\@startsection{subsection}{2}{\z@}{-2.5ex plus -.7ex minus -.2ex}{1.0ex plus .2ex}{\normalfont\large\bfseries}}
\makeatother

\allowdisplaybreaks[1]\widowpenalty=10000\clubpenalty=10000
\title{\textbf{Convex splitting by binary quantum hypothesis testing}}
\author{Mark M.\ Wilde\\[6pt]
 {\small\itshape School of Electrical and Computer Engineering}\\
 {\small\itshape Cornell University, Ithaca, New York 14850, USA}\\[2pt]
 {\small\href{mailto:wilde@cornell.edu}{wilde@cornell.edu}}}
\date{November 18, 2026}
\begin{document}
\maketitle

\begin{abstract}
How many independent quantum systems does it take to hide one system that is correlated with a reference? An observer can measure all the systems together with the reference to detect the correlation. We turn any such measurement into binary tests on a single system and the reference. For each test we bound the probability of declaring a signal when an independent pair was prepared. Averaging the largest signal acceptance probabilities allowed by these bounds controls the normalized trace distance in the convex-split lemma. A second bound gives a sufficient number of systems by minimizing the hypothesis-testing relative entropy over the state of the reference. Both bounds hold for density operators on separable Hilbert spaces. Comparing two distributions for the location of the hidden system gives one-shot bounds for private communication and one-way secret key distillation, including the public information available to an eavesdropper. For a fixed finite-alphabet model with separable quantum outputs and positive information advantage and variances, we recover the known Gaussian corrections to the attainable number of private or key bits from $n$ independent repetitions and bound the remaining loss by $\log_2 n+O(1)$ at fixed errors. The position-based key distillation protocol attains this bound with $O(n)$ public bits.
\end{abstract}

\tableofcontents

\section{Introduction}\label{sec:intro}
\subsection{Background and motivation}
How much randomness does it take to hide a correlation? Suppose that $K$ systems look the same when measured individually, but one of them is correlated with a reference. An observer receives all $K$ systems and the reference, but no record of the chosen position. The observer can measure them jointly to decide whether the correlation is present. The convex-split lemma of Anshu, Devabathini, and Jain~\cite{ADJ} bounds the maximum distinguishing advantage of this measurement. In private communication, Alice, the sender, wants to transmit a message to Bob, the receiver, while keeping it secret from Eve, an eavesdropper who receives part of the channel output. Alice chooses at random among several signals for each message. To protect the message, the state Eve receives for a given message should be close, on average over messages, to the state she receives when the message is unknown.

A one-shot bound describes what one use of a channel or one copy of a shared state can accomplish. Binary hypothesis testing gives such bounds across quantum communication theory. Wang and Renner gave a testing bound for unassisted classical communication~\cite{WangRenner}, and Anshu, Jain, and Warsi treated communication with shared entanglement~\cite{AJWPack}. Cheng sharpened the bounds for both tasks~\cite{Cheng}. Reference~\cite{CW} gives such bounds for quantum communication and one-way entanglement distillation, including a conversion from entanglement generation to entanglement transmission (see also~\cite{ChengQuantum} for a related approach to entanglement generation). Cheng and Gao gave testing bounds for private communication and one-way secret key distillation~\cite[Theorems~14 and~19]{CG}. Thus all six core tasks admit one-shot achievability bounds in terms of binary quantum hypothesis testing. For the privacy tasks, one binary test bounds the error of Bob's decoder. A second certifies how many random choices suffice for secrecy against any measurement Eve might perform.

Shen, Gao, and Cheng analyzed this hiding problem for classical inputs and quantum outputs~\cite{SGC}. Cheng and Gao bounded the number of positions for a general quantum pair in terms of optimized hypothesis-testing mutual information~\cite[Theorem~4]{CG}. Their direct bound uses spectral pinching, which removes the off-diagonal blocks between eigenspaces, and requires finite dimension. Cheng et al.~studied multipartite convex splitting with relative entropy as the error criterion~\cite{CGHHL}. For the collision relative entropy, Gour found an exact identity for uniform convex splitting~\cite[Lemma~1]{GourCollision}.

Our question is how to turn the observer's joint measurement into a binary test on a single pair. We use a cancellation between distinct positions that also appears in the multipartite convex-splitting estimates of Berta, Cheng, and Gao~\cite[Lemma~5 and Remark~6]{CGB}. Keeping the reference system throughout this calculation gives one test that works against every independent reference state. The resulting direct bound has no dimension or spectral factor and holds for normal states on separable Hilbert spaces. The one-shot protocols and, for a finite classical alphabet, the second-order coding bounds also allow separable quantum outputs. Appendix~\ref{sec:separable} shows why the alphabet assumption controls the fluctuations needed for the latter conclusion.

\subsection{Summary of results}
Let $K\in\mathbb N\coloneqq\{1,2,\ldots\}$, let $\rho_{AB}$ be the state to be hidden, and let $\rho_B\coloneqq\Tr_A[\rho_{AB}]$ and $\tau_A$ be the reference marginal and background state, respectively. The background state $\tau_A$ can be arbitrary. Writing $A^K$ for $A_1\cdots A_K$, the observer distinguishes the following two states:
\begin{align}
 \omega_{A^K B}&\coloneqq \frac{1}{K}\sum_{i=1}^K \rho_{A_iB}\otimes\bigotimes_{j\ne i}\tau_{A_j},\label{eq:intro-signal}\\*
 \omega^0_{A^K B}&\coloneqq\tau_A^{\otimes K}\otimes\rho_B.\label{eq:intro-split}
\end{align}
For an effect $Q_{AB}$ accepting the signal, a type~I error rejects $\rho_{AB}$ when it was prepared. A type~II error declares the signal when an independent pair was prepared. Let $G^\eta(\rho_{AB}\Vert\tau_A)$ be the greatest signal acceptance probability at type~II error probability at most $\eta$ against every $\tau_A\otimes\sigma_B$, as defined in \eqref{eq:roc}. Theorem~\ref{thm:ordinary} states that
\begin{equation}
 \frac12\norm{\omega_{A^K B}-\omega^0_{A^K B}}_1
 \le \int_0^1 G^{1/(1+4Kt^2)}(\rho_{AB}\Vert\tau_A)\,dt.\label{eq:main-intro}
\end{equation}
The maximum difference in acceptance probabilities between the two preparations is the normalized trace distance on the left of \eqref{eq:main-intro}. With equally likely preparations, the optimal probability of a correct guess is one half plus half this distance. The integral arises because a positive eigenvalue $x$ of the single-pair observable passes a fraction $x$ of thresholds chosen uniformly from $[0,1]$. At each fixed $t>0$, increasing $K$ lowers the permitted probability of accepting an independent pair, which limits the evidence for a hidden correlation.

A single testing tolerance also suffices to choose $K$. For target distance $0<\varepsilon<1$ and slack $0<\delta<\varepsilon$, Corollary~\ref{cor:scalar} implies
\begin{equation}\label{eq:intro-scalar}
 K\ge \frac{2^{I_H^{1-\varepsilon+\delta}(\rho_{AB}\Vert\tau_A)}}{16\delta}
 \quad\Longrightarrow\quad
 \frac12\norm{\omega_{A^K B}-\omega^0_{A^K B}}_1\le\varepsilon .
\end{equation}
This gives a finite value of $K$ whenever the optimized hypothesis-testing mutual information $I_H^{1-\varepsilon+\delta}$ in \eqref{eq:info} is finite. The proof bounds the distance by $\delta$ plus the signal acceptance probability of a single-pair test. The remaining allowance $\varepsilon-\delta$ explains the complementary type~I error probability tolerance in \eqref{eq:intro-scalar}. For a strictly positive background state $\tau_A$ and the same testing tolerance, \eqref{eq:intro-scalar} removes Cheng and Gao's spectral factor and improves the explicit positive-slack prefactor from order $\delta^{-4}$ to order $\delta^{-1}$ (Section~\ref{ssec:scalar}). At $\delta=0$, \eqref{eq:scalar-zero} gives a finite value of $K$ whenever the corresponding testing probability is positive.

To use convex splitting for privacy, we need to compare two ways of choosing the hidden position. Alice arranges her possible signals in a table, with one row for each message and $K$ entries in each row. Given the message, she chooses an entry uniformly from its row. Averaging over the $M$ equiprobable messages makes the chosen entry uniform over the whole table. For a random codebook, Theorem~\ref{thm:cover} bounds the expected distance between Eve's state given a message and her state averaged over all messages. The secrecy criterion averages this distance over messages for a fixed codebook. Cheng's trace inequality~\cite{Cheng} bounds the probability that Bob's pretty-good measurement decodes the wrong row. Theorem~\ref{thm:wire} selects one codebook by bounding the sum of the mean decoding and secrecy errors for the position-based wiretap code, following the analysis in~\cite{Wilde}.

In one-way key distillation, Alice and Bob start with correlated data and use a public message from Alice to establish a common secret key. Eve can read this message. Theorem~\ref{thm:key} applies the same row comparison to Eve's quantum system together with the public announcement~\cite{KKGW}. For a finite source alphabet $\mathcal X$, Corollary~\ref{cor:transcript} bounds the length of the announcement by $O(\log_2(MK|\mathcal X|))$ bits.

For $n$ independent uses of a known finite-alphabet channel or copies of a known finite-alphabet source, Theorem~\ref{thm:rate} gives the following achievable number of private or key bits:
\begin{equation}\label{eq:intro-rate}
 \log_2 M\ge n(I_B-I_E)+\sqrt{nV_B}\,\Phi^{-1}(\varepsilon_B)+\sqrt{nV_E}\,\Phi^{-1}(\varepsilon_E)-\log_2 n-O(1).
\end{equation}
Here $I_B>I_E$ and $V_B,V_E>0$ are the mutual informations and variances in \eqref{eq:rate-information}--\eqref{eq:rate-moments}, and $\Phi$ is the standard normal cumulative distribution function. The positive constants $\varepsilon_B,\varepsilon_E$ are budgets for the averaged decoding and secrecy estimates and satisfy $\varepsilon_B+\varepsilon_E<1$. Selecting a channel codebook guarantees joint error at most $\varepsilon_B+\varepsilon_E$ in \eqref{eq:private}; its individual decoding and secrecy errors need not meet the respective budgets. Cheng and Gao obtained the two Gaussian terms with an $O(\log n)$ loss~\cite[Theorems~14 and~19]{CG}, improving on earlier squared secrecy tolerances~\cite{Wilde,KKGW}. Equation~\eqref{eq:intro-rate} gives an explicit achievable loss of $\log_2 n+O(1)$. It decreases by $\tfrac12\log_2 n$ when Bob's conditional states are pure or commute with his average state, and independently by another $\tfrac12\log_2 n$ when Eve's conditional states share an eigenbasis.

\subsection{Paper organization}
Section~\ref{sec:prelim} defines the hidden-position mixtures and the binary test of the signal state against independent pairs. Section~\ref{sec:cover} derives sufficient bounds on the number $K$ of positions and compares two necessary bounds in Section~\ref{ssec:converse}. Sections~\ref{sec:packing} and~\ref{sec:key} apply them to a wiretap channel and a shared source. Section~\ref{sec:spectral-curve} derives the second-order rates. Appendix~\ref{sec:split-examples} illustrates covering with a copied bit and a Bell pair. The appendices establish the sharpness and asymptotic refinements, construct a short public announcement, and give the separable-space argument.

\section{The hidden pair and the binary test}\label{sec:prelim}
The setting that we focus on here has
the observer measure all the positions of the states in \eqref{eq:intro-signal} and \eqref{eq:intro-split} together with their common reference. Section~\ref{sec:cover} will bound the distinguishability of these preparations by the performance of tests on just one pair. We specify the states in each experiment and the criterion for a successful test.

Let us first recall some basic concepts and establish notation.
An effect $0\le W\le I$ specifies the binary measurement $(W,I-W)$. If outcome $W$ selects the state $\eta$ over the state $\zeta$, the largest difference of their acceptance probabilities is the normalized trace distance,
\begin{equation}\label{eq:distance}
 \frac12\norm{\eta-\zeta}_1=\max_{0\le W\le I}\Tr[W(\eta-\zeta)].
\end{equation}
A projector is a self-adjoint idempotent. We write $\norm{\cdot}$ for the operator norm or the norm of a vector, and $\norm{\cdot}_1$ for the trace norm of an operator. We take all logarithms of information quantities to base two. For a single abstract system we omit its label. For several systems a subscript identifies the factors on which an operator acts. An observable acts as the identity on every unlisted factor, so $Q_{AB}-Y_B$ means $Q_{AB}-I_A\otimes Y_B$. Tensor factors are ordered by their labels. We give the finite-dimensional arguments first and treat separable Hilbert spaces in Appendix~\ref{sec:separable}.

\subsection{Choosing the position}
Consider $N\in\mathbb N$ positions $A^N\coloneqq A_1\cdots A_N$ and a reference $B$. Placing the signal state $\rho_{AB}$ at position $i$ leaves the other $N-1$ positions in the background state $\tau_A$. An independent preparation uses the same reference marginal $\rho_B\coloneqq\Tr_A[\rho_{AB}]$. These choices give
\begin{align}
 \omega^i_{A^N B}&\coloneqq \rho_{A_iB}\otimes\bigotimes_{j\ne i}\tau_{A_j} \qquad(1\le i\le N),\label{eq:placements}\\
 \omega^0_{A^N B}&\coloneqq\tau_A^{\otimes N}\otimes\rho_B.\label{eq:background}
\end{align}
If the index is drawn from a probability distribution $p$ on $\{0,\ldots,N\}$ and withheld, the observer receives
\begin{equation}\label{eq:mixture}
 \Omega_{A^N B}(p)\coloneqq\sum_{i=0}^N p_i\omega^i_{A^N B}.
\end{equation}
Ordinary convex splitting compares a signal placed uniformly among $K$ positions,
\begin{equation}\label{eq:ordinary}
 \omega_{A^K B}\coloneqq\frac{1}{K}\sum_{i=1}^K\omega^i_{A^K B},
\end{equation}
with the independent preparation $\omega^0_{A^K B}$ in \eqref{eq:background}. When $\tau_A=\rho_A\coloneqq\Tr_B[\rho_{AB}]$, the two states have the same one-system marginals. When $\rho_A$ and $\tau_A$ have orthogonal supports, testing for the support of $\rho_A$ at each site distinguishes them perfectly.

It is helpful to allow both preparations to use an arbitrary distribution on the states in \eqref{eq:placements}--\eqref{eq:background}. To compare $\Omega_{A^NB}(p)$ with $\Omega_{A^NB}(q)$ in \eqref{eq:mixture}, we use two quantities. The first includes the independent preparation, indexed by $0$. The second sums only over the $N$ possible signal positions. Writing $p_{1:N}\coloneqq(p_1,\ldots,p_N)$ and likewise for $q$, the two quantities are
\begin{align}
 \tfrac12\norm{p-q}_1&\coloneqq\tfrac12\sum_{i=0}^N|p_i-q_i|=\sum_{i=0}^N(p_i-q_i)_+,\label{eq:weights}\\
 \norm{p_{1:N}-q_{1:N}}_2^2&\coloneqq\sum_{i=1}^N(p_i-q_i)^2.\label{eq:site-norm}
\end{align}
Here $x_+\coloneqq\max\{x,0\}$. The equality in \eqref{eq:weights} follows from $|x|=2x_+-x$ and $\sum_{i=0}^N(p_i-q_i)=1-1=0$. For \eqref{eq:ordinary} versus \eqref{eq:background}, $p_0=0$, $p_i=1/K$ for $1\le i\le K$, and $q_0=1$. Thus \eqref{eq:weights} equals one and \eqref{eq:site-norm} equals $1/K$.

\subsection{The two error probabilities for a single-pair test}
The test on one pair must detect the signal while seldom accepting an independent pair. First consider two specified states $\rho,\sigma$. For an effect $Q$ that accepts $\rho$, a type~I error rejects $\rho$ and has probability $\Tr[(I-Q)\rho]$. A type~II error declares the signal when $\sigma$ was prepared and has probability $\Tr[Q\sigma]$. For $0\le\varepsilon<1$, the smallest type~II error probability at type~I error probability at most $\varepsilon$ gives the hypothesis-testing relative entropy~\cite{BuscemiDatta,BrandaoDatta,WangRenner},
\begin{align}
 \beta_\varepsilon(\rho\Vert\sigma)&\coloneqq\min_Q\bigl\{\Tr[Q\sigma]:0\le Q\le I,\ \Tr[Q\rho]\ge 1-\varepsilon\bigr\},\label{eq:simple-beta}\\
 D_H^\varepsilon(\rho\Vert\sigma)&\coloneqq-\log_2\beta_\varepsilon(\rho\Vert\sigma).\label{eq:dh}
\end{align}
We take $-\log_2 0\coloneqq+\infty$. The independent preparation uses $\tau_A\otimes\rho_B$. Requiring the same effect to work against every $\tau_A\otimes\sigma_B$ lets us optimize over the reference state in \eqref{eq:info}. To compute its acceptance probability, average the observable over the system $A$ prepared in the state $\tau_A$. This average acts on the system $B$, and its trace adjoint inserts the state $\tau_A$,
\begin{align}
 \E_A^\tau(Q_{AB})&\coloneqq\Tr_A[\tau_A^{1/2}Q_{AB}\tau_A^{1/2}],\label{eq:slice}\\
 (\E_A^\tau)^\dag(\sigma_B)&=\tau_A\otimes\sigma_B.\label{eq:slice-adjoint}
\end{align}
Their defining relation is
\begin{equation}\label{eq:slice-pairing}
 \Tr[\E_A^\tau(Q_{AB})\sigma_B]=\Tr[Q_{AB}(\E_A^\tau)^\dag(\sigma_B)]
 =\Tr[Q_{AB}(\tau_A\otimes\sigma_B)].
\end{equation}
For an effect $Q_{AB}$, the supremum of these probabilities over states $\sigma_B$ equals $\norm{\E_A^\tau(Q_{AB})}$. A type~II error probability at most $\eta$ for every $\sigma_B$ is therefore equivalent to $\E_A^\tau(Q_{AB})\le\eta I_B$.

Two optimizations describe this test. The first maximizes the signal acceptance probability subject to a bound on every type~II error probability. The second minimizes the worst-case type~II error probability subject to a bound on the type~I error probability. Its negative logarithm is the generalized hypothesis-testing mutual information of Cheng and Gao~\cite[Eq.~(68)]{CG}, also called the smoothed min-generalized mutual information by Ji \textit{et al.}~\cite[Eqs.~(9) and~(16)]{JGW}. For $0\le\eta,\varepsilon\le1$, define
\begin{align}
 G^\eta(\rho_{AB}\Vert\tau_A)&\coloneqq\max_{Q_{AB},\,\gamma}\!
 \left\{\gamma:
 \begin{gathered}
 0\le Q_{AB}\le I_{AB},\quad 0\le\gamma\le\Tr[Q_{AB}\rho_{AB}],\\[-2pt]
 \E_A^\tau(Q_{AB})\le\eta I_B
 \end{gathered}\right\},\label{eq:roc}\\
 b_\varepsilon(\rho_{AB}\Vert\tau_A)&\coloneqq\min_{Q_{AB}}
 \bigl\{\norm{\E_A^\tau(Q_{AB})}:0\le Q_{AB}\le I_{AB},\ \Tr[Q_{AB}\rho_{AB}]\ge1-\varepsilon\bigr\}.\label{eq:beta}
\end{align}
Both sides of the following equivalence express the existence of one effect meeting both constraints,
\begin{equation}\label{eq:roc-beta-equivalence}
 G^\eta(\rho_{AB}\Vert\tau_A)\ge1-\varepsilon
 \quad\Longleftrightarrow\quad b_\varepsilon(\rho_{AB}\Vert\tau_A)\le\eta .
\end{equation}
Thus either function determines the other, including at the endpoints. Both are semidefinite programs in finite dimensions. For $0\le\varepsilon<1$ the information quantity in bits and its relation to \eqref{eq:beta} are
\begin{align}
 I_H^\varepsilon(\rho_{AB}\Vert\tau_A)
   &\coloneqq\inf_{\sigma_B}D_H^\varepsilon(\rho_{AB}\Vert\tau_A\otimes\sigma_B),\label{eq:info}\\
 b_\varepsilon(\rho_{AB}\Vert\tau_A)&=2^{-I_H^\varepsilon(\rho_{AB}\Vert\tau_A)}.\label{eq:beta-info}
\end{align}
To see \eqref{eq:beta-info}, minimize the last trace in \eqref{eq:slice-pairing} over the effects in \eqref{eq:beta} and maximize over states $\sigma_B$. In finite dimensions both feasible sets are compact and convex, and the trace is bilinear. The minimax theorem then permits the interchange. For separable systems it suffices that the effect set is compact in the topology specified in Appendix~\ref{sec:separable}. Equation~\eqref{eq:dh} then gives \eqref{eq:beta-info}.

\section[The convex-split bound]{A binary testing bound for convex splitting}\label{sec:cover}
What would a successful test for a hidden pair imply about a test on one pair alone? We first answer for a uniformly chosen position. Lemma~\ref{lem:moment} shows how to extract a single-pair observable from a global measurement. Thresholds of that observable then furnish the tests in Theorem~\ref{thm:ordinary}. We subsequently use the same construction for two position distributions and for a bound in terms of one hypothesis-testing relative entropy.

\begin{theorem}[A uniformly hidden pair]\label{thm:ordinary}
For states $\rho_{AB}$ and $\tau_A$ and an integer $K\ge1$, let $\omega_{A^K B}$ be the uniformly hidden state in \eqref{eq:ordinary} and $\omega^0_{A^K B}=\tau_A^{\otimes K}\otimes\rho_B$ the independent state in \eqref{eq:background}. Then
\begin{equation}\label{eq:ordinary-curve}
 \frac12\norm{\omega_{A^K B}-\omega^0_{A^K B}}_1
 \le\int_0^1 G^{1/(1+4Kt^2)}(\rho_{AB}\Vert\tau_A)\,dt ,
\end{equation}
where $G^\eta(\rho_{AB}\Vert\tau_A)$ is defined in \eqref{eq:roc}.
\end{theorem}

Theorem~\ref{thm:ordinary} treats the standard convex-split experiment,
comparing a uniformly hidden pair with the independent preparation.
It is a special case of Theorem~\ref{thm:cover} below.
We prove it first to present the argument based on integrating binary
threshold tests in its simplest form.
The corresponding integral estimate for general position distributions
is subsequently used in Appendix~\ref{sec:gaussian-cover} to establish
the covering bound~\eqref{eq:third-cover} underlying
Theorem~\ref{thm:rate}.

The right side of \eqref{eq:ordinary-curve} lies in $[0,1]$ and is nonincreasing in $K$. Indeed, increasing the permitted type~II error probability $\eta$ enlarges the set of feasible tests in \eqref{eq:roc}, so $G^\eta(\rho_{AB}\Vert\tau_A)$ is nondecreasing in $\eta$. For each $t>0$, increasing $K$ decreases the tolerance $1/(1+4Kt^2)$, so the integrand cannot increase.

To make the dependence on $K$ explicit, fix $0<\eta<1$. Split the integral at $t=\sqrt{(1-\eta)/(4K\eta)}$, where its tolerance equals $\eta$. Below this threshold the integrand is at most one, and above it the integrand is at most $G^\eta(\rho_{AB}\Vert\tau_A)$. Consequently,
\begin{equation}\label{eq:ordinary-tolerance}
 \int_0^1G^{1/(1+4Kt^2)}(\rho_{AB}\Vert\tau_A)\,dt
 \le G^\eta(\rho_{AB}\Vert\tau_A)
      +\frac{1}{2\sqrt K}\sqrt{\frac{1-\eta}{\eta}} .
\end{equation}
If the splitting threshold exceeds one, \eqref{eq:ordinary-tolerance} follows from the integral being at most one. The first term is the greatest permitted single-pair signal acceptance probability. Quadrupling $K$ halves the second term. When $\operatorname{supp}\rho_A\subseteq\operatorname{supp}\tau_A$, with $\rho_A\coloneqq\Tr_B[\rho_{AB}]$, the quantity $G^\eta(\rho_{AB}\Vert\tau_A)$ tends to zero as $\eta$ decreases to zero, by \eqref{eq:roc-zero-limit}--\eqref{eq:roc-zero-support}. One can therefore choose $\eta$ to make the first term small and then increase $K$ to control the second. This shows that the upper bound itself vanishes under the support condition.

\subsection{Extracting a single-pair observable}\label{ssec:single-pair}
An observer receives $A^NB$ in one of the states $\Omega_{A^NB}(p)$ and $\Omega_{A^NB}(q)$ of \eqref{eq:mixture}. A measurement on all these systems can use correlations between the reference $B$ and every site. How can we recover its distinguishing advantage from an observable on just one pair $AB$? We leave one site open and prepare all others in their background states. Repeating this for each site isolates its contribution to the measurement. At their original sites, the centered contributions have vanishing cross terms after the background average. On a common pair $AB$, their weighted sum has the desired signal expectation. Its second moment will bound the type~II error probabilities of the threshold tests in Theorem~\ref{thm:ordinary}.

The orthogonality of centered contributions on a product state appears in quantum $U$-statistics~\cite[Lemma~5.5 and Theorem~5.6]{GB}, multipartite convex splitting~\cite[Lemma~5 and Remark~6]{CGB}, and telescoping estimates for covering~\cite{SenTel}. Here the bound on the second moment is an operator inequality on $B$, so it controls the expectation for every state $\sigma_B$ of the reference.

\begin{figure}[!t]
\centering
\begin{tikzpicture}[x=1cm,y=1cm,>=stealth,font=\small]
\definecolor{proofblue}{RGB}{36,69,101}
\definecolor{proofpale}{RGB}{227,239,247}
\node[anchor=west,font=\bfseries] at (0,0) {(a) Leave one site open};
\node[anchor=west,font=\bfseries] at (5.15,0) {(b) Keep the site labels};
\node[anchor=west,font=\bfseries] at (10.45,0) {(c) Relabel on one pair};
\foreach \x/\a in {0.5/A_1,1.4/A_2,2.3/A_3,3.2/B}{\node at (\x,-.7) {$\a$};}
\foreach \i/\y in {1/-1.35,2/-1.96,3/-2.57}{
 \foreach \j/\x in {1/.5,2/1.4,3/2.3}{
  \ifnum\i=\j\relax
   \node[draw=proofblue,fill=proofpale,minimum width=.77cm,minimum height=.44cm,inner sep=1pt] at (\x,\y) {$A_{\i}$};
  \else
   \node[draw=gray!65,minimum width=.77cm,minimum height=.44cm,inner sep=1pt] at (\x,\y) {$\tau_{A_{\j}}$};
  \fi}
 \node[draw=proofblue,fill=proofpale,minimum width=.68cm,minimum height=.44cm,inner sep=1pt] at (3.2,\y){$B$};}
\node[anchor=north,align=center,text width=4.7cm] at (1.92,-3.02)
 {Prepare the white sites in their background states. Subtract the all-background response.};
\draw[gray!45] (4.62,-.4)--(4.62,-4.32);
\node at (7.35,-1.27){$H^1_{A_1B},\quad H^2_{A_2B},\quad H^3_{A_3B}$};
\node at (7.35,-2.13){$\E^\tau_{A^3}(H^i_{A_iB} H^j_{A_jB})=0$};
\node at (7.35,-2.60){$(i\ne j)$};
\node[anchor=north,align=center,text width=4.72cm] at (7.33,-3.02)
 {Distinct sites have zero cross terms after the background average.};
\draw[gray!45] (9.96,-.4)--(9.96,-4.32);
\node at (12.73,-1.03){$H_{AB}=\tfrac13\sum_{i=1}^3 H^i_{AB}$};
\draw[->,black!65] (10.60,-2.13)--(15.10,-2.13);
\foreach \x in {10.95,11.43,12.15,12.53}{\fill[black!60] (\x,-2.13) circle (2.0pt);}
\foreach \x in {13.69,14.29}{\fill[proofblue] (\x,-2.13) circle (2.3pt);}
\draw[proofblue,thick] (13.0,-2.37)--(13.0,-1.75);
\node[below] at (13.0,-2.36){$t$};
\node[above,text=proofblue] at (14.08,-2.1){accept};
\node[anchor=north,align=center,text width=4.65cm] at (12.78,-3.02)
 {Average the three observables on $AB$. Accept eigenvalues greater than $t$.};
\end{tikzpicture}
\caption{One pair hidden uniformly among three sites, compared with the all-background preparation. In (a), a white box denotes preparation of that site in $\tau_A$, while a shaded box indicates a system left open in the effect. This gives $X^i_{A_iB}$ in \eqref{eq:XY}. Subtracting the all-background response gives $H^i_{A_iB}$. Panel (b) keeps each observable on its original site to bound the sum of its second moments by \eqref{eq:bessel}. Panel (c) instead takes the three centered observables from (a) and relabels them on a common pair. The second moment of their average controls the type~II error probability of each threshold test in \eqref{eq:threshold-prob}. The dots illustrate possible eigenvalues.}\label{fig:proof}
\end{figure}
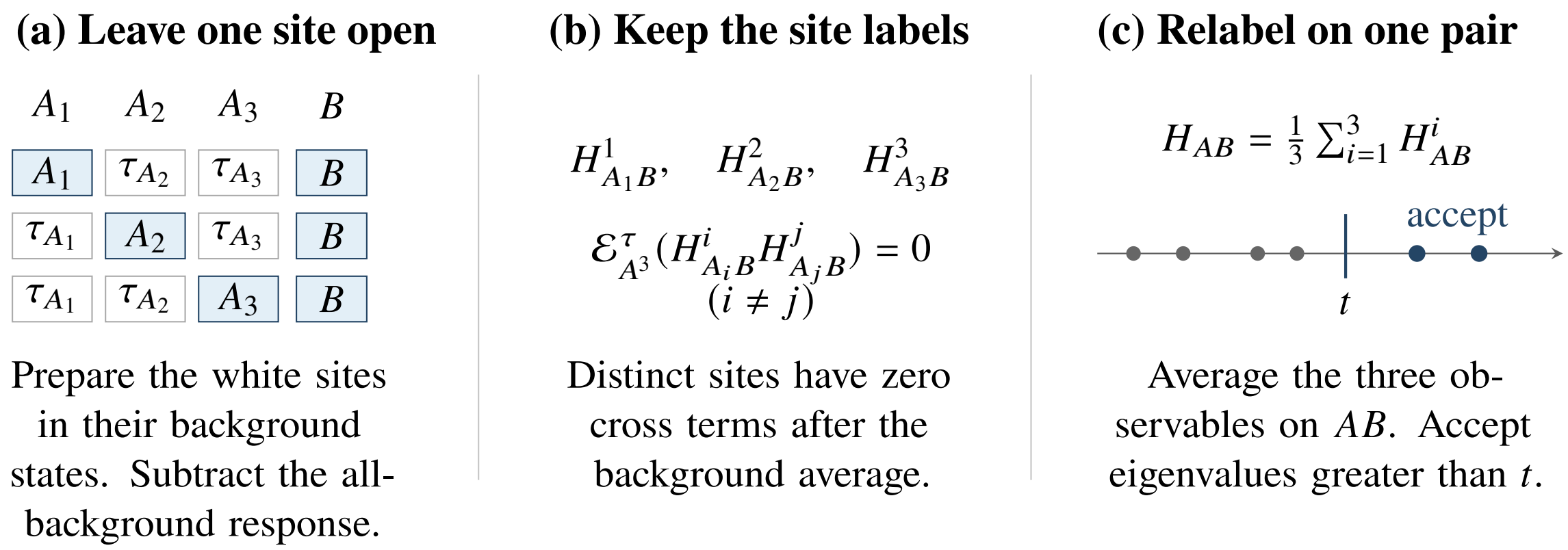

\Needspace{10\baselineskip}
\begin{lemma}[The global measurement on a single pair]\label{lem:moment}
Let $\rho_{AB},\tau_A$ be states, $N\in\mathbb N$, and $p,q$ be probability distributions on $\{0,\ldots,N\}$. Set $c\coloneqq\tfrac12\norm{p-q}_1$ and $v\coloneqq\norm{p_{1:N}-q_{1:N}}_2^2$ as in \eqref{eq:weights}--\eqref{eq:site-norm}. For the mixtures in \eqref{eq:mixture} and the background average in \eqref{eq:slice}, there exists a Hermitian observable $H_{AB}$ satisfying
\begin{align}
 \Tr[H_{AB}\rho_{AB}]&=\tfrac12\norm{\Omega_{A^N B}(p)-\Omega_{A^N B}(q)}_1,\label{eq:moment-mean}\\
 -cI_{AB}&\le H_{AB}\le cI_{AB},\label{eq:moment-range}\\
 \E_A^\tau(H_{AB})&=0,\label{eq:moment-center}\\
 \E_A^\tau(H_{AB}^2)&\le\tfrac{v}{4} I_B.\label{eq:moment}
\end{align}
\end{lemma}
\begin{proof}
Figure~\ref{fig:proof} illustrates the construction. Choose $W_{A^NB}$ to be the projector onto the positive spectrum of $\Omega_{A^NB}(p)-\Omega_{A^NB}(q)$. It attains the maximum in \eqref{eq:distance}. To compute its acceptance probability when some sites are prepared in the background state, let $A_S$ consist of the sites indexed by $S\subseteq\{1,\ldots,N\}$, let $A_{S^c}$ comprise the remaining sites, and set $\tau_{A_S}\coloneqq\bigotimes_{j\in S}\tau_{A_j}$ with empty product $1$. The background average \eqref{eq:slice} and its trace adjoint \eqref{eq:slice-adjoint} act as
\begin{align}
 \E_{A_S}^\tau(Z_{A^NB})&\coloneqq\Tr_{A_S}\!\left[\tau_{A_S}^{1/2}Z_{A^NB}\tau_{A_S}^{1/2}\right],\label{eq:multi-slice}\\
 (\E_{A_S}^\tau)^{\dag}(\eta_{A_{S^c}B})&=\tau_{A_S}\otimes\eta_{A_{S^c}B}.\label{eq:multi-adjoint}
\end{align}
For every retained state $\eta_{A_{S^c}B}$, the defining trace pairing is
\begin{equation}\label{eq:local-acceptance}
 \Tr\!\left[W_{A^NB}(\E_{A_S}^\tau)^{\dag}(\eta_{A_{S^c}B})\right]
 =\Tr\!\left[\E_{A_S}^\tau(W_{A^NB})\,\eta_{A_{S^c}B}\right].
\end{equation}
Let $A_{\ne i}$ denote the sites other than $A_i$. With $A_i$ and $B$ left open we have the effect $X^i_{A_iB}$ below, and with only $B$ open we have $Y_B$. Subtract their responses to isolate the change associated with $A_i$. Relabel $A_i$ as $A$ when summing the resulting observables on a common pair.
\begin{align}
 X^i_{A_iB}&\coloneqq\E_{A_{\ne i}}^\tau(W_{A^NB}),\label{eq:XY}\\
 Y_B&\coloneqq\E_{A^N}^\tau(W_{A^NB}),\label{eq:Y}\\
 H^i_{A_iB}&\coloneqq X^i_{A_iB}-Y_B,\label{eq:site-definition}\\
 H_{AB}&\coloneqq\sum_{i=1}^N(p_i-q_i)H^i_{AB}.\label{eq:Hi}
\end{align}

\emph{The measurement advantage.}
The weights obey $p_0-q_0=-\sum_{i=1}^N(p_i-q_i)$ since $p$ and $q$ each sum to one. Expanding \eqref{eq:mixture} and isolating the independent preparation at $i=0$ gives
\begin{align}
 &\tfrac12\norm{\Omega_{A^NB}(p)-\Omega_{A^NB}(q)}_1\notag \\
 & =\Tr\!\left[W_{A^NB}\bigl(\Omega_{A^NB}(p)-\Omega_{A^NB}(q)\bigr)\right]\label{eq:optimal-global}\\
 &=\sum_{i=0}^N(p_i-q_i)\Tr[W_{A^NB}\omega^i_{A^NB}]\label{eq:all-placements}\\
 &=(p_0-q_0)\Tr[W_{A^NB}\omega^0_{A^NB}]
     +\sum_{i=1}^N(p_i-q_i)\Tr[W_{A^NB}\omega^i_{A^NB}]\label{eq:zero-separate}\\
 &=\sum_{i=1}^N(p_i-q_i)\left(\Tr[W_{A^NB}\omega^i_{A^NB}]-\Tr[W_{A^NB}\omega^0_{A^NB}]\right)\label{eq:expand-placements}\\
 &=\sum_{i=1}^N(p_i-q_i)\left(\Tr\!\left[W_{A^NB}(\E_{A_{\ne i}}^\tau)^{\dag}(\rho_{A_iB})\right]
               -\Tr\!\left[W_{A^NB}(\E_{A^N}^\tau)^{\dag}(\rho_B)\right]\right)\label{eq:use-adjoint}\\
 &=\sum_{i=1}^N(p_i-q_i)\left(\Tr[X^i_{A_iB}\rho_{A_iB}]-\Tr[Y_B\rho_B]\right)\label{eq:paired-tests}\\
 &=\Tr\!\left[\left(\sum_{i=1}^N(p_i-q_i)(X^i_{AB}-Y_B)\right)\rho_{AB}\right]\label{eq:relabel-trace}\\
 &=\Tr[H_{AB}\rho_{AB}].\label{eq:advantage}
\end{align}
Equation~\eqref{eq:use-adjoint} inserts the placement states \eqref{eq:placements}--\eqref{eq:background} using \eqref{eq:multi-adjoint}. Equation~\eqref{eq:paired-tests} uses \eqref{eq:local-acceptance}, and \eqref{eq:relabel-trace} uses $\rho_B=\Tr_A[\rho_{AB}]$.

Averaging the last site in \eqref{eq:XY} prepares the full background in \eqref{eq:Y}. Thus
\begin{align}
 \E_{A_i}^\tau(X^i_{A_iB})&=Y_B,\label{eq:site-response}\\
 \E_{A_i}^\tau(H^i_{A_iB})&=0.\label{eq:site-centered}
\end{align}
Linearity gives \eqref{eq:moment-center}. For \eqref{eq:moment-range}, write $H_{AB}=\sum_{i=1}^N(p_i-q_i)X^i_{AB}+(p_0-q_0)Y_B$. Every term multiplies an effect by a real coefficient. By \eqref{eq:weights}, the positive coefficients sum to $\tfrac12\norm{p-q}_1$, as do the absolute values of the negative ones.

\emph{Second moments on the original sites.}
For operators $C_{A_{S^c}B},D_{A_{S^c}B}$ on the retained systems, the partial trace in \eqref{eq:multi-slice} gives
\begin{equation}\label{eq:slice-bimodule}
 \E_{A_S}^\tau(C_{A_{S^c}B}Z_{A^NB}D_{A_{S^c}B})
 =C_{A_{S^c}B}\,\E_{A_S}^\tau(Z_{A^NB})\,D_{A_{S^c}B}.
\end{equation}
Each retained operator stays on its original side because both may act on $B$. Center the global effect about its all-background response as follows.
\begin{equation}\label{eq:embeds}
 W^\circ_{A^NB}\coloneqq W_{A^NB}-I_{A^N}\otimes Y_B .
\end{equation}
For $i\ne j$, average first over $A_i$ while holding $H^j_{A_jB}$ on the right. Equations~\eqref{eq:slice-bimodule} and \eqref{eq:site-centered} give
\begin{align}
 \E_{A^N}^\tau(H^i_{A_iB}H^j_{A_jB})
 &=\E_{A_{\ne i}}^\tau\!\left(\E_{A_i}^\tau(H^i_{A_iB})H^j_{A_jB}\right)\label{eq:orth-before}\\
 &=0.\label{eq:cross-zero}
\end{align}
For a product involving $W^\circ_{A^NB}$, average instead over $A_{\ne i}$ first. Equations~\eqref{eq:XY}--\eqref{eq:site-definition} give $\E_{A_{\ne i}}^\tau(W^\circ_{A^NB})=H^i_{A_iB}$. Consequently
\begin{align}
 \E_{A^N}^\tau(W^\circ_{A^NB}H^i_{A_iB})&=\E_{A_i}^\tau((H^i_{A_iB})^2),\label{eq:orth}\\
 \E_{A^N}^\tau(H^i_{A_iB}W^\circ_{A^NB})&=\E_{A_i}^\tau((H^i_{A_iB})^2).\label{eq:orth-reverse}
\end{align}
The second identity follows by taking adjoints. We now subtract the site contributions from the centered global effect and expand the squared residual. In the next display all site sums run from $1$ to $N$. The off-diagonal terms vanish by \eqref{eq:cross-zero}. Each of the two mixed products in \eqref{eq:orth}--\eqref{eq:orth-reverse} gives the same diagonal second moment.
\begin{align}
 0&\le\E_{A^N}^\tau\!\left(\left[W^\circ_{A^NB}-\sum_i H^i_{A_iB}\right]^2\right)\label{eq:residual-positive}\\
 &=\E_{A^N}^\tau((W^\circ_{A^NB})^2)
 -\sum_i\E_{A^N}^\tau(W^\circ_{A^NB}H^i_{A_iB}+H^i_{A_iB}W^\circ_{A^NB})\notag\\*
 &\hspace{2em}+\sum_i\E_{A_i}^\tau((H^i_{A_iB})^2)
 +\sum_{i\ne j}\E_{A^N}^\tau(H^i_{A_iB}H^j_{A_jB})\label{eq:residual-all}\\
 &=\E_{A^N}^\tau((W^\circ_{A^NB})^2)
 -2\sum_i\E_{A_i}^\tau((H^i_{A_iB})^2)+\sum_i\E_{A_i}^\tau((H^i_{A_iB})^2).\label{eq:residual-diagonal}
\end{align}
Expanding the first term with \eqref{eq:embeds} and using \eqref{eq:Y} and \eqref{eq:slice-bimodule} gives
\begin{align}
 \E_{A^N}^\tau((W^\circ_{A^NB})^2)&=\E_{A^N}^\tau(W_{A^NB}^2)-\E_{A^N}^\tau(W_{A^NB})Y_B-Y_B\E_{A^N}^\tau(W_{A^NB})+Y_B^2\\
 &=\E_{A^N}^\tau(W_{A^NB}^2)-Y_B^2.\label{eq:centered-variance}
\end{align}
Thus the two negative diagonal sums and one positive diagonal sum in \eqref{eq:residual-diagonal} leave
\begin{equation}\label{eq:residual}
 0\le\E_{A^N}^\tau(W_{A^NB}^2)-Y_B^2-\sum_{i=1}^N\E_{A_i}^\tau((H^i_{A_iB})^2).
\end{equation}
The sum of the one-site second moments is therefore bounded by the variance of the joint effect under the background preparation.
\begin{align}
 \sum_{i=1}^N\E_A^\tau((H^i_{AB})^2)&\le\E_{A^N}^\tau(W_{A^NB}^2)-Y_B^2\label{eq:bessel-start}\\
 &\le Y_B-Y_B^2\label{eq:bessel-variance}\\
 &\le\tfrac14 I_B.\label{eq:bessel}
\end{align}
In \eqref{eq:bessel-start}, relabeling $A_i$ as $A$ leaves the resulting operator on $B$ unchanged. Equation~\eqref{eq:bessel-variance} uses $W_{A^NB}^2\le W_{A^NB}$ and the definition of $Y_B$ in \eqref{eq:Y}. For \eqref{eq:bessel}, the scalar inequality $y-y^2=1/4-(y-1/2)^2\le1/4$ holds on the spectrum of $Y_B$.

\emph{Second moment after relabeling on a common pair.}
When the $H^i_{AB}$ all act on $AB$, their cross terms can survive. Apply the triangle inequality to the vectors $H^i_{AB}|\psi\rangle_{AB}$ and then the ordinary scalar Cauchy--Schwarz inequality to the two sequences of nonnegative numbers on the right:
\begin{align}
 \norm{H_{AB}|\psi\rangle_{AB}}
 &=\left\|\sum_{i=1}^N(p_i-q_i)H^i_{AB}|\psi\rangle_{AB}\right\|\label{eq:common-vector}\\
 &\le\sum_{i=1}^N |p_i-q_i|\,\norm{H^i_{AB}|\psi\rangle_{AB}}\label{eq:common-triangle}\\
 &\le\left(\sum_{i=1}^N(p_i-q_i)^2\right)^{1/2}
       \left(\sum_{i=1}^N\norm{H^i_{AB}|\psi\rangle_{AB}}^2\right)^{1/2}.\label{eq:common-cs}
\end{align}
Square both sides. Since the resulting inequality holds for every $|\psi\rangle_{AB}$ and the observables are Hermitian, it holds as a quadratic-form inequality:
\begin{equation}\label{eq:common-operator}
 H_{AB}^2\le\norm{p_{1:N}-q_{1:N}}_2^2\sum_{i=1}^N(H^i_{AB})^2.
\end{equation}
Apply the positive linear average $\E_A^\tau$ to \eqref{eq:common-operator} and insert \eqref{eq:bessel} to obtain \eqref{eq:moment}.
\end{proof}

\subsection{From the observable to a binary decision}
We can now prove the integral bound in Theorem~\ref{thm:ordinary}. Lemma~\ref{lem:moment} expresses the global distinguishing advantage as the signal expectation of $H_{AB}$. For a uniformly hidden position compared with the independent preparation, $p_i=1/K$ and $q_i=0$ at the $K$ physical sites, while $p_0=0$ and $q_0=1$. The second moment in \eqref{eq:moment} uses only the physical-site differences from \eqref{eq:site-norm}, giving
\begin{equation}\label{eq:uniform-site-variance}
 \E_A^\tau(H_{AB}^2)\le\frac14\sum_{i=1}^K(1/K)^2 I_B
 =\frac{1}{4K}I_B .
\end{equation}
Thus the factor $1/K$ is the sum of $K$ squared weights $1/K^2$ after the cancellation and Cauchy--Schwarz steps of Lemma~\ref{lem:moment}. A threshold turns this second-moment bound into a bound on the type~II error probability of a binary test.

\begin{proof}[Proof of Theorem~\ref{thm:ordinary}]
For the uniform and independent placements in \eqref{eq:ordinary} and \eqref{eq:background}, respectively, the spectrum of $H_{AB}$ lies in $[-1,1]$ by \eqref{eq:moment-range}. Averaging over $A$ in the state $\tau_A$ gives $\E_A^\tau(H_{AB})=0$ and $\E_A^\tau(H_{AB}^2)\le I_B/(4K)$ by \eqref{eq:moment-center}--\eqref{eq:moment}. Let $T^t_{AB}\coloneqq\{H_{AB}>tI_{AB}\}$ be the spectral projector onto eigenvalues of $H_{AB}$ greater than $t$. For all $t>0$ and $r\ge0$, the scalar inequality $\boldsymbol1_{\{x>t\}}\le(x+r)^2/(t+r)^2$ holds because $x+r>t+r$ on the acceptance interval and the square is nonnegative elsewhere. Apply it to each eigenvalue, then apply the positive average~\eqref{eq:slice}:
\begin{align}
 \E_A^\tau(T^t_{AB})
 &\le\frac{\E_A^\tau((H_{AB}+rI_{AB})^2)}{(t+r)^2}\label{eq:threshold-square}\\
 &=\frac{\E_A^\tau(H_{AB}^2)+2r\E_A^\tau(H_{AB})+r^2 I_B}{(t+r)^2}\label{eq:threshold-expand}\\
 &\le\frac{1/(4K)+r^2}{(t+r)^2} I_B.\label{eq:threshold}
\end{align}
The derivative of the scalar coefficient with respect to $r$ is $2(rt-1/(4K))/(t+r)^3$, so its minimum occurs at $r=1/(4Kt)$. Taking the expectation on an arbitrary state $\sigma_B$ now gives the following bound on the type~II error probability:
\begin{align}
 \Tr[T^t_{AB}(\tau_A\otimes\sigma_B)]&=\Tr[\E_A^\tau(T^t_{AB})\sigma_B]\label{eq:threshold-pair}\\
 &\le\frac{1}{1+4Kt^2}.\label{eq:threshold-prob}
\end{align}
At $t=0$ the same bound holds because $T^0_{AB}$ is an effect.

Write $H_{AB}=\sum_x x P^x_{AB}$ in terms of its distinct eigenvalues and spectral projectors. For $x\in[-1,1]$, write $x_+\coloneqq\max\{x,0\}$. The integral of the acceptance decision is
\begin{align}
 \int_0^1\boldsymbol1_{\{x>t\}}\,dt
 &=\begin{cases}0,&-1\le x\le0,\\ \int_0^x 1\,dt,&0<x\le1,\end{cases}\label{eq:threshold-length}\\
 &=x_+.\label{eq:threshold-length-value}
\end{align}
It follows by summing over the spectrum that
\begin{align}
 \int_0^1 T^t_{AB}\,dt
 &=\sum_x\left(\int_0^1\boldsymbol1_{\{x>t\}}\,dt\right)P^x_{AB}\label{eq:threshold-spectral}\\
 &=\sum_x x_+P^x_{AB}\label{eq:positive-decomposition}\\
 &=(H_{AB})_+.\label{eq:threshold-positive}
\end{align}
By \eqref{eq:advantage}, the operator inequality $H_{AB}\le(H_{AB})_+$, and the type~II error probability constraint \eqref{eq:threshold-prob}, we conclude that
\begin{align}
 \tfrac12\norm{\omega_{A^K B}-\omega^0_{A^K B}}_1
 &=\Tr[H_{AB}\rho_{AB}]\label{eq:threshold-advantage}\\
 &\le\Tr[(H_{AB})_+\rho_{AB}]\label{eq:threshold-positive-trace}\\
 &=\int_0^1\Tr[T^t_{AB}\rho_{AB}]\,dt\label{eq:threshold-layer}\\
 &\le\int_0^1 G^{1/(1+4Kt^2)}(\rho_{AB}\Vert\tau_A)\,dt.\label{eq:threshold-int}
\end{align}
In the last line, each $T^t_{AB}$ is a feasible test for the optimization in \eqref{eq:roc}.
\end{proof}

\subsection{Comparing two position distributions}
For the mixtures $\Omega_{A^NB}(p)$ and $\Omega_{A^NB}(q)$ in \eqref{eq:mixture}, the hidden position can have a different distribution in each preparation. For example, it might be uniform within one row of a table in one preparation and uniform over the whole table in the other. Lemma~\ref{lem:moment} expresses the distinguishing advantage as the mean of one observable $H_{AB}$ on the signal. There are two ways to turn this observable into a test. The projector from \eqref{eq:threshold-square} accepts each spectral value $x$ above a chosen threshold $t$. Alternatively, we can let the probability of acceptance increase with the amount by which $x$ exceeds $t$.

For this second test, let $a\coloneqq\tfrac12\norm{p-q}_1$ and $s\coloneqq\norm{p_{1:N}-q_{1:N}}_2^2$ as in \eqref{eq:weights}--\eqref{eq:site-norm}. The spectrum of $H_{AB}$ lies in $[-a,a]$ by \eqref{eq:moment-range}. Averaging over $A$ in the background state $\tau_A$ leaves an operator on $B$, as in \eqref{eq:slice}. Lemma~\ref{lem:moment} gives $\E_A^\tau(H_{AB})=0$ and $\E_A^\tau(H_{AB}^2)\le(s/4)I_B$ in \eqref{eq:moment-center}--\eqref{eq:moment}. By \eqref{eq:slice-pairing}, measuring $H_{AB}$ on any independent pair $\tau_A\otimes\sigma_B$ thus gives an outcome with mean zero and second moment at most $s/4$.

At a threshold $0\le t<a$, accept $x$ with probability $(x-t)_+/(a-t)$. On the signal, the inequality $x\le t+(x-t)_+$ relates acceptance to the mean we want to bound. On an independent pair, we need an upper bound on the mean of $(x-t)_+$. The proof obtains one by placing a quadratic above $(x-t)_+$: its expectation depends only on the known mean and second moment. For larger $t$, using the additional information $x\le a$ gives a smaller bound. The resulting upper bound on the mean excess, for $a>0$ and $0<s\le2a^2$, is
\begin{equation}\label{eq:hinge}
 h(t;a,s)\coloneqq\begin{cases}
  (\sqrt{t^2+s/4}-t)/2, &0\le t\le(a^2-s/4)/(2a),\\[2pt]
  (s/4)(a-t)/(a^2+s/4), &(a^2-s/4)/(2a)<t<a.
 \end{cases}
\end{equation}
Dividing \eqref{eq:hinge} by $a-t$ gives the type~II error probability allowed for this test. Theorem~\ref{thm:cover} below takes the better of the bound from one such test and the integral bound from all the threshold projectors.

\begin{theorem}[Two distributions for the hidden position]\label{thm:cover}
Let $N\in\mathbb N$, let $\rho_{AB}$ and $\tau_A$ be states, and let $p$ and $q$ be probability distributions on $\{0,\ldots,N\}$. The mixtures in \eqref{eq:mixture} and the norms in \eqref{eq:weights}--\eqref{eq:site-norm} satisfy
\begin{equation}\label{eq:two-distances}
 \tfrac12\norm{\Omega_{A^N B}(p)-\Omega_{A^N B}(q)}_1
 \le \mathcal C\!\left(\tfrac12\norm{p-q}_1,\norm{p_{1:N}-q_{1:N}}_2^2;\rho_{AB},\tau_A\right).
\end{equation}
Here $\mathcal C(0,0;\rho_{AB},\tau_A)\coloneqq0$. For $a>0$ and $0<s\le 2a^2$, the testing curve in \eqref{eq:roc} and the excess bound in \eqref{eq:hinge} determine
\begin{multline}\label{eq:layer}
 \mathcal C(a,s;\rho_{AB},\tau_A)\coloneqq\\
 \min\biggl\{
 \inf_{0\le t<a}\!\left[t+(a-t)G^{h(t;a,s)/(a-t)}(\rho_{AB}\Vert\tau_A)\right],\int_0^a G^{s/(s+4t^2)}(\rho_{AB}\Vert\tau_A)\,dt\biggr\}.
\end{multline}
\end{theorem}
\begin{proof}
When $p=q$ the two mixtures coincide. For $p\ne q$ set $a\coloneqq\tfrac12\norm{p-q}_1$ and $s\coloneqq\norm{p_{1:N}-q_{1:N}}_2^2$. Some positive-index weight must differ, since $p_0$ and $q_0$ are determined by the other weights. Thus $s>0$. To check the upper limit $2a^2$ in the domain of \eqref{eq:hinge}, separate the positive and negative differences. Within either group the sum of their squares is at most the square of the sum of their absolute values. Equation~\eqref{eq:weights} then gives
\begin{align}
 s&\le\sum_{i=0}^N(p_i-q_i)^2\\
 &\le\left(\sum_{i=0}^N(p_i-q_i)_+\right)^2+
      \left(\sum_{i=0}^N(q_i-p_i)_+\right)^2=2a^2.\label{eq:placement-domain}
\end{align}
Take the specific observable $H_{AB}$ given by Lemma~\ref{lem:moment} for $p,q,\rho_{AB},\tau_A$. By \eqref{eq:moment-mean}, the left side of \eqref{eq:two-distances} is exactly $\Tr[H_{AB}\rho_{AB}]$. Its spectrum lies in $[-a,a]$, and \eqref{eq:moment-center}--\eqref{eq:moment} give
\begin{align}
 \E_A^\tau(H_{AB})&=0,\label{eq:two-center}\\
 \E_A^\tau(H_{AB}^2)&\le (s/4) I_B.\label{eq:two-moment}
\end{align}

\emph{One test at a chosen threshold.}
For $0\le t<a$, the acceptance rule described above gives the effect
\begin{equation}\label{eq:excess-effect}
 Q^t_{AB}\coloneqq(H_{AB}-tI_{AB})_+/(a-t).
\end{equation}
Indeed, its eigenvalues lie in $[0,1]$, and the scalar inequality $x\le t+(x-t)_+$ gives
\begin{equation}\label{eq:excess-signal}
 \Tr[H_{AB}\rho_{AB}]\le t+(a-t)\Tr[Q^t_{AB}\rho_{AB}].
\end{equation}
It remains to bound the probability that \eqref{eq:excess-effect} accepts an independent pair. We first use only \eqref{eq:two-center} and \eqref{eq:two-moment}. The function $(x-t)_+$ is zero up to $t$ and has slope one above $t$. Choose a parabola that vanishes at $t-r$ and is tangent to $x-t$ at $t+r$. Its slope at $t+r$ fixes the coefficient $1/(4r)$. For all $r>0$ this gives
\begin{equation}\label{eq:excess-parabola}
 (x-t)_+\le\frac{(x-t+r)^2}{4r}.
\end{equation}
For $x\le t$ the left side is zero. For $x\ge t$, subtracting $x-t$ from the right side leaves $(x-t-r)^2/(4r)$. Apply \eqref{eq:excess-parabola} by spectral calculus and average. The cross term vanishes by \eqref{eq:two-center}, so
\begin{align}
 \E_A^\tau((H_{AB}-tI_{AB})_+)
 &\le \frac{\E_A^\tau(H_{AB}^2)+(r-t)^2 I_B}{4r}\label{eq:excess-jensen}\\
 &\le \frac{1}{4}\left(r+\frac{t^2+s/4}{r}-2t\right)I_B .
 \label{eq:excess-r}
\end{align}
The arithmetic--geometric mean inequality shows that \eqref{eq:excess-r} is smallest at $r=\sqrt{t^2+s/4}$. Substitution yields
\begin{equation}\label{eq:excess}
 \E_A^\tau((H_{AB}-tI_{AB})_+)\le\tfrac12(\sqrt{t^2+s/4}-t)I_B.
\end{equation}

The quadratic in \eqref{eq:excess-parabola} touches $x-t$ at $x=t+r$. With the minimizing $r$, this point exceeds the largest allowed spectral value $a$ exactly when $t>(a^2-s/4)/(2a)$. In that range, use a quadratic that agrees with $x-t$ at $x=a$. For $u\ge0$, consider $(a-t)(x+u)^2/(a+u)^2$. It is nonnegative below $t$. On $[t,a]$ its derivative is at most $2(a-t)/(a+u)$, so the quadratic lies above $x-t$ throughout this interval whenever $2(a-t)\le a+u$. Its background expectation is at most $(a-t)(s/4+u^2)/(a+u)^2$ by \eqref{eq:two-center}--\eqref{eq:two-moment}. Differentiating the last ratio gives
\begin{equation}
 \frac{d}{du}\frac{s/4+u^2}{(a+u)^2}=\frac{2(au-s/4)}{(a+u)^3}.
\end{equation}
Its minimizing choice $u=s/(4a)$ satisfies the derivative condition precisely for $t\ge(a^2-s/4)/(2a)$. Consequently,
\begin{align}
 (x-t)_+&\le (a-t)\frac{(x+s/(4a))^2}{(a+s/(4a))^2},\qquad -a\le x\le a,\label{eq:quadratic}\\
 \E_A^\tau((H_{AB}-tI_{AB})_+)&\le
 (a-t)\frac{s/4+(s/(4a))^2}{(a+s/(4a))^2}I_B
 =\frac{(s/4)(a-t)}{a^2+s/4}I_B.\label{eq:excess-endpoint}
\end{align}
Equations~\eqref{eq:excess} and \eqref{eq:excess-endpoint} give the two expressions in \eqref{eq:hinge}. They agree at $t=(a^2-s/4)/(2a)$, where both equal $s/(8a)$. For every independent reference state $\sigma_B$, the trace pairing \eqref{eq:slice-pairing} now gives
\begin{align}
 \Tr[Q^t_{AB}(\tau_A\otimes\sigma_B)]
 &=\frac{\Tr[\E_A^\tau((H_{AB}-tI_{AB})_+)\sigma_B]}{a-t}\\
 &\le \frac{h(t;a,s)}{a-t}.\label{eq:excess-type-two}
\end{align}
The tolerance $h(t;a,s)/(a-t)$ is in $[0,1]$: when $t+\sqrt{t^2+s/4}\le a$, \eqref{eq:hinge} gives $h(t;a,s)\le(a-2t)/2\le a-t$. For the other thresholds, the ratio is $(s/4)/(a^2+s/4)<1$. Thus $Q^t_{AB}$ is feasible for \eqref{eq:roc} at this tolerance. Substitution into \eqref{eq:excess-signal} bounds the trace distance by $t+(a-t)G^{h(t;a,s)/(a-t)}(\rho_{AB}\Vert\tau_A)$ for every $0\le t<a$. Taking the infimum proves the first estimate in \eqref{eq:layer}.

\emph{Integrating the threshold projectors.}
For $0<t\le a$, use $T^t_{AB}\coloneqq\{H_{AB}>tI_{AB}\}$. In \eqref{eq:threshold-square}--\eqref{eq:threshold}, insert the moment bound $s/4$ from \eqref{eq:two-moment} and choose $r=s/(4t)$. This gives
\begin{equation}
 \E_A^\tau(T^t_{AB})\le\frac{s/4+r^2}{(t+r)^2}I_B=\frac{s}{s+4t^2}I_B.
\end{equation}
At $t=0$ the same conclusion follows from $T^0_{AB}\le I_{AB}$. Since the spectrum of $H_{AB}$ lies in $[-a,a]$, integrating $T^t_{AB}$ from $0$ to $a$ gives $(H_{AB})_+$ as in \eqref{eq:threshold-length}--\eqref{eq:threshold-positive}. Therefore
\begin{align}
 \Tr[H_{AB}\rho_{AB}]&\le\int_0^a\Tr[T^t_{AB}\rho_{AB}]\,dt\\
 &\le \int_0^a G^{s/(s+4t^2)}(\rho_{AB}\Vert\tau_A)\,dt.
\end{align}
Both estimates bound the same distinguishing advantage \eqref{eq:moment-mean}. Taking their minimum proves \eqref{eq:two-distances}.
\end{proof}

\subsection{Choosing the number of positions}\label{ssec:scalar}
The following corollary specializes the positive-part argument in the proof of Theorem~\ref{thm:cover} to a uniformly hidden position. Its input is the moment estimate \eqref{eq:excess}, together with the observable supplied by Lemma~\ref{lem:moment}. To obtain a sufficient condition on $K$ in terms of the hypothesis-testing mutual information in \eqref{eq:info}, fix a target normalized trace distance $\varepsilon$. Choose $0\le\delta<\varepsilon$ as a threshold for the observable from Lemma~\ref{lem:moment}. It suffices to bound the signal expectation of $(H_{AB}-\delta I_{AB})_+$ by $\varepsilon-\delta$. This explains the choice $1-\varepsilon+\delta$ for the type~I error probability tolerance in \eqref{eq:beta}.

\begin{corollary}[A sufficient number of positions]\label{cor:scalar}
Let $\rho_{AB},\tau_A$ be states, $0<\varepsilon<1$, and $0\le\delta<\varepsilon$. With the definitions in \eqref{eq:beta}--\eqref{eq:beta-info}, suppose that
\begin{equation}\label{eq:scalar-beta}
 b\coloneqq b_{1-\varepsilon+\delta}(\rho_{AB}\Vert\tau_A)
 =2^{-I_H^{1-\varepsilon+\delta}(\rho_{AB}\Vert\tau_A)}>0.
\end{equation}
For the uniformly hidden and independent states in \eqref{eq:ordinary} and \eqref{eq:background}, every positive integer $K$ satisfies
\begin{equation}\label{eq:clean}
 K\ge\frac{1}{16b(\delta+b)}
 \quad\Longrightarrow\quad
 \frac12\norm{\omega_{A^K B}-\omega^0_{A^K B}}_1\le\varepsilon .
\end{equation}
For $\delta>0$, a sufficient condition in bits is
\begin{equation}\label{eq:scalar-bits}
 \log_2 K\ge I_H^{1-\varepsilon+\delta}(\rho_{AB}\Vert\tau_A)+\log_2(1/\delta)-4 .
\end{equation}
For $\delta=0$, \eqref{eq:clean} gives the sufficient condition
\begin{equation}\label{eq:scalar-zero}
 K\ge\frac{1}{16b_{1-\varepsilon}(\rho_{AB}\Vert\tau_A)^2}.
\end{equation}
\end{corollary}
\begin{proof}
For the uniformly hidden and independent states in \eqref{eq:ordinary} and \eqref{eq:background}, choose the observable $H_{AB}$ from Lemma~\ref{lem:moment}. Since $H_{AB}\le I_{AB}$ by \eqref{eq:moment-range}, its positive part above the threshold $\delta$ defines an effect $Q_{AB}\coloneqq(H_{AB}-\delta I_{AB})_+$. The scalar inequality $x\le\delta+(x-\delta)_+$ and \eqref{eq:moment-mean} give
\begin{align}
 \tfrac12\norm{\omega_{A^KB}-\omega^0_{A^KB}}_1&=\Tr[H_{AB}\rho_{AB}]\\
 &\le\delta+\Tr[Q_{AB}\rho_{AB}].\label{eq:scalar-target}
\end{align}
Thus it suffices to show that $Q_{AB}$ accepts the signal with probability at most $\varepsilon-\delta$. The testing probability $b$ in \eqref{eq:scalar-beta} lets us establish this by bounding the probability that $Q_{AB}$ accepts an independent pair.

To see why, let $T_{AB}$ be any effect satisfying $\norm{\E_A^\tau(T_{AB})}\le b$ and write $\gamma\coloneqq\Tr[T_{AB}\rho_{AB}]$. If $\gamma\ge\varepsilon-\delta$, the rescaled effect $(\varepsilon-\delta)T_{AB}/\gamma$ accepts the signal with probability exactly $\varepsilon-\delta$ and is feasible in the minimization \eqref{eq:beta} defining $b$. Hence
\begin{equation}
 b\le\frac{\varepsilon-\delta}{\gamma}\norm{\E_A^\tau(T_{AB})}
   \le\frac{\varepsilon-\delta}{\gamma}b .
\end{equation}
Because $b>0$, it follows that $\gamma\le\varepsilon-\delta$. Together with the case $\gamma<\varepsilon-\delta$, this proves
\begin{equation}\label{eq:scalar-rescale}
 \norm{\E_A^\tau(T_{AB})}\le b\quad\Longrightarrow\quad\Tr[T_{AB}\rho_{AB}]\le\varepsilon-\delta .
\end{equation}

For the particular effect $T_{AB}=Q_{AB}$ in \eqref{eq:scalar-target}, the left side of the implication \eqref{eq:scalar-rescale} is now the condition we need to verify. By \eqref{eq:slice-pairing}, $\norm{\E_A^\tau(Q_{AB})}$ is the supremum of its acceptance probabilities on $\tau_A\otimes\sigma_B$ over all states $\sigma_B$. For uniform placement the weights in \eqref{eq:weights}--\eqref{eq:site-norm} give $a=\tfrac12\norm{p-q}_1=1$ and $s=\norm{p_{1:K}-q_{1:K}}_2^2=1/K$. Setting $t=\delta$ in the moment estimate \eqref{eq:excess} therefore gives
\begin{equation}\label{eq:scalar-excess}
 \norm{\E_A^\tau(Q_{AB})}\le\tfrac12\bigl(\sqrt{\delta^2+1/(4K)}-\delta\bigr).
\end{equation}
The choice $K\ge1/[16b(\delta+b)]$ in \eqref{eq:clean} makes the right side at most $b$. Indeed,
\begin{align}
 \delta^2+1/(4K)&\le\delta^2+4b(\delta+b)=(\delta+2b)^2,\\
 \tfrac12\bigl(\sqrt{\delta^2+1/(4K)}-\delta\bigr)&\le b.\label{eq:scalar-compare}
\end{align}
Equations~\eqref{eq:scalar-excess} and \eqref{eq:scalar-compare} establish the hypothesis of \eqref{eq:scalar-rescale} for $Q_{AB}$. Substituting its signal acceptance bound $\Tr[Q_{AB}\rho_{AB}]\le\varepsilon-\delta$ into \eqref{eq:scalar-target} proves \eqref{eq:clean}.

For $\delta>0$, the stronger condition $K\ge1/(16b\delta)$ implies the hypothesis of \eqref{eq:clean}. Taking binary logarithms and using \eqref{eq:scalar-beta} gives \eqref{eq:scalar-bits}. Setting $\delta=0$ in \eqref{eq:clean} gives \eqref{eq:scalar-zero}.
\end{proof}
The same implication \eqref{eq:scalar-rescale} also applies to the normalized effect $Q^t_{AB}=(H_{AB}-tI_{AB})_+/(1-t)$ from \eqref{eq:excess-effect}. If this effect accepts the signal with probability at most $\varepsilon-\delta$, then \eqref{eq:excess-signal} bounds the trace distance by $t+(1-t)(\varepsilon-\delta)$. Equating this expression to $\varepsilon$ gives $t\coloneqq\delta/(1-\varepsilon+\delta)\in[0,1)$. To verify the hypothesis of \eqref{eq:scalar-rescale}, divide \eqref{eq:excess} by $1-t$ and use $s=1/K$. The resulting type~II error probability bound is at most $b$ whenever $\sqrt{t^2+1/(4K)}\le t+2(1-t)b$. Squaring gives the sufficient condition
\begin{equation}\label{eq:normalized}
 K\ge [16(1-t)b\{t+(1-t)b\}]^{-1}.
\end{equation}
At $\delta=0$, we have $t=0$ and \eqref{eq:normalized} agrees with \eqref{eq:scalar-zero}. Appendix~\ref{sec:sharpness} shows that the coefficient $1/16$ in \eqref{eq:scalar-zero} cannot be decreased uniformly. One copied fair bit at $K=1$ already witnesses this sharpness.

For a comparison in finite dimensions at the same type~I error probability, take strictly positive $\tau_A$ and $0<\delta<\varepsilon$. Equation~(67) of Cheng and Gao~\cite[Theorem~4]{CG}, at their slack $\delta/3$, bounds the required population by $81\nu^2/(b\delta^4)$, where $\nu$ is the number of distinct eigenvalues of $\tau_A$ times $\dim B$. Equation~\eqref{eq:clean} uses the same $b$ and removes the spectral factor while improving the slack dependence.

\subsection{A converse for convex splitting}\label{ssec:converse}
Corollary~\ref{cor:scalar} gives a sufficient number of positions for hiding a correlated pair. A converse asks how many positions are necessary. The sufficient bound in \eqref{eq:clean} uses one test against every independent pair $\tau_A\otimes\sigma_B$. For a converse it suffices to detect the signal against the particular pair $\tau_A\otimes\rho_B$, which is the marginal at each site of the independent preparation \eqref{eq:background}. We construct a joint test from this single-pair test and compare the resulting necessary condition with that of Cheng and Gao~\cite[Theorem~5]{CG}. The joint test must respond to a signal at any site, even though the site tests share $B$ and may fail to commute. Fix $0\le\varepsilon<1$ and choose $Q_{AB}$ attaining $\beta_\varepsilon(\rho_{AB}\Vert\tau_A\otimes\rho_B)$ in \eqref{eq:simple-beta}. Embed it at each site and set
\begin{align}
 S_{A^K B}&\coloneqq\sum_{i=1}^K Q_{A_iB},\label{eq:reverse-sum}\\
 T^r_{A^K B}&\coloneqq S_{A^K B}(S_{A^K B}+rI_{A^K B})^{-1}\quad(r>0).\label{eq:reverse-effect}
\end{align}
The joint test accepts a spectral value $x$ of \eqref{eq:reverse-sum} with probability $x/(x+r)$, so $r$ sets the scale at which the test responds. The function is operator monotone because $x/(x+r)=1-r/(x+r)$ and inversion reverses order on positive invertible operators. Thus $S_{A^KB}\ge Q_{A_iB}$ gives $T^r_{A^KB}\ge Q_{A_iB}(Q_{A_iB}+rI)^{-1}\ge Q_{A_iB}/(1+r)$. The last inequality uses $0\le Q_{A_iB}\le I$. Also $T^r_{A^KB}\le S_{A^KB}/r$. Averaging the first estimate over the signal position, and using the second on the independent state, gives
\begin{align}
 \Tr[T^r_{A^KB}\omega_{A^KB}]&\ge\frac{1}{K(1+r)}\sum_{i=1}^K\Tr[Q_{A_iB}\omega^i_{A^KB}]
 \ge\frac{1-\varepsilon}{1+r},\label{eq:reverse-signal}\\
 \Tr[T^r_{A^KB}\omega^0_{A^KB}]&\le\frac1r\sum_{i=1}^K\Tr[Q_{A_iB}\omega^0_{A^KB}]
 =\frac{K\beta_\varepsilon(\rho_{AB}\Vert\tau_A\otimes\rho_B)}r.\label{eq:reverse-noise}
\end{align}
The equalities of the single-pair marginals follow from \eqref{eq:placements}--\eqref{eq:background}. By \eqref{eq:distance}, \eqref{eq:reverse-signal}, and \eqref{eq:reverse-noise},
\begin{align}
 \tfrac12\norm{\omega_{A^KB}-\omega^0_{A^KB}}_1
 &\ge \Tr[T^r_{A^KB}(\omega_{A^KB}-\omega^0_{A^KB})]\\
 &\ge\frac{1-\varepsilon}{1+r}-\frac{K\beta_\varepsilon(\rho_{AB}\Vert\tau_A\otimes\rho_B)}r.
 \label{eq:reverse-difference}
\end{align}
For nonnegative reals $A,B$, the supremum of $A/(1+r)-B/r$ over $r>0$ is $(\sqrt A-\sqrt B)_+^2$. For $0<B<A$, differentiation gives $r=\sqrt B/(\sqrt A-\sqrt B)$; for $B=0$ or $B\ge A$, take the limit at zero or infinity, respectively. Thus \eqref{eq:reverse-difference} becomes
\begin{equation}\label{eq:converse}
 \frac12\norm{\omega_{A^K B}-\omega^0_{A^K B}}_1
 \ge \left(\sqrt{1-\varepsilon}-\sqrt{K\,\beta_\varepsilon(\rho_{AB}\Vert\tau_A\otimes\rho_B)}\right)_+^2 .
\end{equation}
To compare the two converses at the same testing tolerance, let $0<d<1$ be the target normalized trace distance and choose $0<\delta<1-d$. Set $\varepsilon=1-d-\delta$ in \eqref{eq:converse} and write $\beta\coloneqq\beta_{1-d-\delta}(\rho_{AB}\Vert\tau_A\otimes\rho_B)$ as in \eqref{eq:simple-beta}. If the distance is at most $d$, then \eqref{eq:converse} gives $\sqrt{K\beta}\ge\sqrt{d+\delta}-\sqrt d$, or equivalently
\begin{equation}\label{eq:converse-sites}
 K\beta\ge (\sqrt{d+\delta}-\sqrt d)^2
 =\frac{\delta^2}{(\sqrt{d+\delta}+\sqrt d)^2}.
\end{equation}
Theorem~5, Eq.~(101), of Cheng and Gao~\cite{CG} uses the same $\beta$ and contains a free parameter $c>0$. For $c$ with $\delta-c(1-d-\delta)>0$, set $u\coloneqq c/(1+c)$. Exponentiating their bound gives \eqref{eq:cg-converse-param}, and maximizing its right side gives \eqref{eq:cg-converse-opt}.
\begin{align}
 K\beta&\ge\frac{c[\delta-c(1-d-\delta)]}{(1+c)^2}
 =\delta u-(1-d)u^2,\label{eq:cg-converse-param}\\
 K\beta&\ge\frac{\delta^2}{4(1-d)}.\label{eq:cg-converse-opt}
\end{align}
The quadratic in \eqref{eq:cg-converse-param} attains its maximum at $u=\delta/[2(1-d)]$. Both necessary conditions hold for the same test. At $d=1/4$, the denominator in \eqref{eq:converse-sites} is less than $(1+1/2)^2=9/4$, while $4(1-d)=3$. Equation~\eqref{eq:converse-sites} then gives the larger lower bound on $K\beta$. For $d\ge1/2$, the denominator in \eqref{eq:converse-sites} exceeds $4d\ge4(1-d)$, so \eqref{eq:cg-converse-opt} gives the larger lower bound. If $\beta=0$, \eqref{eq:converse-sites} rules out every finite $K$ at this target distance. 

\section[Private communication]{Private messages over a quantum wiretap channel}\label{sec:packing}
In private communication, Alice, the sender, wants Bob, the receiver, to recover her message while Eve, the eavesdropper, learns little about it. In order to achieve this goal, Alice assigns a row of codewords to each message. To send that message, she draws a column at random and transmits the entry in the selected row and column. Bob knows the table and attempts to identify the row. Eve also knows the table, so secrecy must hold even with this knowledge. Randomized encoders occur in the original quantum wiretap coding theorems~\cite{CWY,Devetak}. Renes and Renner gave one-shot bounds via reconciliation and privacy amplification~\cite{RRWire}. Here we follow the position-based construction in~\cite[Sec.~4]{Wilde}. Theorem~\ref{thm:cover} lets us compare Eve's state for one message with her actual state averaged over all messages.

A classical--quantum wiretap channel $\mathcal N_{X\to BE}$ prepares $\rho^x_{BE}$ when Alice supplies a letter $x$ from a finite alphabet $\mathcal X$. Bob receives $B$ and Eve receives $E$. For an input distribution $p_X$, define
\begin{equation}\label{eq:cq}
 \rho_{XBE}\coloneqq\sum_{x\in\mathcal X}p_X(x)|x\rangle\langle x|_X\otimes\rho^x_{BE}.
\end{equation}
The output marginals for letter $x$ are $\rho^x_B\coloneqq\Tr_E[\rho^x_{BE}]$ and $\rho^x_E\coloneqq\Tr_B[\rho^x_{BE}]$. Alice uses private randomness. The protocol uses one channel use, without a shared key or a public message.

\begin{figure}[!t]
\centering
\begin{tikzpicture}[x=1cm,y=1cm,font=\small,>=Stealth]
 \node[font=\small\bfseries,anchor=west] at (-.18,1.96) {Alice's codebook};
 \foreach \k in {1,...,4}{
   \node[font=\footnotesize] at (1.25*\k-.625,1.52) {$k=\k$};}
 \foreach \m in {1,...,3}{
   \node[anchor=east,font=\footnotesize] at (-.10,1.49-.54*\m) {$m=\m$};
   \foreach \k in {1,...,4}{
     \ifnum\m=2 \fill[softblue] (1.25*\k-1.25,1.76-.54*\m) rectangle ++(1.25,-.54);\fi
     \draw[black!50] (1.25*\k-1.25,1.76-.54*\m) rectangle ++(1.25,-.54);
     \node at (1.25*\k-.625,1.49-.54*\m) {$t_{\m\k}$};}}
 \draw[inkblue,very thick] (0,.68) rectangle (5.00,.14);
 \draw[inkgreen,very thick,rounded corners=2pt] (2.58,.61) rectangle (3.67,.21);
 \node[align=center,font=\footnotesize] at (2.50,-.95) {Message $m=2$ fixes the row.\\Private randomness selects $k=3$.};
 \draw[->,thick] (5.13,.41)--node[above,font=\footnotesize] {send $t_{23}$}(6.83,.41);
 \node[draw,rounded corners=3pt,fill=black!4,minimum width=1.60cm,minimum height=.90cm,align=center] (channel) at (7.80,.41) {wiretap\\channel};
 \node[draw=inkblue,rounded corners=3pt,anchor=west,minimum width=3.62cm,minimum height=.90cm,align=left] (bob) at (10.30,1.15) {Bob measures $B$\\and reports a row $\widehat m$.};
 \node[draw=inkgreen,rounded corners=3pt,anchor=west,minimum width=3.62cm,minimum height=.90cm,align=left] (eve) at (10.30,-.42) {Eve receives $E$.\\Alice does not announce $k$.};
 \draw[->,thick] (channel.east)--node[above,sloped,font=\footnotesize] {$B$}(bob.west);
 \draw[->,thick] (channel.east)--node[below,sloped,font=\footnotesize] {$E$}(eve.west);
\end{tikzpicture}
\caption{A private code with three message rows and four columns. Bob and Eve both know the fixed codebook. For message $2$, each of the four entries in the outlined row is equally likely to be sent. The marked entry illustrates one draw. Bob succeeds if he reports row $2$, regardless of his estimate of the column. Averaging Eve's output over the four column choices gives her state for this message, as in \eqref{eq:wire-preparation}.}
\label{fig:wire-code}
\end{figure}
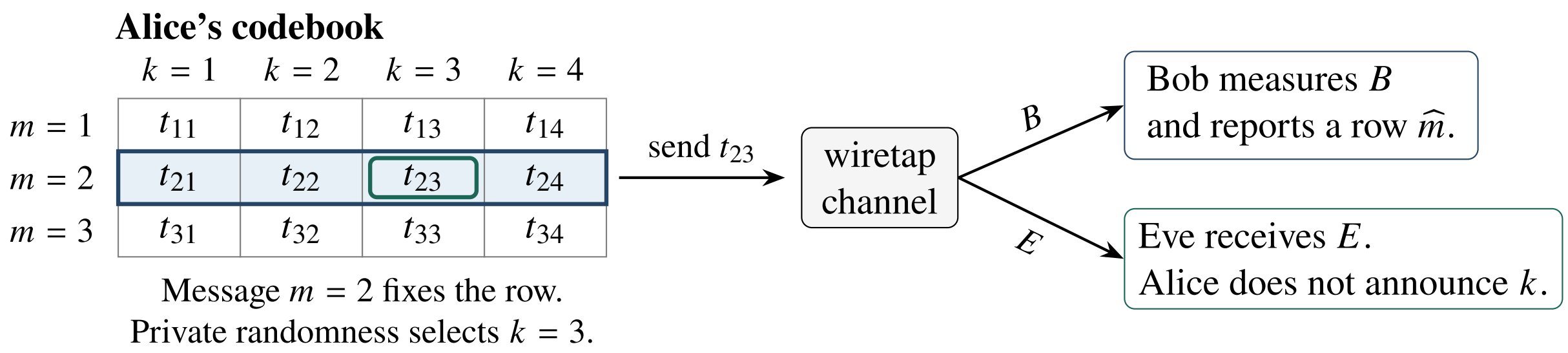

\subsection{A measurement for decoding the row}
We first draw a random codebook and analyze Bob's average error probability. The proof of Theorem~\ref{thm:wire} will then select one fixed codebook that also protects the message from Eve. For integers $M\ge2$ and $K\ge1$, let $T$ be an $M$-by-$K$ array whose entries are independent draws from $p_X$. A lowercase $t$ denotes a realization:
\begin{align}
 T&\coloneqq(T_{mk})_{1\le m\le M,\,1\le k\le K},\qquad t=(t_{mk})_{m,k}\in\mathcal X^{M\times K},\label{eq:random-codebook}\\
 p_T(t)&\coloneqq\Pr\{T=t\}=\prod_{m=1}^{M}\prod_{k=1}^{K}p_X(t_{mk}).\label{eq:table-security}
\end{align}
Figure~\ref{fig:wire-code} illustrates the choice of the transmitted entry. Alice's message $m$ and her private column $k$ are uniform and independent of each other and of $T$. Bob applies a pretty-good measurement to estimate the pair $(m,k)$ and keeps only the row. Cheng's trace inequality bounds the error probability of this measurement using a binary test between a candidate state and the sum of its competitors~\cite[Appendix, Lemma~1]{Cheng}. Lemma~\ref{lem:packing} uses this principle to bound the probability of a wrong row.

For positive semidefinite operators $A,B$ of finite trace on a common system, write the optimal binary testing cost as
\begin{align}
 e(A,B)&\coloneqq\min_{0\le Q\le I}\bigl\{\Tr[(I-Q)A]+\Tr[QB]\bigr\}\label{eq:cost}\\
 &=\tfrac12\bigl(\Tr[A+B]-\norm{A-B}_1\bigr).\label{eq:cost-norm}
\end{align}
The projector onto the positive spectrum of $A-B$ attains the minimum. When $A$ and $B$ are states, the two terms in the objective are the type~I and type~II error probabilities. For a codebook the second hypothesis carries weight $MK-1$, the number of competing positions. With $\rho_B\coloneqq\sum_x p_X(x)\rho_B^x$ from \eqref{eq:cq}, define
\begin{align}
 e_x&\coloneqq e\bigl(\rho_B^x,(MK-1)\rho_B\bigr),\label{eq:letter-cost}\\
 \mathcal P_{M,K}(X;B)&\coloneqq
 \frac{K(M-1)}{MK-1}\sum_x p_X(x)e_x\left(1-\frac{e_x}{MK}\right).
 \label{eq:pack-bound}
\end{align}
The fraction $K(M-1)/(MK-1)$ counts the $K(M-1)$ positions in other rows among the $MK-1$ incorrect positions. When $M=1$, Bob can always return the correct row.

\begin{lemma}[Decoding the row of a random codebook]\label{lem:packing}
Let $x\mapsto\rho^x_B$ be a classical--quantum channel with input distribution $p_X$ as in \eqref{eq:cq}. For integers $M\ge2$, $K\ge1$, draw $T$ according to \eqref{eq:random-codebook}--\eqref{eq:table-security}, choose a position uniformly and independently of $T$, and transmit its letter. The pretty-good measurement estimating the position and reporting its row has average row error probability at most $\mathcal P_{M,K}(X;B)$ in \eqref{eq:letter-cost}--\eqref{eq:pack-bound}. The average includes the codebook, position, and measurement outcome.
\end{lemma}
\begin{proof}
Fix a realization $t$ of the codebook in \eqref{eq:random-codebook}. With $S_B(t)^{-1/2}$ interpreted on its support, Bob's pretty-good measurement has the effects
\begin{align}
 S_B(t)&\coloneqq\sum_{j=1}^M\sum_{\ell=1}^K\rho^{t_{j\ell}}_B,\label{eq:pgm-sum}\\
 \Pi_B(t)&\coloneqq\operatorname{proj}_{\operatorname{supp} S_B(t)},\label{eq:pgm-support}\\
 \Lambda_B^{(m,k),t}&\coloneqq S_B(t)^{-1/2}\rho_B^{t_{mk}}S_B(t)^{-1/2}
                         +\frac{I_B-\Pi_B(t)}{MK}.\label{eq:pgm-site}
\end{align}
The complement term makes the effects sum to $I_B$ and has zero probability on every signal state. Collecting the outcomes in each row gives
\begin{align}
 \Lambda_B^{m,t}&\coloneqq\sum_{k=1}^{K}\Lambda_B^{(m,k),t},\label{eq:pgm-row}\\
 \sum_{m=1}^M\Lambda_B^{m,t}&=I_B.\label{eq:row-povm}
\end{align}
When Alice sends $t_{mk}$, the other positions contribute $S_B(t)-\rho_B^{t_{mk}}$. Cheng's trace inequality~\cite[Appendix, Lemma~1, Eq.~(A2)]{Cheng} bounds the probability of an incorrect position by
\begin{align}
 &\Tr[(I_B-\Lambda_B^{(m,k),t})\rho_B^{t_{mk}}]\notag\\
 &=\Tr\!\left[S_B(t)^{-1/2}(S_B(t)-\rho_B^{t_{mk}})S_B(t)^{-1/2}\rho_B^{t_{mk}}\right]\label{eq:pgm-identity}\\
 &\le e\bigl(\rho_B^{t_{mk}},S_B(t)-\rho_B^{t_{mk}}\bigr)
 \left(1-\frac{e(\rho_B^{t_{mk}},S_B(t)-\rho_B^{t_{mk}})}{MK}\right).
 \label{eq:pgm-cond}
\end{align}
The denominator is $\Tr[S_B(t)]=MK$. An alternative proof of Cheng's trace inequality appears in the analysis of quantum simultaneous decoding~\cite[Lemma~2.1]{WildeSimul}.

Condition on the transmitted entry $T_{mk}=x$. By \eqref{eq:table-security}, the other entries still have distribution $p_X$, and hence
\begin{equation}\label{eq:independent-competitors}
 \mathbb E\bigl[S_B(T)-\rho_B^{T_{mk}}\mid T_{mk}=x\bigr]=(MK-1)\rho_B .
\end{equation}
For fixed unit-trace $A$, $e(A,B)$ lies in $[0,1]$ and is concave in $B$, being the minimum of affine functions of $B$ in \eqref{eq:cost}. The function $y\mapsto y(1-y/(MK))$ is concave and nondecreasing on this interval. Their composition is therefore concave in the competing operator. Applying conditional Jensen's inequality to \eqref{eq:pgm-cond} and then \eqref{eq:independent-competitors} bounds the conditional position error by $e_x(1-e_x/(MK))$.

For fixed $(m,k)$, permuting the other positions preserves their joint sampling distribution in \eqref{eq:table-security} and permutes the corresponding effects in \eqref{eq:pgm-site}. All $MK-1$ incorrect positions consequently have equal average probability. Exactly $K(M-1)$ give a wrong row. Multiplying the position-error bound by $K(M-1)/(MK-1)$ and averaging over $x$ gives \eqref{eq:pack-bound}.
\end{proof}
By the classical block structure in \eqref{eq:cq}, the unconditional cost $e(\rho_{XB},(MK-1)\rho_X\otimes\rho_B)$ equals $\sum_xp_X(x)e_x$. Compared with applying the quadratic in \eqref{eq:pgm-cond} to this single cost~\cite[Remark~2]{Cheng}, conditioning reduces the position-error upper bound by $\operatorname{Var}_{p_X}(e_x)/(MK)$. 

\subsection{Comparing the actual and ideal output states}
Lemma~\ref{lem:packing} controls Bob's error averaged over codebooks. To include secrecy, we keep track of Eve's state conditional on Alice's message. For a fixed table $t=(t_{mk})_{m,k}$ from \eqref{eq:random-codebook}, Alice transmits its entry in row $m$ and a uniform column. Averaging over that column gives the state $\omega^{m,t}_{BE}$ before decoding. Eve receives its marginal $\omega_E^{m,t}\coloneqq\Tr_B[\omega_{BE}^{m,t}]$. Averaging over messages gives her unconditional state $\omega_E^t$. Thus
\begin{align}
 \omega^{m,t}_{BE}&\coloneqq\frac1K\sum_{k=1}^K\rho_{BE}^{t_{mk}},\label{eq:wire-preparation}\\
 \omega_E^t&\coloneqq\frac1M\sum_{m=1}^M\omega_E^{m,t}.\label{eq:wire-mean}
\end{align}
Bob's decoding channel $\mathcal D^t_{B\to\widehat M}$ measures the row effects $\Lambda_B^{j,t}$ from \eqref{eq:pgm-row} and records the outcome $j$ in the classical register $\widehat M$. Its action on an input operator $Z_B$ is
\begin{align}
 \mathcal D^t_{B\to\widehat M}(Z_B)
 &\coloneqq\sum_{j=1}^M\Tr[\Lambda_B^{j,t}Z_B]|j\rangle\langle j|_{\widehat M}.\label{eq:wire-decoder}
\end{align}
On a state of $BE$, $\mathcal D^t_{B\to\widehat M}$ acts on $B$ and leaves $E$ untouched. In the ideal experiment Bob receives the correct label and Eve receives the same state $\omega_E^t$ for every message.

\begin{theorem}[One-shot private communication]\label{thm:wire}
Let $\mathcal N_{X\to BE}$ be a classical--quantum wiretap channel with finite input alphabet, input distribution $p_X$, and associated state \eqref{eq:cq}. For integers $M\ge2$ and $K\ge1$, there exists a fixed codebook $t$ of the form \eqref{eq:random-codebook} for one use of the channel. With a uniform message, a private uniform choice of column, and the preparation and decoder in \eqref{eq:wire-preparation}--\eqref{eq:wire-decoder}, it satisfies
\begin{equation}\label{eq:private}
 \frac1{2M}\sum_m\norm{\mathcal D^t_{B\to\widehat M}(\omega^{m,t}_{BE})-|m\rangle\langle m|_{\widehat M}\otimes\omega_E^t}_1
 \le\mathcal P_{M,K}(X;B)+\mathcal C(a_M,a_M/K;\rho_{XE},\rho_X).
\end{equation}
Here $a_M\coloneqq1-1/M$, the sum runs over the $M$ messages, and the decoding and covering terms are defined in \eqref{eq:pack-bound} and \eqref{eq:layer}, respectively. For the fixed reference $\rho_E$ in \eqref{eq:cq}, there also exists a codebook, possibly different, satisfying \eqref{eq:private} with $\rho_E$ in place of $\omega_E^t$ on the left and $\mathcal C(1,1/K;\rho_{XE},\rho_X)$ as the covering term on the right.
\end{theorem}
\begin{proof}
We use the position registers and the selection of a fixed codebook from~\cite[Sec.~4.2]{Wilde}. In that argument Eve's state is compared with a fixed reference. Here we first compare it with her actual state for the same codebook, averaged over the messages. This requires two distributions of the hidden position.

\emph{The random codebook and the output received by Eve.}
Write $X_{j,\ell}$ for the classical register containing entry $(j,\ell)$ of the table in \eqref{eq:random-codebook}. In the random-code analysis $T$ records which codebook was drawn. It consists of all $X_{j,\ell}$ and $|t\rangle_T=\bigotimes_{j,\ell}|t_{j\ell}\rangle_{X_{j,\ell}}$. If Alice transmits entry $(j,\ell)$, the joint state of the random table and Eve's output is
\begin{align}
 \zeta^{j,\ell}_{TE}
 &\coloneqq\sum_t p_T(t)|t\rangle\langle t|_T\otimes\rho_E^{t_{j\ell}}\label{eq:wire-site-block}\\
 &=\left(\sum_xp_X(x)|x\rangle\langle x|_{X_{j,\ell}}\otimes\rho_E^x\right)
 \otimes\!\bigotimes_{(a,b)\ne(j,\ell)}
 \left(\sum_y p_X(y)|y\rangle\langle y|_{X_{a,b}}\right)\label{eq:wire-site-expanded}\\
 &=\rho_{X_{j,\ell}E}\otimes\!\bigotimes_{(a,b)\ne(j,\ell)}\rho_{X_{a,b}}.\label{eq:wire-site-product}
\end{align}
Equation~\eqref{eq:wire-site-expanded} follows by summing separately over each entry in the product distribution \eqref{eq:table-security}. Only $t_{j\ell}$ determines Eve's state. Equation~\eqref{eq:wire-site-product} is precisely \eqref{eq:placements} with $MK$ positions, signal pair $\rho_{XE}$ from \eqref{eq:cq}, and background state $\rho_X$.

Given message $m$, Alice chooses a uniform column in row $m$. Averaging also over a uniform message gives a uniform position in the whole table. Substituting \eqref{eq:wire-preparation} and \eqref{eq:wire-mean} gives
\begin{align}
 \omega_{TE}^m&\coloneqq\sum_t p_T(t)|t\rangle\langle t|_T\otimes\omega_E^{m,t}
 =\frac1K\sum_{\ell=1}^K\zeta^{m,\ell}_{TE},\label{eq:row-table-state}\\
 \omega_{TE}&\coloneqq\sum_t p_T(t)|t\rangle\langle t|_T\otimes\omega_E^t
 =\frac1{MK}\sum_{j=1}^M\sum_{\ell=1}^K\zeta^{j,\ell}_{TE}.\label{eq:full-table-state}
\end{align}
These are $\Omega_{TE}(p)$ and $\Omega_{TE}(q)$ in \eqref{eq:mixture}, respectively. Both give zero weight to the independent preparation. At the physical positions their weights and differences are
\begin{align}
  p_{j\ell}&=\frac{\boldsymbol1_{\{j=m\}}}{K},\qquad q_{j\ell}=\frac1{MK},\label{eq:wire-site-weights}\\
  p_{j\ell}-q_{j\ell}&=
  \begin{cases}(M-1)/(MK),&j=m,\\ -1/(MK),&j\ne m.\end{cases}\label{eq:wire-site-diff}
\end{align}
There are $K$ positive differences and $(M-1)K$ negative ones. By \eqref{eq:weights} the total variation distance is the sum of the positive differences. By \eqref{eq:site-norm} all the physical positions contribute their squared differences. Thus
\begin{align}
 \tfrac12\norm{p-q}_1&=K\frac{M-1}{MK}=1-\frac1M=a_M,\label{eq:row-variation}\\
 \norm{p_{1:MK}-q_{1:MK}}_2^2
 &=K\left(\frac{M-1}{MK}\right)^2+(M-1)K\left(\frac1{MK}\right)^2
 =\frac{a_M}{K}.\label{eq:row-squares}
\end{align}
The trace norm adds over the orthogonal table labels in \eqref{eq:row-table-state} and \eqref{eq:full-table-state}. Applying Theorem~\ref{thm:cover} with \eqref{eq:row-variation} and \eqref{eq:row-squares} now bounds the mean secrecy error for each $m$:
\begin{align}
 \sum_t p_T(t)\tfrac12\norm{\omega_E^{m,t}-\omega_E^t}_1
 &=\tfrac12\norm{\omega^m_{TE}-\omega_{TE}}_1\label{eq:wire-classical-average}\\
 &\le\mathcal C(a_M,a_M/K;\rho_{XE},\rho_X).\label{eq:wire-ensemble}
\end{align}

\emph{Reliability and secrecy for a single codebook.}
To include reliability, fix a table $t$ and the true row $m$. Eve's subnormalized state when Bob decodes row $j$ is
\begin{align}
 \theta_E^{j|m,t}&\coloneqq\Tr_B\!\left[(\Lambda_B^{j,t})^{1/2}\omega^{m,t}_{BE}(\Lambda_B^{j,t})^{1/2}\right],\label{eq:wire-blocks}\\
 \mathcal D^t_{B\to\widehat M}(\omega_{BE}^{m,t})
 &=\sum_{j=1}^M |j\rangle\langle j|_{\widehat M}\otimes\theta_E^{j|m,t}.\label{eq:wire-decoded-blocks}
\end{align}
Each block in \eqref{eq:wire-blocks} is positive semidefinite. By cyclicity of the partial trace on $B$ and \eqref{eq:row-povm}, $\sum_j\theta_E^{j|m,t}=\omega_E^{m,t}$. Compare Bob's decoded state with the state in which we give him the correct label while retaining Eve's actual conditional state. In each classical block $j\ne m$ the difference is $\theta_E^{j|m,t}$. In block $m$ it is $-\sum_{j\ne m}\theta_E^{j|m,t}$. Adding the trace norms of the blocks gives
\begin{align}
 \tfrac12\norm{\mathcal D^t_{B\to\widehat M}(\omega^{m,t}_{BE})
 -|m\rangle\langle m|_{\widehat M}\otimes\omega_E^{m,t}}_1
 &=\sum_{j\ne m}\Tr[\theta_E^{j|m,t}]\label{eq:wire-block-sum}\\
 &=\Pr\{\widehat M\ne m\mid m,t\}.\label{eq:wire-block-distance}
\end{align}
Insert $|m\rangle\langle m|_{\widehat M}\otimes\omega_E^{m,t}$ between the actual and ideal states in \eqref{eq:private}. The first distance is the decoding error by \eqref{eq:wire-block-distance}. In the second, the tensor factor $|m\rangle\langle m|_{\widehat M}$ has trace norm one. The triangle inequality gives, for each $m,t$,
\begin{equation}\label{eq:wire-pointwise}
 \tfrac12\norm{\mathcal D^t_{B\to\widehat M}(\omega^{m,t}_{BE})-
 |m\rangle\langle m|_{\widehat M}\otimes\omega_E^t}_1
 \le\Pr\{\widehat M\ne m\mid m,t\}+\tfrac12\norm{\omega_E^{m,t}-\omega_E^t}_1 .
\end{equation}
Average first over the $M$ equiprobable messages and then over $t$. Lemma~\ref{lem:packing} bounds the average of the first term, and \eqref{eq:wire-ensemble} bounds the second:
\begin{align}
 &\sum_t p_T(t)\frac1{2M}\sum_{m=1}^M
 \norm{\mathcal D^t_{B\to\widehat M}(\omega^{m,t}_{BE})-|m\rangle\langle m|_{\widehat M}\otimes\omega_E^t}_1\notag\\
 &\quad\le \sum_{m,t}\frac{p_T(t)}M\Pr\{\widehat M\ne m\mid m,t\}
 +\frac1M\sum_{m=1}^M\sum_t p_T(t)\tfrac12\norm{\omega_E^{m,t}-\omega_E^t}_1\label{eq:wire-split-average}\\
 &\quad\le\mathcal P_{M,K}(X;B)+\mathcal C(a_M,a_M/K;\rho_{XE},\rho_X).\label{eq:wire-select}
\end{align}
Some table in the support of $p_T$ has joint error at most this average, which proves \eqref{eq:private} for one fixed codebook.

For the fixed reference $\rho_E$, replace \eqref{eq:full-table-state} by $\sum_t p_T(t)|t\rangle\langle t|_T\otimes\rho_E$. This is \eqref{eq:background}. Its position distribution has $q_0=1$, so the two norms in \eqref{eq:weights} and \eqref{eq:site-norm} are $1$ and $1/K$, respectively. Replacing $\omega_E^t$ by $\rho_E$ on both sides of \eqref{eq:wire-pointwise} and retaining $\omega_E^{m,t}$ in its secrecy term gives the same averaging argument \eqref{eq:wire-select} for the fixed reference.
\end{proof}
Classical preprocessing of the input is included by applying Theorem~\ref{thm:wire} to the channel $u\mapsto\sum_x p_{X|U}(x|u)\rho^x_{BE}$.

\section[Secret key distillation]{Secret key distillation with a public transcript}\label{sec:key}
In Section~\ref{sec:packing}, Alice chose a letter to send through a channel. In key distillation she already holds a classical sample $X$, correlated with the systems $B$ and $E$ held by Bob and Eve. Their known source state is
\begin{equation}\label{eq:source-state}
 \rho_{XBE}\coloneqq\sum_{x\in\mathcal X}p_X(x)|x\rangle\langle x|_X\otimes\rho^x_{BE},
\end{equation}
where $\mathcal X$ is finite. Alice and Bob wish to agree on a uniform key using a public message from Alice. Eve can read this message, so secrecy requires the key to be independent of Eve's system and the message together. Following~\cite[Sec.~IV]{KKGW}, Alice makes a table containing her sample at a random site, announces the table, and keeps the row as her key. The decoder from Section~\ref{sec:packing} will let Bob estimate this row.

\Needspace{10\baselineskip}
\begin{theorem}[One-way key distillation]\label{thm:key}
Let $\rho_{XBE}$ in \eqref{eq:source-state} be a known source with a finite classical alphabet. Fix integers $M\ge2$ and $K\ge1$. Using local randomness and one authenticated public message $T$ from Alice to Bob, available also to Eve, Alice obtains a uniform $M$-valued key $U$ and Bob produces an estimate $\widehat U$. There exists such a protocol with output state $\omega_{U\widehat UTE}$ satisfying
\begin{align}
 \tfrac12\norm{\omega_{U\widehat UTE}-\overline\Phi^M_{U\widehat U}\otimes\omega_{TE}}_1
 &\le \mathcal P_{M,K}(X;B)+\mathcal C(a_M,a_M/K;\rho_{XE},\rho_X),\label{eq:key-joint}\\
 \overline\Phi^M_{U\widehat U}&\coloneqq\frac1M\sum_{m=1}^M
 |m\rangle\langle m|_U\otimes |m\rangle\langle m|_{\widehat U}.\label{eq:ideal-key}
\end{align}
Here $a_M\coloneqq 1-1/M$ as in \eqref{eq:row-variation}, and $\omega_{TE}$ is the actual marginal of the protocol output. The decoding term $\mathcal P_{M,K}(X;B)$ from \eqref{eq:pack-bound} bounds $\Pr\{\widehat U\ne U\}$. The covering term $\mathcal C$ from \eqref{eq:layer} bounds $\tfrac12\norm{\omega_{UTE}-(I_U/M)\otimes\omega_{TE}}_1$.
\end{theorem}
\begin{proof}
We use the public-table protocol of~\cite[Sec.~IV.B, Eqs.~(231)--(233)]{KKGW}. Its joint state for a chosen position is the same as in the coding experiment in Theorem~\ref{thm:wire}. Here the full table is transmitted publicly and remains part of Eve's information throughout the proof.

\emph{Preparing and announcing the table.}
Alice draws $m,k$ uniformly, independently of her source sample. She places her sample at site $(m,k)$ and draws all remaining entries independently from $p_X$. She keeps $U=m$ and announces $T=t$. A full description takes at most $\lceil MK\log_2|\mathcal X|\rceil$ public bits. For each fixed $(m,k)$ the inserted sample has distribution $p_X$, independently of the entries Alice draws. Hence the conditional probability of announcing $t$ is $p_X(t_{mk})\prod_{(a,b)\ne(m,k)}p_X(t_{ab})=p_T(t)$ by \eqref{eq:table-security}. Conditional on the site and table, the joint state of $BE$ is $\rho^{t_{mk}}_{BE}$. Thus the state before Bob decodes, including the public transcript, is
\begin{equation}\label{eq:source-planted}
 \zeta^{m,k}_{TBE}\coloneqq\sum_t p_T(t)|t\rangle\langle t|_T\otimes\rho^{t_{mk}}_{BE}.
\end{equation}

\begin{figure}[!t]
\centering
\begin{tikzpicture}[x=.88cm,y=.69cm,font=\small]
\foreach \xbase/\ttl in {0/{Given $U=2$},5.75/{Averaged over $U$},11.5/{Difference of weights}}{
 \node[font=\small\bfseries] at (\xbase+2,3.85) {\ttl};}
\foreach \j in {0,1,2}{\foreach \k in {0,1,2,3}{
 \fill[softblue] (5.75+\k,\j) rectangle ++(1,1);
 \node at (6.25+\k,\j+.5) {$\tfrac{1}{12}$};
 \ifnum\j=1
   \fill[inkblue!19] (\k,\j) rectangle ++(1,1);
   \node at (\k+.5,\j+.5) {$\tfrac14$};
   \fill[inkgreen!21] (11.5+\k,\j) rectangle ++(1,1);
   \node at (12+\k,\j+.5) {$+\tfrac16$};
 \else
   \node[text=black!65] at (\k+.5,\j+.5) {$0$};
   \node at (12+\k,\j+.5) {$-\tfrac{1}{12}$};
 \fi}}
\foreach \xbase in {0,5.75,11.5}{
 \foreach \k in {0,...,4}{\draw[black!50] (\xbase+\k,0)--(\xbase+\k,3);}
 \foreach \j in {0,...,3}{\draw[black!50] (\xbase,\j)--(\xbase+4,\j);}}
\draw[inkblue,thick] (0,1) rectangle (4,2);
\draw[inkgreen,thick] (11.5,1) rectangle (15.5,2);
\node at (4.86,1.5) {$-$};\node at (10.62,1.5) {$=$};
\node at (2,-.69) {$\omega^{2}_{TE}$};
\node at (7.75,-.69) {$\omega_{TE}$};
\node at (13.5,-.69) {$p_i-q_i$};
\end{tikzpicture}
\caption{The position distributions in the secrecy test for $M=3$ rows and $K=4$ columns. Each cell in the first two tables gives the probability that the site carries Alice's source sample. The first table is conditional on the second key row, and the second averages over the key. The third table subtracts the second from the first, cell by cell. By \eqref{eq:weights}--\eqref{eq:site-norm}, the total positive weight is $\tfrac12\norm{p-q}_1=2/3$, and the sum of the squares of all twelve differences is $\norm{p_{1:12}-q_{1:12}}_2^2=1/6$. Both states in \eqref{eq:key-placement}--\eqref{eq:key-placement-mean} include the same public table register $T$ as part of Eve's information. A short public index identifying the table preserves the comparison (Corollary~\ref{cor:transcript}).}
\label{fig:rows}
\end{figure}

\emph{Decoding and testing secrecy.}
For an announced table $t$, Bob uses the effects in \eqref{eq:pgm-row} with output register $\widehat U$ in place of $\widehat M$. Conditional on the site and $T=t$, Bob receives $\rho^{t_{mk}}_B$ by \eqref{eq:source-planted}. Averaging over the uniform site and the table probability distribution in \eqref{eq:table-security} therefore gives exactly the decoding experiment in Lemma~\ref{lem:packing}. Figure~\ref{fig:rows} compares the position distributions conditional on a key and averaged over all keys. Tracing out $B$ in \eqref{eq:source-planted} gives the state $\zeta^{m,k}_{TE}$ in \eqref{eq:wire-site-product}. Averaging over the column and then over the key gives the two mixtures
\begin{align}
 \omega^{m}_{TE}&\coloneqq\frac1K\sum_{k=1}^K\zeta^{m,k}_{TE},\label{eq:key-placement}\\
 \omega_{TE}&\coloneqq\frac1{MK}\sum_{j=1}^M\sum_{k=1}^K\zeta^{j,k}_{TE}.\label{eq:key-placement-mean}
\end{align}
The differences of the placement weights have the total variation distance in \eqref{eq:row-variation} and the squared norm in \eqref{eq:row-squares}. Orthogonality of the key labels gives
\begin{align}
 \tfrac12\norm{\omega_{UTE}-(I_U/M)\otimes\omega_{TE}}_1 &=\frac1M\sum_{m=1}^M\tfrac12\norm{\omega^{m}_{TE}-\omega_{TE}}_1\label{eq:key-secrecy-blocks}\\
 &\le\mathcal C(a_M,a_M/K;\rho_{XE},\rho_X).\label{eq:key-secrecy}
\end{align}
To include Bob's estimate, let $\mathcal D^t$ in the next display denote the decoder in \eqref{eq:wire-decoder} with output $\widehat U$. Both the actual state and the ideal state in \eqref{eq:key-joint} are diagonal in $U,T$. In block $(m,t)$, the actual output on $\widehat U E$ is $\mathcal D^t(\omega_{BE}^{m,t})$ and the ideal output is $|m\rangle\langle m|_{\widehat U}\otimes\omega_E^t$. The blocks have the same weight $p_T(t)/M$ in both states. Thus
\begin{align}
 \tfrac12\norm{\omega_{U\widehat UTE}-\overline\Phi^M_{U\widehat U}\otimes\omega_{TE}}_1 & =\sum_{m,t}\frac{p_T(t)}{2M}
 \norm{\mathcal D^t(\omega_{BE}^{m,t})-|m\rangle\langle m|_{\widehat U}\otimes\omega_E^t}_1\label{eq:key-joint-blocks}\\
 &\quad\le\mathcal P_{M,K}(X;B)+\mathcal C(a_M,a_M/K;\rho_{XE},\rho_X),\label{eq:key-joint-combine}
\end{align}
where the last step is exactly the ensemble bound \eqref{eq:wire-split-average}--\eqref{eq:wire-select}. Summation over $t$ in \eqref{eq:key-joint-blocks} accounts for the public transcript in the security criterion.
\end{proof}

In general $\omega_{TE}$ retains correlations between the public
transcript and Eve, even when the transcript entries are independent of
one another. Using the actual marginal in \eqref{eq:key-placement-mean} makes
these correlations part of the security criterion used by Renes and Renner~\cite[Sec.~II and Theorem~2]{RRKey}. Equations~(227)--(228) of~\cite{KKGW} compare optimization over an arbitrary reference state with the actual-marginal criterion. Cheng and Gao specify the actual marginal in Eq.~(146) and use a product reference in the calculation in Eq.~(152) of~\cite{CG}.

\subsection{A smaller public transcript}
Devetak and Winter attained the key rate $I(X;B)-I(X;E)$ with $H(X|B)$ forward public bits per copy~\cite[Theorem~2.1 and Remark~2.2]{DW}. For the position-based protocol, sending the full table can cost exponentially many bits. We preserve its one-shot guarantee by agreeing on a smaller family of tables with independent \emph{pairs} of entries. Anshu, Jain, and Warsi used pairwise-independent positions for convex splitting and position-based decoding and observed the reduction in codebook support for private communication~\cite[Lemma~1, Claim~1, and the Introduction]{AJW}. Here we retain the estimate against the actual joint state $\omega_{TE}$ for arbitrary finite source distributions.

\Needspace{10\baselineskip}
\begin{corollary}[Public communication for key distillation]\label{cor:transcript}
For the source $\rho_{XBE}$ in \eqref{eq:source-state}, let $d\coloneqq|\operatorname{supp}p_X|$. For integers $M\ge2$ and $K\ge1$, the joint, disagreement, and secrecy bounds in Theorem~\ref{thm:key} and \eqref{eq:key-joint} are achievable with at most
\begin{equation}\label{eq:transcript-size}
 \left\lceil\log_2\!\left(1+MK(d-1)+\binom{MK}{2}(d-1)^2\right)\right\rceil
\end{equation}
authenticated public bits and no preshared random seed. For $n$ independent copies of a fixed finite-alphabet source and $\log_2(MK)=O(n)$, this number is $O(n)$. The bound counts the worst-case number of transmitted bits and allows unrestricted local randomness and computation, including exact sampling from the specified distribution.
\end{corollary}
\begin{proof}
We construct a small family of tables and show that the parties can use an index into this family as their entire public message. Number the sites $1,\ldots,MK$. Their assignment to key rows will be fixed after estimating Bob's error.

\emph{Choosing the family.}
Label the symbols in $\operatorname{supp}p_X$ by $1,\ldots,d$. For a table $t=(t_1,\ldots,t_{MK})$, record the following vector of zeros and ones:
\begin{equation}\label{eq:table-moments}
 F(t)\coloneqq\bigl((f_{i,a}(t))_{i,a},\,
       (f_{i,a}(t) f_{j,b}(t))_{1\le i<j\le MK,\,1\le a,b<d}\bigr),
\end{equation}
where the indicator of symbol $a$ at site $i$ is
\begin{equation}\label{eq:table-indicator}
 f_{i,a}(t)\coloneqq\boldsymbol1_{\{t_i=a\}},\qquad
      1\le i\le MK,\quad 1\le a<d .
\end{equation}
The first part has $MK(d-1)$ coordinates, one for each site and retained symbol. In the second part, choose two distinct sites in $\binom{MK}{2}$ ways, then choose the symbol for each site in $(d-1)^2$ ways. Under independent sampling, the mean of the vector is
\begin{equation}\label{eq:table-convex-hull}
 \sum_t \left(\prod_{i=1}^{MK}p_X(t_i)\right) F(t)
 =\bigl((p_X(a))_{i,a},\,(p_X(a)p_X(b))_{i<j,\,a,b<d}\bigr).
\end{equation}
The left side is a convex combination. By Carath\'eodory's theorem it has a representation $\sum_t\mu(t)F(t)$ using at most one more table than the number of coordinates. Let $T=(T_1,\ldots,T_{MK})$ have this probability distribution $\mu$, so $T_i$ is its entry at site $i$. Matching the expectations of the coordinates in \eqref{eq:table-convex-hull} fixes the one-site and two-site probabilities for the retained symbols. The identity $\boldsymbol1_{\{t_i=d\}}=1-\sum_{a<d}f_{i,a}(t)$ recovers the probabilities involving $d$, including a pair of omitted symbols by multiplying the identities for two sites. Thus each $T_i$ has distribution $p_X$, and each pair $T_i,T_j$ with $i\ne j$ is independent. For $d=1$ the family consists of a single table.

\emph{Announcing the index.}
The parties agree on $\mu$ and its indexed support in advance. Alice chooses a uniform site $I$, independently of her sample $X$. Given $I=i$ and $X=x$, she draws a table from $\mu$ conditional on $T_i=x$ and announces its index. At every site, the marginal just established gives
\begin{align}
 \Pr\{T=t\mid I=i,X=x\}&=\frac{\mu(t)\boldsymbol1_{\{t_i=x\}}}{p_X(x)}
       \quad (p_X(x)>0),\label{eq:compressed-sampling}\\
 \Pr\{T=t\mid I=i\}&=\sum_{x:p_X(x)>0}p_X(x)\frac{\mu(t)\boldsymbol1_{\{t_i=x\}}}{p_X(x)}
 =\mu(t).\label{eq:compressed-table-distribution}
\end{align}
The index determines $t$, so we can keep the whole table register in Eve's state. Given $i,t$, her state is $\rho_E^{t_i}$ by \eqref{eq:source-state}.

\emph{Secrecy for every assignment of rows.}
The smaller ensemble preserves all one-site and two-site probabilities, but the full table can have correlations absent from the product distribution in \eqref{eq:table-security}. We repeat the centered-observable calculation of Lemma~\ref{lem:moment} with this new table distribution. It suffices to consider a test $W_{TE}=\sum_t|t\rangle\langle t|_T\otimes W_E(t)$ because the announced table is classical. Throughout this proof $\mathbb E_\mu$ denotes expectation over $T\sim\mu$, the pairwise-independent distribution constructed from \eqref{eq:table-convex-hull}. In analogy with \eqref{eq:Y} and \eqref{eq:site-definition}, define
\begin{align}
 Y_E&\coloneqq\mathbb E_\mu[W_E(T)]=\sum_t\mu(t)W_E(t),\label{eq:source-aware-mean}\\
 H_E^i(x)&\coloneqq\mathbb E_\mu[W_E(T)\mid T_i=x]-Y_E
 =\sum_{t:t_i=x}\frac{\mu(t)}{p_X(x)}W_E(t)-Y_E\quad(p_X(x)>0).\label{eq:source-aware}
\end{align}
Extend $H_E^i(x)$ by zero outside the support. Each $H_E^i(T_i)$ has mean zero. As in \eqref{eq:cross-zero}, pairwise independence gives, in the indicated operator order, for $i\ne j$,
\begin{equation}\label{eq:public-cross}
 \mathbb E_\mu[H_E^i(T_i) H_E^j(T_j)]
 =\left(\sum_x p_X(x)H_E^i(x)\right)\left(\sum_y p_X(y)H_E^j(y)\right)=0.
\end{equation}
Conditioning on $T_i$ gives $\mathbb E_\mu[(W_E(T)-Y_E)H_E^i(T_i)]=\mathbb E_\mu[H_E^i(T_i)^2]$. Taking adjoints gives the reversed product as in \eqref{eq:orth}--\eqref{eq:orth-reverse}. Expanding the positive squared residual as in \eqref{eq:residual-positive}--\eqref{eq:residual-diagonal} therefore yields
\begin{align}
 0&\le\mathbb E_\mu\!\left[\left(W_E(T)-Y_E-\sum_i H_E^i(T_i)\right)^2\right]\notag\\
 &=\mathbb E_\mu[W_E(T)^2]-Y_E^2-\sum_i\sum_x p_X(x)H_E^i(x)^2,\label{eq:public-residual}\\
 \sum_i\sum_xp_X(x)H_E^i(x)^2 &\le\mathbb E_\mu[W_E(T)^2]-Y_E^2\label{eq:pair-independent-residual}\\
 &\le Y_E-Y_E^2\le I_E/4.\label{eq:pair-independent-moment}
\end{align}
The last line uses $W_E(t)^2\le W_E(t)$ and \eqref{eq:bessel-variance}--\eqref{eq:bessel}.

Fix any row, let $p_i$ be uniform on its $K$ sites, and let $q_i=1/(MK)$ for every site. Take the single-pair observable
\begin{equation}\label{eq:source-single}
 H_{XE}\coloneqq\sum_x|x\rangle\langle x|_X\otimes\sum_i(p_i-q_i)H_E^i(x).
\end{equation}
Using \eqref{eq:compressed-sampling} and $\sum_i(p_i-q_i)=0$, its signal mean is precisely the row-to-average acceptance difference:
\begin{equation}\label{eq:public-source-acceptance}
 \Tr[H_{XE}\rho_{XE}]=\sum_i(p_i-q_i)\sum_t\mu(t)\Tr[W_E(t)\rho_E^{t_i}].
\end{equation}
Equation~\eqref{eq:source-aware} gives $\E_X^{\rho_X}(H_{XE})=\sum_i(p_i-q_i)\sum_xp_X(x)H_E^i(x)=0$. In each $x$-block the $Y_E$ terms cancel, and the remaining conditional averages of $W_E$ are effects. The positive and negative coefficients therefore bound the spectrum by $[-a_M,a_M]$ as in \eqref{eq:row-variation}. Weighted Cauchy--Schwarz as in \eqref{eq:common-cs}--\eqref{eq:common-operator}, followed by \eqref{eq:pair-independent-moment} and \eqref{eq:row-squares}, gives $\E_X^{\rho_X}(H_{XE}^2)\le a_M I_E/(4K)$. Take $W_{TE}$ optimal. Applying both tests from the proof of Theorem~\ref{thm:cover}, the spectral projectors and the normalized positive-part effect, now gives \eqref{eq:key-secrecy} for every partition into rows of size $K$.

\emph{Choosing a partition for Bob.}
Given Alice's uniform site $I$, let $J$ be the outcome of Bob's site measurement \eqref{eq:pgm-site}, before reporting a row. Let $\Gamma_B^{j,t}$ denote the effect in \eqref{eq:pgm-site} for site $j$ in the numbering $1,\ldots,MK$. The sampling rule \eqref{eq:compressed-table-distribution} gives
\begin{equation}\label{eq:public-site-joint}
 \Pr\{I=i,J=j\}=\frac1{MK}\sum_t\mu(t)\Tr\!\left[\Gamma_B^{j,t}\rho_B^{t_i}\right].
\end{equation}
The site probabilities are fixed before the assignment of rows. Pairwise independence gives $\mathbb E_\mu[\rho_B^{T_j}\mid T_i=x]=\rho_B$ for $j\ne i$. Thus the conditional-mean and Jensen steps \eqref{eq:pgm-cond}--\eqref{eq:independent-competitors} give
\begin{equation}\label{eq:public-site-error}
 \Pr\{J\ne I\}\le\sum_x p_X(x)e_x\bigl(1-e_x/(MK)\bigr)
\end{equation}
with $e_x$ from \eqref{eq:letter-cost}. To count assignments giving a wrong row, consider a uniform assignment $R$ of sites to $M$ labeled rows of size $K$, independently of $I,J$. For fixed distinct $i,j$, once the slot of $i$ is chosen, $j$ is equally likely to occupy any of the other $MK-1$ slots, of which $K(M-1)$ are outside $i$'s row. Consequently,
\begin{align}
 \Pr_R\{R(j)\ne R(i)\}&=\frac{K(M-1)}{MK-1}\quad(i\ne j),\label{eq:public-other-row}\\
 \mathbb E_R\!\left[\Pr\{R(J)\ne R(I)\mid R\}\right]
 &=\sum_{i\ne j}\Pr\{I=i,J=j\}\Pr_R\{R(j)\ne R(i)\}\notag\\
 &=\frac{K(M-1)}{MK-1}\Pr\{J\ne I\}\le\mathcal P_{M,K}(X;B),\label{eq:public-partition-error}
\end{align}
where the inequality uses \eqref{eq:public-site-error} and \eqref{eq:pack-bound}. At least one fixed assignment has disagreement probability at most this average. Alice and Bob agree on it in advance and use $U=R(I)$ and $\widehat U=R(J)$. Equal row sizes make $U$ uniform, and the secrecy estimate \eqref{eq:key-secrecy} holds for this assignment. The same triangle inequality as in Theorem~\ref{thm:key} now gives the joint bound \eqref{eq:key-joint}.

Only the table index is transmitted. The support bound following \eqref{eq:table-convex-hull} gives precisely the worst-case bit bound \eqref{eq:transcript-size}, including zero bits when $d=1$. For $n$ copies use $d^n\le|\mathcal X|^n$ in place of $d$. The argument of the logarithm in \eqref{eq:transcript-size} is at most $(1+MK|\mathcal X|^n)^2$, whose logarithm is $O(n)$ when $\log_2(MK)=O(n)$.
\end{proof}
Appendix~\ref{sec:explicit-table} gives an explicit family with the same asymptotic order of public communication, together with the conditional sampling rule that Alice needs.

\subsection{Choosing integer message and randomization sizes}
Theorems~\ref{thm:wire} and~\ref{thm:key} use the entire testing curve \eqref{eq:roc}. Two single-pair tests suffice for a direct choice of the code sizes. Eve's test determines how many columns $K$ suffice for secrecy. Bob's test determines how many positions $MK$ can be decoded. The remaining choice of $M$ rows determines the amount of private information.

\begin{corollary}[A code from two binary tests]\label{cor:wire-simple}
For one use of the wiretap channel \eqref{eq:cq}, or one copy of the source \eqref{eq:source-state} with an authenticated one-way public channel, fix $0<\delta_B<\varepsilon_B$, $0<\delta_E<\varepsilon_E$, and $\varepsilon_B+\varepsilon_E<1$. Evaluate
\begin{align}
 I_{H,B}&\coloneqq D_H^{\varepsilon_B-\delta_B}(\rho_{XB}\Vert\rho_X\otimes\rho_B),\label{eq:bob-divergence}\\
 b_E&\coloneqq 2^{-I_H^{1-\varepsilon_E+\delta_E}(\rho_{XE}\Vert\rho_X)}.
 \label{eq:two-tests}
\end{align}
Here $I_{H,B}$ is Bob's hypothesis-testing mutual information with the fixed product state $\rho_X\otimes\rho_B$, using \eqref{eq:simple-beta}--\eqref{eq:dh}. Thus $2^{-I_{H,B}}$ is the least probability of accepting the product $\rho_X\otimes\rho_B$ while accepting $\rho_{XB}$ with probability at least $1-\varepsilon_B+\delta_B$. The quantity $b_E$ is Eve's minimum worst-case type~II error probability at allowed type~I error probability $1-\varepsilon_E+\delta_E$, by \eqref{eq:beta} and \eqref{eq:beta-info}. If integers $M\ge2$, $K\ge1$ satisfy
\begin{align}
 16K b_E(\delta_E+b_E)&\ge1,\label{eq:integer-cover}\\
 MK-1&\le\delta_B 2^{I_{H,B}},\label{eq:integer-rule}
\end{align}
then a code or protocol exists, respectively, with joint error at most $\varepsilon_B+\varepsilon_E$ in \eqref{eq:private} or \eqref{eq:key-joint}. The source protocol can use the public bit count in \eqref{eq:transcript-size}. With Eve's actual marginal as reference, \eqref{eq:integer-cover} can be relaxed to $16K b_E(\delta_E+b_E)\ge1-1/M$.

In particular, choose powers of two for the numbers of columns and rows. Whenever the right side of \eqref{eq:round-messages} is at least one, the following numbers of bits are achievable:
\begin{align}
 \log_2 K&=\max\!\left\{0,\left\lceil\log_2\frac{1}{16b_E(\delta_E+b_E)}\right\rceil\right\},\label{eq:round-one-shot}\\
 \log_2 M&=\left\lfloor\log_2\!\left(1+\delta_B2^{I_{H,B}}\right)\right\rfloor-\log_2 K.\label{eq:round-messages}
\end{align}
Alice uses $\log_2 K$ independent fair bits to choose the column of the wiretap code and sends $\log_2 M$ private message bits. For the source, $\log_2 M$ is the length of the distilled key. The source protocol also uses local randomness to generate the row and the public table. If the right-hand side of \eqref{eq:round-messages} is less than one, $M=K=1$ yields zero bits with zero joint error relative to Eve's actual marginal.
\end{corollary}
\begin{proof}
First, the finite alphabet ensures that $b_E$ in \eqref{eq:two-tests} is positive. Let $d\coloneqq|\operatorname{supp}p_X|$ and $\sigma_E\coloneqq d^{-1}\sum_{x:p_X(x)>0}\rho_E^x$. Since $\rho_E^x\le d\sigma_E$ for every input in the support, the classical $x$-blocks give
\begin{align}
 \rho_{XE}&\le d\,\rho_X\otimes\sigma_E,\label{eq:bounded-alphabet}\\
 b_E&\ge (\varepsilon_E-\delta_E)/d>0.\label{eq:positive-beta}
\end{align}
Indeed, a feasible effect in \eqref{eq:beta} accepts the left side of \eqref{eq:bounded-alphabet} with probability at least $\varepsilon_E-\delta_E$. Bob's information in \eqref{eq:bob-divergence} is also finite: for every $x$ in the support, $\rho_B\ge p_X(x)\rho_B^x$, so $\rho_X\otimes\rho_B\ge(\min_{x:p_X(x)>0}p_X(x))\rho_{XB}$.

For the secrecy bound, let $H_{XE}$ be the observable in Lemma~\ref{lem:moment} for the row and full-table distributions, whose squared difference is $(1-1/M)/K$ by \eqref{eq:row-squares}. Its signal mean equals the row's distinguishing advantage. If $\delta_E\ge a_M=1-1/M$, the range estimate \eqref{eq:moment-range} gives $H_{XE}\le a_M I_{XE}$, so its mean is at most $\delta_E<\varepsilon_E$. The excess effect then vanishes. For $\delta_E<a_M$, the effect $Q_{XE}\coloneqq(H_{XE}-\delta_E I_{XE})_+$ has worst-case type~II error probability bounded by \eqref{eq:excess}. The relaxed size condition in the statement gives
\begin{equation}\label{eq:row-scalar-test}
 \norm{\E_X^{\rho_X}(Q_{XE})}
 \le \frac{\sqrt{\delta_E^2+(1-1/M)/(4K)}-\delta_E}{2}\le b_E .
\end{equation}
The comparison with $b_E$ in \eqref{eq:row-scalar-test} is equivalent to $1-1/M\le16K b_E(\delta_E+b_E)$. The testing implication \eqref{eq:scalar-rescale} from Corollary~\ref{cor:scalar}, together with $b_E>0$ in \eqref{eq:positive-beta}, now bounds $\Tr[Q_{XE}\rho_{XE}]$ by $\varepsilon_E-\delta_E$. Since $H_{XE}\le \delta_E I_{XE}+Q_{XE}$, its signal mean is at most $\varepsilon_E$. The comparison with the independent preparation in Corollary~\ref{cor:scalar} uses $1/K$ in place of $(1-1/M)/K$ and gives the condition in \eqref{eq:integer-cover}.

For Bob let $Q_{XB}$ attain the testing divergence in \eqref{eq:bob-divergence}, so that
\begin{align}
 \Tr[(I_{XB}-Q_{XB})\rho_{XB}]&\le\varepsilon_B-\delta_B,\label{eq:bob-miss}\\
 \Tr[Q_{XB}(\rho_X\otimes\rho_B)]&=2^{-I_{H,B}}.\label{eq:bob-tests}
\end{align}
Using $Q_{XB}$ in \eqref{eq:cost} upper-bounds $e(\rho_{XB},(MK-1)\rho_X\otimes\rho_B)$ by $\varepsilon_B-\delta_B+(MK-1)2^{-I_{H,B}}$. The factors $1-e_x/(MK)$ and $K(M-1)/(MK-1)$ in \eqref{eq:pack-bound} are at most one, and the binary cost in \eqref{eq:cost} separates across the classical $x$-blocks of \eqref{eq:cq}. Thus the decoding term is at most $\varepsilon_B$ by \eqref{eq:integer-rule}. The protocol theorems and \eqref{eq:integer-cover}--\eqref{eq:integer-rule} give the joint bound. For the compressed transcript of Corollary~\ref{cor:transcript}, the observable in \eqref{eq:source-single} has the same range and second-moment bounds, so the range argument for $\delta_E\ge a_M$ and the test in \eqref{eq:row-scalar-test} for $\delta_E<a_M$ both apply. Equations~\eqref{eq:public-site-error}--\eqref{eq:public-partition-error} give the same decoding term for the selected partition.

To obtain \eqref{eq:round-one-shot}, choose the least nonnegative integer $\log_2 K$ for which \eqref{eq:integer-cover} holds. Equation~\eqref{eq:integer-rule} then allows $\log_2 M+\log_2 K\le\log_2(1+\delta_B2^{I_{H,B}})$. The largest integer $\log_2 M$ satisfying this condition is the right-hand side of \eqref{eq:round-messages}.
\end{proof}
For arbitrary integer sizes the choice $K=\lceil1/[16b_E(\delta_E+b_E)]\rceil$, $M=\lfloor(1+\delta_B2^{I_{H,B}})/K\rfloor$ is also available whenever $M\ge2$. A direct optimization of \eqref{eq:private} can improve upon these sufficient sizes.

\section[Second-order coding rates]{Second-order rates for private messages and secret keys}\label{sec:spectral-curve}
Theorems~\ref{thm:wire} and~\ref{thm:key} give one-shot codes for a known channel and key protocols for a shared source. We now ask how many private bits each protocol yields from $n$ independent repetitions. Our decoder estimates a site among $MK$ positions and keeps its row. For secrecy we compare a preparation with the site hidden in one row of $K$ positions to the preparation averaged over the full table. Taking the difference of the logarithms of these two population sizes determines the number of usable bits.

The mean and variance of a log-likelihood ratio describe how binary testing changes with $n$. For states $\rho,\sigma$ on one finite-dimensional system, with $\operatorname{supp}(\rho)\subseteq\operatorname{supp}(\sigma)$, define the relative entropy of Umegaki~\cite{Umegaki} and the relative entropy variance used in second-order quantum hypothesis testing by Tomamichel and Hayashi~\cite{TH} and Li~\cite{Li} as follows:
\begin{align}
 D(\rho\Vert\sigma)&\coloneqq\Tr\!\left[(\log_2\rho-\log_2\sigma)\rho\right],\label{eq:relative-entropy}\\
 V(\rho\Vert\sigma)&\coloneqq\Tr\!\left[(\log_2\rho-\log_2\sigma-D(\rho\Vert\sigma)I)^2\rho\right].
 \label{eq:relative-moments}
\end{align}
The traces are taken on the support of $\sigma$, and we may extend $\log_2\rho$ by zero on the kernel of $\rho$ in these expressions. For the joint state $\rho_{XBE}$ of the channel in \eqref{eq:cq} or the source in \eqref{eq:source-state}, define
\begin{align}
 I_J&\coloneqq D(\rho_{XJ}\Vert\rho_X\otimes\rho_J),\label{eq:rate-information}\\
 V_J&\coloneqq V(\rho_{XJ}\Vert\rho_X\otimes\rho_J),\qquad J\in\{B,E\}.\label{eq:rate-moments}
\end{align}
For separable quantum outputs, use the spectral mean and variance in Appendix~\ref{sec:separable}, Eqs.~\eqref{eq:separable-rate-information}--\eqref{eq:separable-rate-moments}. In finite dimensions these agree with the traces above. Thus $I_J$ is the mutual information $I(X;J)_\rho$. The leading rate is Bob's information advantage $I_B-I_E$. The variances $V_B$ and $V_E$ determine the fluctuations at large but finite $n$. The standard normal cumulative distribution function is denoted by~$\Phi$.

\Needspace{10\baselineskip}
\begin{theorem}[Second-order achievability for private messages and secret keys]\label{thm:rate}
Consider a known classical--quantum wiretap channel with a fixed input distribution, or a known source, on separable quantum systems, with one-copy state $\rho_{XBE}$ as in \eqref{eq:cq} or \eqref{eq:source-state}, respectively, and finite classical alphabet $\mathcal X$. Suppose $I_B>I_E$ and $V_B,V_E>0$ as in \eqref{eq:relative-entropy}--\eqref{eq:rate-moments}, using \eqref{eq:separable-rate-information}--\eqref{eq:separable-rate-moments} for infinite-rank outputs, and let $\Phi$ denote the standard normal cumulative distribution function. Fix $\varepsilon_B,\varepsilon_E>0$ such that $\varepsilon_B+\varepsilon_E<1$. For all sufficiently large $n$, there exists a code for $n$ independent uses of this channel, or a one-way key protocol for $n$ independent copies of this source, respectively, with $M$ messages or key values satisfying
\begin{equation}\label{eq:third-net}
 \log_2 M\ge n(I_B-I_E)+\sqrt{nV_B}\,\Phi^{-1}(\varepsilon_B)+\sqrt{nV_E}\,\Phi^{-1}(\varepsilon_E)-\log_2 n-O(1).
\end{equation}
The channel code uses $n$ parallel inputs and a collective measurement on $B^n$. The source protocol uses local randomness and an authenticated public message. The joint error in \eqref{eq:private} or \eqref{eq:key-joint} is at most $\varepsilon_B+\varepsilon_E$. Its reference uses Eve's actual marginal, including the public transcript in the source protocol. All bounded terms can depend on the specified state and error parameters.

The achievable loss $\log_2 n+O(1)$ in \eqref{eq:third-net} decreases by $\tfrac12\log_2 n$ if $[\rho_{XB},\rho_X\otimes\rho_B]=0$ or if each $\rho_B^x$ with $p_X(x)>0$ is pure. Independently it decreases by $\tfrac12\log_2 n$ if the states $\rho_E^x$ with $p_X(x)>0$ are diagonal in a common basis. The source protocol uses $O(n)$ authenticated public bits under the general hypotheses above, including both refinements.
\end{theorem}
Cheng and Gao obtained the Gaussian terms in \eqref{eq:third-net} with an $O(\log n)$ loss at these error tolerances~\cite[Theorems~14 and~19]{CG}. Earlier private and key analyses used a squared secrecy tolerance~\cite{Wilde,KKGW}. Here the integral over the one-pair tests determines the remainder associated with hiding the position.

\begin{proof}
First consider Eve. In the integral bound of Theorem~\ref{thm:cover}, every test satisfying the constraint for all $\sigma_E$ also satisfies it for $\sigma_E=\rho_E$. Allowing all tests for this fixed pair can only increase the maximum signal acceptance probability, so the resulting integral still upper-bounds the covering error. Appendix~\ref{sec:gaussian-cover} applies the normal approximation to this fixed pair of hypotheses, uniformly over all thresholds in the integral. Equations~\eqref{eq:fullroc-integral} and~\eqref{eq:integrated-normal} show that both the uniform placement and the comparison between a row and the full array have covering error at most $\varepsilon_E$ whenever
\begin{equation}
 \log_2 K\ge nI_E+\sqrt{nV_E}\,\Phi^{-1}(1-\varepsilon_E)+\tfrac12\log_2 n+c_0
 \label{eq:third-cover}
\end{equation}
for a sufficiently large constant $c_0$. A smaller secrecy tolerance requires more positions per row.

Next consider Bob. We choose a binary test of $\rho_{XB}^{\otimes n}$ against $(\rho_X\otimes\rho_B)^{\otimes n}$ at type~I error probability at most $\varepsilon_B-1/\sqrt n$. Li's testing estimate~\cite[Theorem~5, Eq.~(44)]{Li} and its separable-space version~\cite[Proposition~A.2]{OMW} give a constant $C$ independent of $n$ such that
\begin{align}
&D_H^{\varepsilon_B-1/\sqrt n}\!\left(\rho_{XB}^{\otimes n}\middle\Vert
 (\rho_X\otimes\rho_B)^{\otimes n}\right)\notag\\
&\qquad\ge nI_B+\sqrt{nV_B}\,\Phi^{-1}\!\left(\varepsilon_B-\frac{1+C}{\sqrt n}\right)
 \label{eq:li-quantile}\\
 &\qquad\ge nI_B+\sqrt{nV_B}\,\Phi^{-1}(\varepsilon_B)-O(1).\label{eq:moving-tolerance}
\end{align}
The second inequality uses the bounded derivative of $\Phi^{-1}$ near the fixed interior value $\varepsilon_B$. The competing $MK-1$ positions then add at most $1/\sqrt n$ to the binary error cost \eqref{eq:cost} if
\begin{equation}
 \log_2(MK)\le nI_B+\sqrt{nV_B}\,\Phi^{-1}(\varepsilon_B)-\tfrac12\log_2 n-c_1
 \label{eq:third-pack}
\end{equation}
for a sufficiently large fixed $c_1$. This is Cheng's packing analysis before selecting the codebook~\cite[Proposition~2]{Cheng}. See also Tomamichel and Hayashi for the second-order testing expansion~\cite{TH}.

Set $K$ to the least positive integer satisfying \eqref{eq:third-cover}, and choose $M$ to be the greatest integer satisfying \eqref{eq:third-pack}. Since $I_B>I_E$, the number of rows grows exponentially and integer rounding loses at most a bounded number of bits. Subtract \eqref{eq:third-cover} from the packing threshold and use $\Phi^{-1}(1-\varepsilon_E)=-\Phi^{-1}(\varepsilon_E)$ to obtain \eqref{eq:third-net}. Theorems~\ref{thm:wire} and~\ref{thm:key} give the joint error bound. For the compressed table, \eqref{eq:pair-independent-moment}--\eqref{eq:public-source-acceptance} give the same range and second-moment bounds for Eve's single-pair observable. Its threshold tests therefore obey \eqref{eq:fullroc-integral}, and \eqref{eq:public-site-error}--\eqref{eq:public-partition-error} retain Bob's packing estimate. Since the support of $X^n$ has size at most $|\mathcal X|^n$ and $\log_2(MK)=O(n)$, \eqref{eq:transcript-size} gives $O(n)$ public bits.

We finish with the two improvements. When Bob's testing states commute, diagonalize them simultaneously. Let $Z_n$ be the log-likelihood ratio in their common eigenbasis under the signal hypothesis and set $z\coloneqq\log_2(MK-1)$. The binary cost in \eqref{eq:cost} is
\begin{equation}
 \mathbb E\min\{1,2^{z-Z_n}\}
 =\Pr\{Z_n\le z\}+\mathbb E\!\left[2^{z-Z_n}\boldsymbol 1_{\{Z_n>z\}}\right].
 \label{eq:commute-pack-cost}
\end{equation}
The last term is $O(n^{-1/2})$ by summing unit-interval probabilities above $z$ with geometric weights, as in \eqref{eq:spectral-local}. Decreasing the normal quantile $nI_B+\sqrt{nV_B}\Phi^{-1}(\varepsilon_B)$ by a sufficiently large constant offsets this error. This removes the logarithmic term from \eqref{eq:third-pack}. For pure conditional states Bob can use the projected test in Appendix~\ref{sec:pure-pack} to obtain the same improvement, including for noncommuting outputs.

When Eve's conditional states are simultaneously diagonal, her normalized trace distances equal the total variation distances in their common eigenbasis. Theorem~\ref{thm:classical-rate} removes the logarithmic term from the required $\log_2 K$, and \eqref{eq:classical-all-exponent} controls the comparison with Eve's actual marginal for either private messages or secret keys. The same variance calculation applies to the pairwise-independent public arrays, as explained after \eqref{eq:classical-all-exponent}. The reductions on Bob's and Eve's sides can therefore be used together.
\end{proof}
\section{Conclusion}\label{sec:conclusion}
\subsection{Summary}
Convex splitting hides a correlation by randomizing its position. Lemma~\ref{lem:moment} converts a measurement seeking this correlation into a single-pair observable with a controlled second moment. Theorem~\ref{thm:ordinary} integrates its binary threshold tests. The positive-part argument in Theorem~\ref{thm:cover} gives Corollary~\ref{cor:scalar}, improving the slack dependence and removing the spectral factor in Cheng and Gao's direct bound~\cite[Theorem~4]{CG}.

Together with earlier results, binary hypothesis testing provides a common formulation of one-shot achievability for unassisted classical communication, entanglement-assisted classical communication, quantum communication, one-way entanglement distillation, private communication, and secret key distillation~\cite{WangRenner,AJWPack,Cheng,CW,CG}. The present analysis strengthens the privacy bounds within this framework. The convex-split estimates hold for normal states on separable Hilbert spaces, and the finite-alphabet channel and source protocols permit quantum outputs on separable Hilbert spaces (Appendix~\ref{sec:separable}).

Comparing two position distributions retains Eve's actual marginal in the joint reliability and secrecy criterion, including the public announcement in key distillation. Corollary~\ref{cor:transcript} reduces that announcement to a linear number of public bits at fixed rates. Theorem~\ref{thm:rate} recovers the Gaussian coding terms at the prescribed error tolerances with an explicit logarithmic bound on the remaining achievable loss.

\subsection{Future directions}
Appendix~\ref{sec:classical-cover-order} determines the size of the independent classical convex-split experiment to within a bounded number of bits. For noncommuting states our achievability proof allows an additional half-logarithmic term. Removing this loss would improve both achievable coding rates. A matching covering converse would identify a limitation of this construction. A separate question concerns the precise coefficient in a one-divergence bound with positive slack: Appendix~\ref{sec:sharpness} bounds that coefficient from both sides. For key distillation, the support bound in Corollary~\ref{cor:transcript} controls transmitted bits. Constructing the corresponding encoder and decoder with computation polynomial in the source blocklength remains open for general input distributions.

\paragraph{Acknowledgments.}
The author acknowledges funding from the Cornell University School of Electrical and Computer Engineering.

\paragraph{Statement on AI usage.}
ChatGPT assisted with proof formulation and exploration, exposition, literature checks, figure preparation, and numerical tests during the development of this manuscript. It also helped draft and revise the manuscript text. The checks performed during this assistance were internal reviews, not independent external peer review or formal verification. Responsibility for any submitted version and its claims rests with the authors.

\phantomsection\addcontentsline{toc}{section}{References}
\begingroup
\fontsize{9.5}{10.4}\selectfont
\setlength{\bibsep}{.5pt}
\interlinepenalty=10000
\bibliographystyle{unsrtnat}
\bibliography{convex_split_references}
\endgroup

\clearpage
\appendix
\linespread{.94}\selectfont
\addtolength{\abovedisplayskip}{-0.7pt}\addtolength{\belowdisplayskip}{-0.7pt}
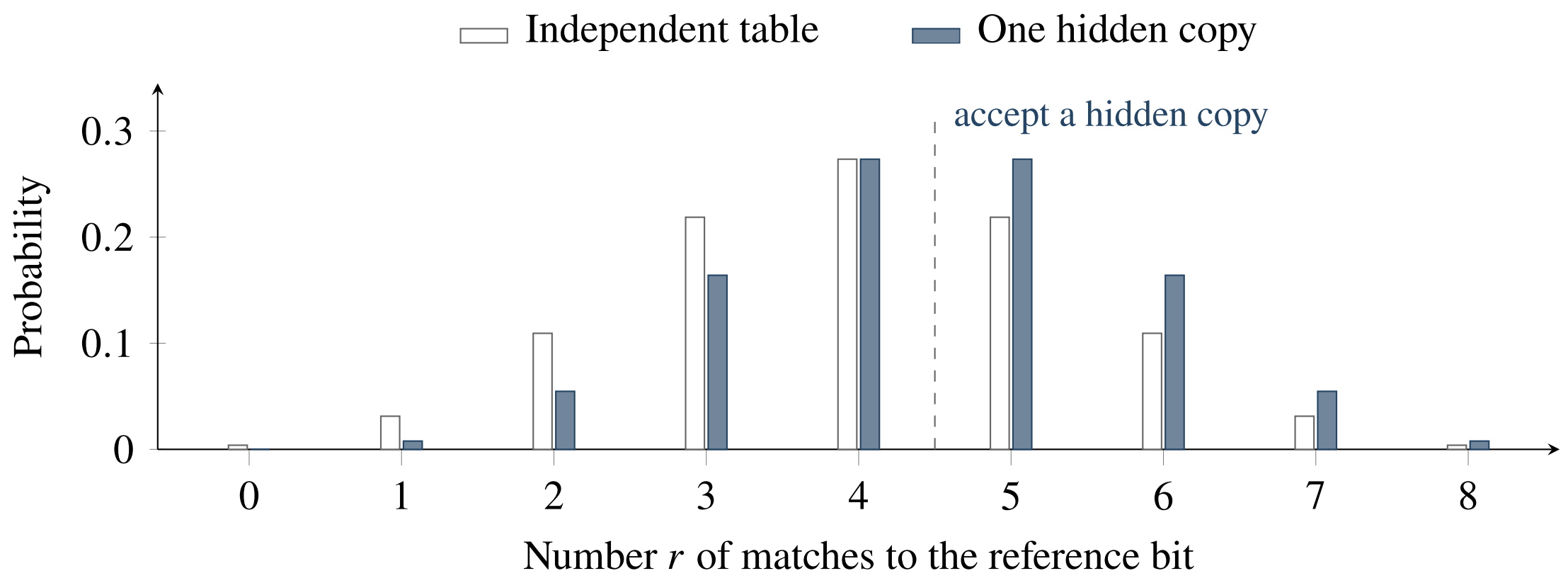
\begin{figure}[!t]
\centering
\begin{tikzpicture}
\begin{axis}[width=.85\linewidth,height=5.0cm,
 xmin=-.6,xmax=8.6,ymin=0,ymax=.345,
 axis x line=bottom,axis y line=left,
 xlabel={Number $r$ of matches to the reference bit},ylabel={Probability},
 xtick={0,...,8},ytick={0,.1,.2,.3},tick label style={font=\small},
 label style={font=\small},clip=false,
 ybar=0pt,bar width=5pt]
\addplot[draw=black!60,fill=white,bar shift=-3pt] coordinates
 {(0,.00390625)(1,.03125)(2,.109375)(3,.21875)(4,.2734375)(5,.21875)(6,.109375)(7,.03125)(8,.00390625)};
\addplot[draw=inkblue,fill=inkblue!65,bar shift=3pt] coordinates
 {(0,0)(1,.0078125)(2,.0546875)(3,.1640625)(4,.2734375)(5,.2734375)(6,.1640625)(7,.0546875)(8,.0078125)};
\node[anchor=south,font=\small] at (rel axis cs:.5,1.045)
 {\setlength{\fboxsep}{0pt}\fcolorbox{black!60}{white}{\rule{0pt}{.68ex}\hspace{1.1em}}\, Independent table
  \qquad\fcolorbox{inkblue}{inkblue!65}{\rule{0pt}{.68ex}\hspace{1.1em}}\, One hidden copy};
\draw[dashed,black!60] (axis cs:4.5,0)--(axis cs:4.5,.318);
\node[anchor=west,font=\footnotesize,text=inkblue] at (axis cs:4.52,.313) {accept a hidden copy};
\end{axis}
\end{tikzpicture}
\caption{Detecting a copy of a fair bit among eight positions. The height of each open bar is $2^{-8}\binom 8r$. For a planted copy the probability is $2r/8$ times as large. The optimal decision accepts a hidden copy when $r>4$; either decision at $r=4$ gives the same error for equal priors. Summing the excess of the filled over the open bars to the right of the dashed line gives the normalized trace distance in \eqref{eq:intro-bit}.}\label{fig:bit-test}
\end{figure}
\section[Two examples]{Two examples of hiding a correlation}\label{sec:split-examples}
A copied bit and a Bell pair have the same one-qubit marginals. Their correlations nevertheless call for different measurements. We evaluate the one-pair testing curve \eqref{eq:roc} and compare the bounds of Section~\ref{sec:cover} with the exact distance in each example.

\subsection{A copied fair bit}
Suppose that $A$ and $B$ store the same fair bit and $\tau_A=I_A/2$. The hidden preparation in \eqref{eq:ordinary} overwrites one uniformly chosen table entry with a copy of $B$. In the independent preparation \eqref{eq:background}, every entry is a fresh fair bit. Figure~\ref{fig:bit-test} compares the two distributions for $K=8$.

Let $R$ count the table entries that match $B$. Under independent preparation it has the binomial distribution $\operatorname{Bin}(K,1/2)$. Conditional on the reference bit, a table with $R$ matches is $2R/K$ times as likely to occur when a copy was inserted. Indeed, any of its $R$ matching sites could have been overwritten, and each such choice doubles the probability of that table. The best yes-or-no test therefore accepts the hidden copy for $R>K/2$. Its acceptance advantage, which equals the normalized trace distance by \eqref{eq:distance}, is
\begin{equation}
 \mathbb E\!\left[\left(\frac{2R}{K}-1\right)_+\right]
     =\frac{1}{K}\mathbb E\!\left[|R-K/2|\right],
 \label{eq:intro-bit}
\end{equation}
Here the expectation is under independent preparation. The equality follows because $2R/K-1$ has mean zero. At $K=1000$ the difference in \eqref{eq:intro-bit} is $0.01261$. With equal prior probabilities, the optimal probability of a correct guess is one half plus half of \eqref{eq:intro-bit}, or about $50.63\%$.

For a one-pair test, reject mismatches and accept matches with probability $\min\{1,2\eta\}$. Its type~II error probability is at most $\eta$ for every independent reference because the fair bit in $A$ matches $B$ with probability $1/2$. Conversely, for the uniform independent reference the signal density is at most twice the background density. Hence
\begin{equation}
 G^\eta(\rho_{AB}\Vert\tau_A)=\min\{1,2\eta\}.
\end{equation}
At $K=1000$ the upper bound in \eqref{eq:ordinary-curve} is $0.04015$. The exact distance in \eqref{eq:intro-bit} is $0.01261$.

For the copied fair bit at $\varepsilon=1/5$, $\delta=1/10$, the testing probability in \eqref{eq:scalar-beta} is $b=1/20$. The background has one distinct eigenvalue and $\dim B=2$, so the spectral factor uses $\nu=2$. At these parameters Cheng and Gao's bound~\cite[Theorem~4]{CG} and \eqref{eq:clean} give sufficient populations of $64{,}800{,}000$ and $9$, respectively, after rounding up. There is a simpler bound for this particular distribution. Since $R$ in \eqref{eq:intro-bit} has variance $K/4$, Cauchy--Schwarz gives
\begin{equation}
 \frac12\norm{\omega_{A^K B}-\omega^0_{A^K B}}_1
 =\frac{\mathbb E|R-K/2|}{K}\le\frac{1}{2\sqrt K}.
\end{equation}
This certifies $K=7$, as does the excess test in \eqref{eq:layer} at $t=0$ for this testing curve. The exact minimum is $4$ by \eqref{eq:sharp-binomial}.

\subsection{A Bell pair at two possible sites}
Classical bit comparisons can be performed simultaneously. To see what changes if the two tests for a match are incompatible, take $A$ and $B$ to be qubits and set
\begin{align}
 |\Phi^2\rangle_{AB}&\coloneqq (|00\rangle_{AB}+|11\rangle_{AB})/\sqrt2,\label{eq:bell-ket}\\
 \Phi^2_{AB}&\coloneqq|\Phi^2\rangle\langle\Phi^2|_{AB},\label{eq:bell-example}\\
 \tau_A&\coloneqq I_A/2.\label{eq:bell-tau}
\end{align}
For $\eta\in[0,1]$, the effect $\min\{1,4\eta\}\Phi^2_{AB}$ has type~II error probability at most $\eta$ on every $\tau_A\otimes\sigma_B$ since $\E_A^\tau(\Phi^2_{AB})=I_B/4$. Conversely, testing against the particular alternative $I_{AB}/4$ forces $\Tr[Q_{AB}]\le4\eta$, so $\Tr[Q_{AB}\Phi^2_{AB}]\le\min\{1,4\eta\}$. We obtain $G^\eta(\Phi^2_{AB}\Vert I_A/2)=\min\{1,4\eta\}$.

For $K=2$ let $P^i_{A^2B}\coloneqq\Phi^2_{A_iB}$ for $i\in\{1,2\}$, with the identity on the other site understood. These candidate match tests share $B$ and do not commute. The hidden and independent states are $(P^1_{A^2B}+P^2_{A^2B})/4$ and $I_{A^2B}/8$. Their distance can still be computed exactly. The respective ranges of the projections have orthonormal bases
\begin{align}
 |a_j\rangle_{A^2B}&\coloneqq |\Phi^2\rangle_{A_1B}|j\rangle_{A_2},\label{eq:bell-a}\\
 |b_j\rangle_{A^2B}&\coloneqq |j\rangle_{A_1}|\Phi^2\rangle_{A_2B},\qquad j\in\{0,1\}.\label{eq:bell-b}
\end{align}
The overlaps $\langle a_j|b_k\rangle=\delta_{jk}/2$ give eigenvectors $|a_j\rangle_{A^2B}\pm|b_j\rangle_{A^2B}$ of $P^1_{A^2B}+P^2_{A^2B}$ with eigenvalues $1\pm1/2$. The other four eigenvalues vanish. Thus
\begin{align}
 \tfrac12\norm{\omega_{A^2B}-\omega^0_{A^2B}}_1&=\tfrac12,\label{eq:bell-two}\\
 \tfrac12\norm{\omega_{AB}-\omega^0_{AB}}_1&=\tfrac34.\label{eq:bell-one}
\end{align}
For the one-site value in \eqref{eq:bell-one}, $\Phi^2_{AB}-I_{AB}/4$ has eigenvalue $3/4$ once and $-1/4$ three times. The proof of Theorem~\ref{thm:ordinary} allows such incompatibility because all of its site observables are derived from one joint effect $W_{A^NB}$. For the two-site instance, substituting the Bell testing curve into \eqref{eq:ordinary-curve} gives an upper bound of about $0.8723$. The excess test in Theorem~\ref{thm:cover} yields a tighter estimate here. With $t=1/\sqrt{24}$ in \eqref{eq:layer} it certifies the smaller value $\sqrt{3/8}\simeq0.6124$, compared with the exact $1/2$.

\section[Sharpness of the constants]{How large can the improvement in the constants be?}\label{sec:sharpness}
Corollary~\ref{cor:scalar} gives a sufficient number of positions using one hypothesis-testing relative entropy. Here $\varepsilon$ is the desired covering error and $1-\varepsilon+\delta$ is the type~I error probability tolerance, where $\delta$ is the slack. The composite type~II error probability $b_{1-\varepsilon+\delta}(\rho_{AB}\Vert\tau_A)$ is defined in \eqref{eq:beta}. We give two classical examples. The first determines the numerical coefficient when $\delta=0$. The second shows that at fixed $\varepsilon$ the dependence on positive $\delta$ can be necessary.

\subsection{A uniformly copied symbol}
We can evaluate the one-pair test and the entire $K$-position state exactly when the two records store the same uniform symbol. This turns a universal claim about all quantum states into an explicit restriction on its coefficient. For an alphabet of size $D\ge2$, define
\begin{align}
 \rho_{AB}^{[D]}&\coloneqq D^{-1}\sum_{j=1}^D |j\rangle\langle j|_A\otimes|j\rangle\langle j|_B,\label{eq:copy-D}\\
 \tau_A^{[D]}&\coloneqq I_A/D.\label{eq:copy-background}
\end{align}
The effect that accepts a match with probability $1-\varepsilon$ has type~I error probability $\varepsilon$ and type~II error probability $(1-\varepsilon)/D$ against every alternative. Against the particular alternative $I_{AB}/D^2$, every test of type~I error probability at most $\varepsilon$ has type~II error probability at least $(1-\varepsilon)/D$: indeed $\rho_{AB}^{[D]}\le D(I_{AB}/D^2)$. Consequently
\begin{equation}
 b_\varepsilon(\rho_{AB}^{[D]}\Vert\tau_A^{[D]})=(1-\varepsilon)/D,\qquad 0\le\varepsilon<1 .
 \label{eq:copy-beta}
\end{equation}
Let $\omega^{[D]}_{A^K B}$ and $\omega^{0,[D]}_{A^K B}$ be the uniform placement and independent states in \eqref{eq:ordinary} and \eqref{eq:background}, respectively, for \eqref{eq:copy-D}--\eqref{eq:copy-background}. Under the independent preparation, condition on the symbol at $B$ and let $N$ count the matches among $K$ positions. The likelihood ratio of the hidden-copy preparation to the independent preparation is $DN/K$. Since $N\sim\operatorname{Binomial}(K,1/D)$ and $\mathbb E[DN/K]=1$, the expected deficit gives
\begin{align}
 \tfrac12\norm{\omega^{[D]}_{A^K B}-\omega^{0,[D]}_{A^K B}}_1
 &=\mathbb E[(1-DN/K)_+]\label{eq:binomial-deficit}\\
 &=\frac{\mathbb E|N-K/D|}{2K/D}.
 \label{eq:sharp-binomial}
\end{align}
For $K\le D$ only $N=0$ contributes to the positive part, yielding $(1-1/D)^K$.

Consider the population rule \eqref{eq:scalar-zero} at $\delta=0$. A universal coefficient $C$ in a sufficient condition $K\ge C/b_{1-\varepsilon}^2$ must satisfy $C\ge\sup K b_{1-\varepsilon}^2$, where the supremum runs over instances with distance greater than $\varepsilon$. For $D=2$ and $K=1$, \eqref{eq:sharp-binomial} gives distance $1/2$, and \eqref{eq:copy-beta} gives $b_{1-\varepsilon}=\varepsilon/2$. Taking the supremum over $0<\varepsilon<1/2$ yields
\begin{equation}
 C\ge\sup_{0<\varepsilon<1/2}(\varepsilon/2)^2=1/16.
\end{equation}
Corollary~\ref{cor:scalar} attains this coefficient.

For positive slack, define $C_{\mathrm{opt}}$ to be the infimum of all universal coefficients $C$ for which the sufficient population
\begin{equation}
 K\ge\frac{C}{\delta\, b_{1-\varepsilon+\delta}(\rho_{AB}\Vert\tau_A)},\qquad 0<\delta<\varepsilon<1,
 \label{eq:linear-form}
\end{equation}
guarantees covering error at most $\varepsilon$ whenever the denominator is positive. The uniform copying example and Corollary~\ref{cor:scalar} imply
\begin{equation}
 \frac{1}{8\mathrm e}\le C_{\mathrm{opt}}\le\frac1{16}.
 \label{eq:sharp-linear}
\end{equation}
Indeed, take $K\coloneqq\lfloor D/2\rfloor$, choose tolerances $\varepsilon$ just below $(1-1/D)^K$ tending to $e^{-1/2}$, and set $\delta\coloneqq\varepsilon/2$. Their covering errors exceed $\varepsilon$, yet
\begin{equation}
 K\delta\,b_{1-\varepsilon+\delta}(\rho_{AB}^{[D]}\Vert\tau_A^{[D]})
 =\frac KD\frac{\varepsilon^2}{4}\longrightarrow\frac1{8\mathrm e}.
\end{equation}
This proves the lower bound in \eqref{eq:sharp-linear}. The upper bound follows from \eqref{eq:clean}.

\subsection{Positive slack at a fixed covering error}
The uniformly copied symbol in \eqref{eq:copy-D} allowed the target error to vary. We now fix $\varepsilon\in(0,1)$ and let the single-pair testing probability become small. The example separates a rarely selected, almost perfectly distinguishable signal from one whose empirical match count fluctuates near its mean.

Register $A$ contains two independent uniform symbols with respective alphabet sizes $D$ and $L$. An independent flag chooses the $L$-symbol with probability $q$ and the $D$-symbol otherwise. Register $B$ records both the flag and the chosen symbol. Denote their joint state by $\rho_{AB}$ and use $\tau_A=\rho_A$. Conditioned on the flag this is the experiment in \eqref{eq:copy-D} with the unused symbol independent. Thus the trace distance of the split states is the corresponding mixture of the two quantities in \eqref{eq:sharp-binomial}. At uniform type~II error probability at most $\eta$, the largest signal acceptance probability is
\begin{equation}
 G^\eta(\rho_{AB}\Vert\rho_A)
 =q\min\{1,L\eta\}+(1-q)\min\{1,D\eta\}.
 \label{eq:sharp-flagged-roc}
\end{equation}
For a fixed $B$ value, the density of the matching input relative to the uniform background is bounded by the size of its alphabet. A test accepting a match with probability $\min\{1,L\eta\}$ or $\min\{1,D\eta\}$, according to the flag, attains these two bounds simultaneously. Testing against each classical $B$ value bounds any quantum effect in the same way, since only its diagonal entries enter these states.

The copied bit in Appendix~\ref{sec:split-examples} gives a finite illustration when a rare flag reveals a larger alphabet. Take $D=2$, $q=1/20$, $\varepsilon=1/5$, and $\delta=1/10$. For $L\ge38$, \eqref{eq:sharp-flagged-roc} gives $b_{1-\varepsilon+\delta}(\rho_{AB}\Vert\rho_A)=1/38$; \eqref{eq:clean} certifies $K=19$ independently of $L$. By contrast, the variance of the single-site likelihood ratio under independent preparation is $(L+18)/20$. Conditioned on a copied alphabet of size $d$, that ratio equals $d$ with probability $1/d$ and is zero otherwise. The sites have zero covariance even though they share the flag and reference symbol, since their conditional means all equal one. A Cauchy--Schwarz bound using just the resulting overall variance certifies $K\ge\lceil5(L+18)/16\rceil$. The rare flag can raise the variance even when its probability fits within the allowed testing error. Conditioning on the flag instead gives the bound $q+(1-q)/(2\sqrt K)$: the distance is at most one for the $L$-symbol and at most $1/(2\sqrt K)$ for the fair bit. This certifies $K=11$.

Set $K\coloneqq D^3$, $L\coloneqq K^3$, and let $N\sim\operatorname{Binomial}(K,1/D)$ as in \eqref{eq:sharp-binomial}. Choose
\begin{align}
 z&\coloneqq\frac{\mathbb E|N-K/D|}{2K/D},\label{eq:flag-binomial-error}\\
 \delta&\coloneqq(1-\varepsilon)z/2,\label{eq:flag-slack}\\
 q&\coloneqq\varepsilon-2\delta.\label{eq:sharp-sequence}
\end{align}
The centered binomial count divided by its standard deviation converges in distribution to a standard normal variable. Its second moment is one, giving uniform integrability of its absolute value. Hence $z\sim1/\sqrt{2\pi K/D}$, and the parameters in \eqref{eq:flag-binomial-error}--\eqref{eq:sharp-sequence} are admissible for large $D$. Since $1-(1-1/L)^K\le K/L=o(z^2)$, the covering error of the flagged state satisfies
\begin{align}
 \tfrac12\norm{\omega_{A^K B}-\omega^0_{A^K B}}_1
 &=q(1-1/L)^K+(1-q)z\\
 &=\varepsilon+(1-\varepsilon)z^2-o(z^2)>\varepsilon.
\end{align}
To invert \eqref{eq:sharp-flagged-roc}, require acceptance $\varepsilon-\delta=q+\delta$. At $\eta=\delta/[(1-q)D]$ we have $L\eta\ge1$ and $D\eta<1$ for large $D$. This gives
\begin{align}
 b_{1-\varepsilon+\delta}(\rho_{AB}\Vert\rho_A)&=\frac{\delta}{(1-q)D},\\
 \frac{b_{1-\varepsilon+\delta}(\rho_{AB}\Vert\rho_A)}{\delta}&\longrightarrow0,\\
 K\delta\,b_{1-\varepsilon+\delta}(\rho_{AB}\Vert\rho_A)&\longrightarrow\frac{1-\varepsilon}{8\pi}.
 \label{eq:sharp-fixed-epsilon}
\end{align}
Thus a sufficient population $c_\varepsilon(\delta)/b_{1-\varepsilon+\delta}(\rho_{AB}\Vert\rho_A)$ cannot hold with $c_\varepsilon(\delta)=o(1/\delta)$ in all dimensions. For a quantitative upper comparison define $C_\varepsilon(\delta)$ as the supremum of $K\delta\,b_{1-\varepsilon+\delta}(\rho_{AB}\Vert\tau_A)$ over instances with covering error above $\varepsilon$. The normalized test estimate \eqref{eq:normalized}, applied at $t=\delta/(1-\varepsilon+\delta)$, gives $C_\varepsilon(\delta)\le(1-\varepsilon+\delta)^2/[16(1-\varepsilon)]$. Therefore
\begin{equation}
 \frac{1-\varepsilon}{8\pi}\le\limsup_{\delta\downarrow0}C_\varepsilon(\delta)
 \le\frac{1-\varepsilon}{16}.
 \label{eq:fixed-epsilon-coeff}
\end{equation}

\section[Gaussian covering estimate]{The Gaussian estimate for quantum convex splitting}\label{sec:gaussian-cover}
This appendix proves the quantum covering estimate \eqref{eq:third-cover}. The type~II error probability constraint in the integral of Theorem~\ref{thm:cover} changes with the spectral threshold. A bound uniform in that allowance lets us integrate the normal approximation without spending a second logarithmic slack.

First consider states $\rho$ and $\sigma$ on one finite-dimensional system. For $0\le\eta\le1$, define
\begin{equation}
 U^\eta(\rho\Vert\sigma)\coloneqq
 \max_{T}\bigl\{\Tr[T\rho]:0\le T\le I,\ \Tr[T\sigma]\le\eta\bigr\}.
 \label{eq:fixed-roc-full}
\end{equation}
Every $Q_{AB}$ admissible in \eqref{eq:roc} is admissible in \eqref{eq:fixed-roc-full} for the particular alternative $\tau_A\otimes\rho_B$. Thus $G^\eta(\rho_{AB}\Vert\tau_A)\le U^\eta(\rho_{AB}\Vert\tau_A\otimes\rho_B)$. For ordinary splitting with $K$ positions, or a row of $K$ compared with the entire array, $\tfrac12\norm{p-q}_1\le1$ and $\norm{p_{1:N}-q_{1:N}}_2^2\le1/K$ in \eqref{eq:weights}--\eqref{eq:site-norm}. Equation~\eqref{eq:layer} consequently gives
\begin{equation}
 \tfrac12\norm{\Omega_{A^N B}(p)-\Omega_{A^N B}(q)}_1
 \le\int_0^1 U^{1/(1+4Kt^2)}(\rho_{AB}\Vert\tau_A\otimes\rho_B)\,dt .
 \label{eq:fullroc-integral}
\end{equation}

We give the required single-pair estimate for arbitrary $\rho,\sigma$ with $\operatorname{supp}(\rho)\subseteq\operatorname{supp}(\sigma)$. Write their spectral decompositions as
\begin{align}
 \rho&=\sum_i r_i|u_i\rangle\langle u_i|,\\
 \sigma&=\sum_j s_j|v_j\rangle\langle v_j|.
\end{align}
Under the Nussbaum--Szko\l{}a distribution~\cite{NS}, the pair $(i,j)$ has probability $r_i|\langle u_i|v_j\rangle|^2$. Set $Z\coloneqq\log_2(r_i/s_j)$ on this distribution's support. Expanding the two traces defining the relative entropy and its variance in Section~\ref{sec:spectral-curve} shows that $Z$ has mean $D\coloneqq D(\rho\Vert\sigma)$ and variance $V\coloneqq V(\rho\Vert\sigma)$. For every effect $T$ and every $s>0$,
\begin{align}
 \Tr[(I-T)\rho]+s\Tr[T\sigma]&\ge\sum_{ij}\frac{r_i s s_j}{r_i+s s_j}|\langle u_i|v_j\rangle|^2
 \\
 &=\mathbb E\!\left[\frac1{1+2^Z/s}\right].
 \label{eq:harmonic-test}
\end{align}
The expectation in \eqref{eq:harmonic-test} is over spectral indices $(i,j)$ drawn with the Nussbaum--Szko\l{}a probabilities $r_i|\langle u_i|v_j\rangle|^2$, with $Z=\log_2(r_i/s_j)$ on their support. A term with zero denominator contributes zero. For a proof replace $I-T$ and $T$ by their squares to lower-bound the left side. After inserting the two eigenbases, apply the scalar inequality $r|a|^2+s'|b|^2\ge rs'|a+b|^2/(r+s')$ with $a\coloneqq\langle u_i|(I-T)|v_j\rangle$, $b\coloneqq\langle u_i|T|v_j\rangle$, $r\coloneqq r_i$, and $s'\coloneqq s s_j$. This bound was proved by Jak\v{s}i\'c et al.~\cite[Proposition~3.2(2)]{JOPS} and by Datta, Pautrat, and Rouz\'e~\cite[Lemma~2, Eq.~(3.11)]{DPR}. Cheng and Liu recover it within a more general interpolation~\cite[Theorem~4.2 and Remark~B.2]{CL}.

Write $Z_n\coloneqq\sum_{\ell=1}^n Z^{(\ell)}$ for the sum of $n$ independent copies of the spectral log-likelihood $Z$, drawn from the probability distribution specified in \eqref{eq:harmonic-test}. All probabilities and expectations involving $Z_n$ in this appendix use this product distribution. Suppose $V>0$. The third absolute centered moment is finite on the finite-dimensional spectral support and, for the separable-output models here, by \eqref{eq:separable-moments}. Berry--Esseen gives a Gaussian approximation to $\Pr\{Z_n\le z\}$ uniformly in $z$, with an error at most $C/\sqrt n$. Subtracting two such estimates bounds every unit-interval probability. Summing these probabilities with geometric weights yields, uniformly in $z$,
\begin{align}
 \Pr\{z-1<Z_n\le z\}&\le C'/\sqrt n,\label{eq:spectral-interval}\\
 \mathbb E\!\left[2^{Z_n-z}\boldsymbol 1_{\{Z_n\le z\}}\right]&\le 2C'/\sqrt n,\label{eq:spectral-lower-tail}\\
 \mathbb E\!\left[2^{z-Z_n}\boldsymbol 1_{\{Z_n>z\}}\right]&\le 2C'/\sqrt n .
 \label{eq:spectral-local}
\end{align}
Apply \eqref{eq:harmonic-test} to $n$ copies at $s=2^z$. The quantity $1-(1+2^{Z_n-z})^{-1}$ is at most one above $z$ and at most $2^{Z_n-z}$ below $z$. Thus \eqref{eq:fixed-roc-full} satisfies
\begin{equation}
 U^\eta(\rho^{\otimes n}\Vert\sigma^{\otimes n})
 \le 1-\Phi\!\left(\frac{z-nD}{\sqrt{nV}}\right)+\frac{C''}{\sqrt n}+2^z\eta .
 \label{eq:uniform-roc}
\end{equation}
In \eqref{eq:fullroc-integral} use $n$ independent pairs, set $\eta\coloneqq1/(1+4Kt^2)$, and select
\begin{equation}
 z\coloneqq\log_2(4Kt^2)-\tfrac12\log_2 n,\qquad t>0 .
\end{equation}
This choice ensures $2^z\eta\le1/\sqrt n$ at every threshold. For a fixed $0<\varepsilon<1$, suppose $\log_2K\ge nD+\sqrt{nV}\,\Phi^{-1}(1-\varepsilon)+\tfrac12\log_2 n+c_0$. Applying \eqref{eq:uniform-roc} inside the integral and using $\sup_x\Phi'(x)=1/\sqrt{2\pi}$ gives the following upper bound on the right side of \eqref{eq:fullroc-integral}:
\begin{equation}
 1-\Phi\!\left(\Phi^{-1}(1-\varepsilon)+\frac{c_0+2}{\sqrt{nV}}\right)
 +\frac{C''+1}{\sqrt n}
 +\frac{2\int_0^1|\log_2t|\,dt}{\sqrt{2\pi nV}}.
 \label{eq:integrated-normal}
\end{equation}
The integral equals $1/\ln 2$. The density $\Phi'$ is positive at the fixed quantile, so a sufficiently large $c_0$ makes \eqref{eq:integrated-normal} at most $\varepsilon$ for all sufficiently large $n$. Taking $\rho=\rho_{XE}$, and $\sigma=\rho_X\otimes\rho_E$ proves \eqref{eq:third-cover}.
\section[Classical convex splitting]{A bounded remainder for classical convex splitting}
\label{sec:classical-cover-order}
In the classical case the observation can include the likelihood ratio at every position. Typical fluctuations of their average determine the covering error. We obtain both an upper bound and a lower bound for the prescribed convex-split distribution. We fix the sampling distribution and estimate the mean distance at fixed error. Liu, Cuff, and Verd\'u studied classical resolvability with a likelihood-ratio criterion~\cite{LCV}. Classical and classical--quantum resolvability can instead optimize the codebook for a given output distribution~\cite{WHres,HCG}. Related results concern high-probability guarantees~\cite{Cuff} and the exponential decay of the mean distance away from the threshold~\cite{YC,Yassaee,CGSoft}.

\begin{theorem}[Classical convex-split sample size]\label{thm:classical-rate}
Let $p_{AB}$ be a probability distribution on finite $A$ and finite or countable $B$, $p_B$ its marginal, and $\tau_A$ a distribution with $p_{AB}\ll\tau_A p_B$. Define the information density $Z\coloneqq\log_2[p_{A|B}(A|B)/\tau_A(A)]$ under $p_{AB}$, its mean $D\coloneqq\mathbb E Z$, and its variance $V\coloneqq\operatorname{Var}(Z)>0$. For $n$ independent copies let $Z_n$ be the sum of the $n$ information densities. Let $\omega_{A^{nK}B^n}$ and $\omega^0_{A^{nK}B^n}$ be the states in \eqref{eq:ordinary} and \eqref{eq:background} for the uniform placement of one copy of $p_{AB}^{\otimes n}$ among $K$ blocks of $n$ symbols and for independent preparation, respectively. With diagonal states representing the distributions, the independent state is $\omega^0_{A^{nK}B^n}=(\tau_A^{\otimes n})^{\otimes K}\otimes p_B^{\otimes n}$. Write $K_\varepsilon(n)$ for the least positive integer $K$ with covering error at most the fixed $\varepsilon\in(0,1)$. For $\log_2K=nD+O(\sqrt n)$ the distance obeys \eqref{eq:classical-cover-approximation}, and the least population obeys \eqref{eq:classical-cover-optimum}:
\begin{align}
 \tfrac12\norm{\omega_{A^{nK}B^n}-\omega^0_{A^{nK}B^n}}_1&=\Pr\{Z_n>\log_2K\}+O(n^{-1/2}),\label{eq:classical-cover-approximation}\\
 \log_2K_\varepsilon(n)&=nD+\sqrt{nV}\,\Phi^{-1}(1-\varepsilon)+O(1).
 \label{eq:classical-cover-optimum}
\end{align}
Here $\Phi$ is the standard normal cumulative distribution function. In \eqref{eq:classical-cover-approximation} the error is uniform for $|\log_2K-nD|\le C\sqrt n$ at each fixed $C$. Equation~\eqref{eq:classical-cover-optimum} gives the sample size to within a bounded number of bits.
\end{theorem}

\begin{proof}
We first prove the claim for finite alphabets. Appendix~\ref{sec:separable} gives the uniform conditional moment estimates needed for countable $B$. We condition on the output string $B^n$ and estimate the average of the likelihood ratios. Values much larger than $K$ occur infrequently under the product reference but carry a substantial part of the signal probability. After removing such values, we bound the fluctuations of the average. The cutoff will depend on $B^n$ so that this fluctuation estimate works uniformly enough to average. Our first aim is the upper bound, with $z\coloneqq\log_2K$,
\begin{equation}
 \tfrac12\norm{\omega_{A^{nK}B^n}-\omega^0_{A^{nK}B^n}}_1\le\Pr\{Z_n>z\}+O(n^{-1/2}).
 \label{eq:classical-cover-tail}
\end{equation}
To evaluate \eqref{eq:classical-cover-tail}, condition on $B^n=b$. Under the independent preparation the $K$ position likelihoods have conditional mean one and are conditionally independent. The average of the $K$ likelihoods is the density of the uniform mixture relative to the independent preparation. Consequently,
\begin{align}
 L_i&\coloneqq\frac{p_{A^n|B^n}(A_i^n|b)}{\tau_A^{\otimes n}(A_i^n)},\label{eq:conditional-site-likelihood}\\
 F_b&\coloneqq\mathbb E_{\mathrm{ind}}\!\left[\left(1-K^{-1}\sum_{i=1}^KL_i\right)_+\middle|B^n=b\right],\label{eq:conditional-deficit-definition}\\
 \tfrac12\norm{\omega_{A^{nK}B^n}-\omega^0_{A^{nK}B^n}}_1&=\mathbb E_{p_B^{\otimes n}} F_b .
 \label{eq:conditional-deficit}
\end{align}
The subscript in \eqref{eq:conditional-deficit-definition} indicates sampling from the independent preparation. For the information density $Z_n$, use the signal distribution $p_{AB}^{\otimes n}$ and define
\begin{align}
 q_b(t)&\coloneqq\Pr\{Z_n>t\mid b\},\label{eq:classical-tail-function}\\
 m_b&\coloneqq\mathbb E[Z_n\mid b],\\
 V_b&\coloneqq\operatorname{Var}(Z_n\mid b),\\
 w&\coloneqq\mathbb E\operatorname{Var}(Z\mid B).\label{eq:classical-conditional-variance}
\end{align}
First suppose $w>0$. Except on a set of exponentially
small $p_B^{\otimes n}$ probability, $V_b\ge nw/2$ and $|m_b-z|\le\kappa n$
for any sufficiently small fixed $\kappa>0$. Hoeffding's inequality applies to the sums of the bounded single-letter conditional means and variances, and $z-nD=O(\sqrt n)$. For countable $B$, \eqref{eq:separable-classical-tail} gives these bounds uniformly. On this set
the Berry--Esseen bound for independent, possibly nonidentically distributed
summands gives, uniformly in the location of an interval $I$,
\begin{equation}
 \Pr\{Z_n\in I\mid b\}\le C(1+|I|)/\sqrt n,
 \label{eq:classical-interval}
\end{equation}
where $|I|$ denotes interval length.

We need to control how far the cutoff can move on a typical output string. For this purpose we bound the upper-tail probability $q_b(z)$ from below.
On the same set,
\begin{equation}
 -\log_2 q_b(z)\le C\bigl(1+(z-m_b)_+^2/n\bigr).
 \label{eq:conditional-upper-tail}
\end{equation}
To obtain this estimate, increase the weight of larger values until a whole concentration interval around the new mean lies above $z$. More precisely, under the conditional signal distribution at $B^n=b$, tilt the distribution of the $\ell$th information density $Z^{(\ell)}$ by $\exp(\lambda Z^{(\ell)})/\mathbb E[\exp(\lambda Z^{(\ell)})\mid B_\ell=b_\ell]$ and write
$\psi_b(\lambda)\coloneqq\ln\mathbb E[\exp(\lambda Z_n)\mid b]$. The bounded range of the individual draws gives explicit bounds. Let $R$ be the range of $Z$ on the
support of $p_{AB}$. For $0\le\lambda\le1$ the tilt density of each
coordinate is at least $e^{-R}$. Expressing each conditional variance as half the expected squared difference of two independent draws, and applying this lower bound to both draws, gives
$c n\le\psi_b''(\lambda)\le C_0 n$, with $c\coloneqq e^{-2R}w/2$ and $C_0\coloneqq R^2/4$.
Fix $\kappa\coloneqq c/2$ and set $a\coloneqq2\sqrt{C_0n}$ and
$\psi_b'(\lambda)=\max\{m_b,z+2a\}$. Such $0\le\lambda\le1$ exists for
large $n$, and
$\lambda\le ((z-m_b)_++2a)/(cn)$. Chebyshev's inequality gives tilted
probability at least $3/4$ to the interval
$[\psi_b'(\lambda)-a,\psi_b'(\lambda)+a]$, which lies above $z$.
On that interval the reverse density change is at least
$\exp[-\lambda(\psi_b'(\lambda)+a)+\psi_b(\lambda)]$.
Using $\psi_b(\lambda)\ge\lambda m_b$ proves \eqref{eq:conditional-upper-tail}.

Set
\begin{align}
 t_b&\coloneqq z+\log_2 q_b(z),\\
 \overline L'&\coloneqq K^{-1}\sum_i L_i\mathbf 1\{L_i\le2^{t_b}\}.
\end{align}
The downward shift of the cutoff makes its conditional variance proportional to $q_b(z)$, which will cancel the small denominator in \eqref{eq:classical-deficit}.
The conditional mean of $\overline L'$ is $1-q_b(t_b)$. Its conditional variance satisfies
\begin{align}
 \operatorname{Var}_{\mathrm{ind}}(\overline L'\mid b)&\le \frac{2^{t_b}}K\,
 \mathbb E[2^{Z_n-t_b}\mathbf 1\{Z_n\le t_b\}\mid b]\\
 &\le Cq_b(z)/\sqrt n.
 \label{eq:classical-truncated-variance}
\end{align}
In \eqref{eq:classical-truncated-variance} the unprimed likelihood ratios are sampled under the product reference, while the expectation containing $Z_n$ is under the signal. This change of measure removes one factor of the likelihood. The last inequality follows by summing \eqref{eq:classical-interval} over unit intervals below $t_b$ with geometric weights. Truncation and
$(y-q)_+\le y^2/(4q)$ for real $y$ and $q>0$ imply
\begin{align}
 F_b&\le \mathbb E_{\mathrm{ind}}[(1-\overline L')_+\mid b]\\
 &=q_b(t_b)+\mathbb E_{\mathrm{ind}}[(\overline L'-1)_+\mid b]\\
 &\le q_b(t_b)+\frac{\operatorname{Var}_{\mathrm{ind}}(\overline L'\mid b)}{4q_b(t_b)} .
 \label{eq:classical-deficit}
\end{align}
Since $q_b(t_b)\ge q_b(z)$, the variance contribution is $O(n^{-1/2})$.
By \eqref{eq:classical-interval}, $q_b(t_b)-q_b(z)$ is at most
$C(1+|\log_2 q_b(z)|)/\sqrt n$. Its average over the chosen set is
$O(n^{-1/2})$ by \eqref{eq:conditional-upper-tail} and
$\mathbb E[(z-m_b)^2]=O(n)$. On the exceptional set use $F_b\le1$.

If $w=0$, $Z_n=m_b$ with probability one conditional on $b$.
Under the product reference, $L_i$ equals $2^{m_b}$ on the conditional
support and zero off it. Therefore $F_b\le\min\{1,2^{(m_b-z)/2}/2\}$ by
the second moment of the empirical mean. The random variable $m_b$ is then a
sum of independent variables with variance $nV$. A geometric sum of the
Berry--Esseen unit-interval estimates below $z$ proves
\eqref{eq:classical-cover-tail} in this case as well.

There is a matching lower bound for the ordinary split state. Observe $b$ and
all $K$ positions, and accept if at least one of their likelihood ratios
exceeds $K$. The planted position alone gives acceptance probability at least
$\Pr\{Z_n>z\}$ under the split state. The union bound under the product reference
gives type~II error probability at most
\begin{align}
 K\Pr_{(\tau_A p_B)^{\otimes n}}\{Z_n>z\}
 &=\mathbb E_{p_{AB}^{\otimes n}}[2^{z-Z_n}\mathbf 1\{Z_n>z\}]\\
 &=O(n^{-1/2}).
 \label{eq:classical-false-positive}
\end{align}
Here $Z_n$ denotes the extended log-likelihood ratio under the product reference: it equals $-\infty$ wherever the signal density is zero on that reference support. The last
estimate follows by the same geometric sum of unit-interval probabilities,
now above $z$, and is uniform in $z$. The difference of the two acceptance
probabilities is bounded by the normalized trace distance in \eqref{eq:classical-cover-tail}. Together with \eqref{eq:classical-cover-tail},
this proves \eqref{eq:classical-cover-approximation}. To obtain \eqref{eq:classical-cover-optimum}, shift $nD+\sqrt{nV}\Phi^{-1}(1-\varepsilon)$ by fixed constants in both directions. The unconditional Berry--Esseen bound and the strictly positive normal density at the quantile show that the upper shift suffices, and that every smaller $K$ below the lower shift fails. For the latter claim the lower bound $\tfrac12\norm{\omega_{A^{nK}B^n}-\omega^0_{A^{nK}B^n}}_1\ge\Pr\{Z_n>\log_2 K\}-O(n^{-1/2})$ is uniform at every threshold. Rounding a sufficient population to an integer costs at most one bit. This proves \eqref{eq:classical-cover-optimum}.
\end{proof}

\subsection{Classical side information and the actual marginal}\label{sec:classical-actual}
Theorem~\ref{thm:classical-rate} compares the hidden state with independent preparation. In the key criterion \eqref{eq:key-joint} the comparison is between a row and the entire table. We can use the independent product as an intermediate reference because the full table contains exponentially more positions than a single row when $I_B>I_E$.

For the independent-table source protocol of Section~\ref{sec:key}, fix the row $m$. Write $\omega^m_{TE},\omega_{TE}$ for the two states in \eqref{eq:key-placement}--\eqref{eq:key-placement-mean}. The row-to-product trace distance equals the ordinary split distance for $K$ blocks: the unused entries are independent in both preparations and tensoring them onto both states preserves the trace norm. The full-table distance equals the same quantity with $MK$ blocks. With $A=X$ and $B=E$ in the local covering experiment, the triangle inequality gives
\begin{align}
 \tfrac12\norm{\omega^m_{TE}-\omega_{TE}}_1
 &\le \tfrac12\norm{\omega_{X^{nK}E^n}-\omega^0_{X^{nK}E^n}}_1
 +\tfrac12\norm{\omega_{X^{nMK}E^n}-\omega^0_{X^{nMK}E^n}}_1 .
 \label{eq:classical-row-full}
\end{align}
Here the registers $X^{nK}$ and $X^{nMK}$ store, respectively, the symbols of the chosen row or of the full independent table. The same two distances bound the expected row-to-marginal secrecy error for the wiretap codebook in \eqref{eq:private}.

At the covering choice $\log_2K=nD+O(\sqrt n)$ in \eqref{eq:classical-cover-optimum}, the number $M$ of rows grows at a positive exponential rate because $I_B>I_E$. Thus $\log_2(MK)\ge n(D+\xi)$ for some $\xi>0$ and all sufficiently large $n$. Truncate the likelihoods of \eqref{eq:conditional-site-likelihood} for this full table at $2^u$, where $u\coloneqq n(D+\xi/2)$. Their truncated mean has conditional expectation $1-q_b(u)$ and conditional variance at most $2^u/(MK)$. The mean negative part of a centered random variable is at most half of its standard deviation. An exponential Markov bound on the upper tail of $Z_n$ (see Appendix~\ref{sec:separable} for countable outputs) gives, for some $c'>0$,
\begin{equation}
 \tfrac12\norm{\omega_{X^{nMK}E^n}-\omega^0_{X^{nMK}E^n}}_1
 \le\Pr\{Z_n>u\}+\tfrac12\sqrt{2^u/(MK)}=O(e^{-c'n}).
 \label{eq:classical-all-exponent}
\end{equation}
The fixed constant in the choice of $\log_2K$ in \eqref{eq:classical-cover-optimum} can be increased so that the first distance in \eqref{eq:classical-row-full} lies below $\varepsilon_E$ by a positive multiple of $1/\sqrt n$. This absorbs the second distance.

The compressed public tables of Corollary~\ref{cor:transcript} can have higher-order correlations, so the independent-table equalities need a separate argument. For the compressed table probability distribution $\mu$ in \eqref{eq:compressed-table-distribution}, use the reference $\mu(t)p_{E}^{\otimes n}(e^n)$. In the common eigenbasis of Eve's states, write $p_{E^n|X^n}(e^n|t_i)$ for the probability of observing $e^n$ conditional on the word at site $i$. For row $m$, the likelihood ratio on the reference support is
\begin{equation}\label{eq:compressed-row-likelihood}
 \frac{p_{TE^n|U=m}(t,e^n)}{\mu(t)p_E^{\otimes n}(e^n)}
 =\frac1K\sum_{i\,\in\,\mathrm{row}\,m}
       \frac{p_{E^n|X^n}(e^n|t_i)}{p_E^{\otimes n}(e^n)} .
\end{equation}
It follows by conditioning on Alice's site using \eqref{eq:compressed-sampling}--\eqref{eq:compressed-table-distribution}. The full-table ratio is the average over all $MK$ sites. Thus the expected-deficit identity in \eqref{eq:conditional-deficit} applies to both comparisons. At fixed $e^n$ under the product reference, the summands at two sites are independent because the corresponding table entries are pairwise independent. If $w=0$ in \eqref{eq:classical-conditional-variance}, the conditional variance of the row average is at most $2^{m_b}/K$ by pairwise independence, so the same bound on $F_b$ follows. For $w>0$ the row estimate uses truncation. In particular the variance estimates \eqref{eq:classical-truncated-variance} and \eqref{eq:classical-all-exponent} remain valid. Applying the same upper bounds to the row and to the full table establishes the achievable classical-Eve refinement in Theorem~\ref{thm:rate} at the stated public communication cost.

\section[Pure conditional outputs]{Binary tests for pure conditional outputs}\label{sec:pure-pack}
When Bob's conditional states are pure, a suitable projection on the eigenspaces of his average state gives the improved packing estimate in Theorem~\ref{thm:rate}. Different input words can require different projections. We construct the binary effect for each word and then assemble the conditional effects into one test.

Let $X$ have a finite alphabet, let $B$ be separable, and suppose $V_B>0$ in \eqref{eq:rate-moments}. Suppose $\rho_B^x=|\psi_x\rangle\langle\psi_x|_B$ for every $x$ of positive probability. The span of the conditional vectors has dimension at most $|\mathcal X|$. Restrict $B$ to that span and write $\rho_B=\sum_j s_j|v_j\rangle\langle v_j|_B$. Draw $x$ from $p_X$ and then $j$ with conditional probability $|\langle v_j|\psi_x\rangle|^2$ and set $Z\coloneqq-\log_2 s_j$. Its unconditional mean and variance are $I_B$ and $V_B$ by \eqref{eq:rate-moments}. The output spectrum also enters the pure-state packing analysis of Wilde, Renes, and Guha~\cite{WRG}.

Let $z$ be a real threshold satisfying $|z-nI_B|\le C\sqrt n$ for some fixed $C$. For decoding we will set $z=\log_2(MK-1)$; the testing estimate also applies at other real thresholds. For an input word $x^n$, define
\begin{align}
 |\psi_{x^n}\rangle_{B^n}&\coloneqq\bigotimes_{j=1}^n|\psi_{x_j}\rangle_{B_j},\label{eq:pure-word-state}\\
 \sigma_{B^n}&\coloneqq\rho_B^{\otimes n},\\
 q_{x^n}(t)&\coloneqq\Pr\{Z_n>t\mid X^n=x^n\}.
 \label{eq:pure-quantities}
\end{align}
Here $Z_n\coloneqq\sum_{\ell=1}^n Z^{(\ell)}$ is the sum of the $n$ spectral draws, with $Z^{(\ell)}\coloneqq-\log_2 s_j$ conditional on drawing eigenvector $j$ for input $x_\ell$. We project $|\psi_{x^n}\rangle_{B^n}$ onto low-eigenvalue eigenspaces of $\sigma_{B^n}$. The cutoff must depend on the word because normalizing its projected vector increases type~II error probability when the projection has small norm.

First suppose $\mathbb E\operatorname{Var}(Z\mid X)>0$. At typical words the conditional interval and tail estimates \eqref{eq:classical-interval} and \eqref{eq:conditional-upper-tail} apply to the bounded, conditionally independent spectral draws. Let $\mathcal G_n$ consist of the words for which $\operatorname{Var}(Z_n\mid x^n)\ge n\mathbb E\operatorname{Var}(Z\mid X)/2$ and $|\mathbb E[Z_n\mid x^n]-z|\le\kappa n$, with $\kappa>0$ as in the proof of \eqref{eq:conditional-upper-tail} for this spectral distribution. Its complement has exponentially small probability. On $\mathcal G_n$ the mass of an interval of length $h$ is at most $C(1+h)/\sqrt n$, and $q_{x^n}(z)>0$. Averaging \eqref{eq:conditional-upper-tail} only over these words gives
\begin{equation}\label{eq:pure-good-tail}
 \sum_{x^n\in\mathcal G_n}p_{X}^{\otimes n}(x^n)\,|\log_2 q_{x^n}(z)|=O(1),
\end{equation}
because $\mathbb E[(z-\mathbb E[Z_n\mid X^n])^2]=O(n)$. On the exceptional words the tail can vanish; they will contribute only their probability to the error bound.

For each word $x^n\in\mathcal G_n$ set
\begin{align}
 t&\coloneqq z+\log_2[1/q_{x^n}(z)],\\
 A_{x^n}&\coloneqq C(1+|\log_2 q_{x^n}(z)|)/\sqrt n .
\end{align}
Moving the threshold upward gives $q_{x^n}(t)\ge q_{x^n}(z)-A_{x^n}$ by the interval estimate. If $q_{x^n}(z)>2A_{x^n}$, define the effect
\begin{align}
 P_{B^n}^{t}&\coloneqq\{-\log_2\sigma_{B^n}>t I_{B^n}\},\\
 T_{B^n}^{x^n}&\coloneqq
 \frac{P_{B^n}^{t}|\psi_{x^n}\rangle\langle\psi_{x^n}|_{B^n}P_{B^n}^{t}}{q_{x^n}(t)} .
 \label{eq:pure-effect}
\end{align}
This is a rank-one projector. Its type~I error probability is $1-q_{x^n}(t)$, and its type~II error probability satisfies
\begin{equation}
 2^z\Tr[T_{B^n}^{x^n}\sigma_{B^n}]
 =\frac{q_{x^n}(z)}{q_{x^n}(t)}\,
 \mathbb E[2^{t-Z_n}\boldsymbol 1_{\{Z_n>t\}}\mid x^n]\le C'/\sqrt n .
 \label{eq:pure-truncated-test}
\end{equation}
Indeed $q_{x^n}(t)>q_{x^n}(z)/2$, and the remaining expectation is bounded by a geometric sum of unit-interval probabilities. For a typical word with $q_{x^n}(z)\le2A_{x^n}$ set $T_{B^n}^{x^n}\coloneqq0$. Its type~I error probability then obeys $1\le1-q_{x^n}(z)+2A_{x^n}$. For $x^n\notin\mathcal G_n$, the zero effect gives type~I error probability one and type~II error probability zero. Its contribution to the average error is exponentially small.

The joint effect and its two error probabilities are thus
\begin{align}
 T_{X^nB^n}&\coloneqq\sum_{x^n}|x^n\rangle\langle x^n|_{X^n}\otimes T_{B^n}^{x^n},\\
 \Tr[(I_{X^nB^n}-T_{X^nB^n})\rho_{XB}^{\otimes n}]
 &\le \Pr\{Z_n\le z\}+O(n^{-1/2}),\label{eq:pure-type-i}\\
 \Tr[T_{X^nB^n}(\rho_X\otimes\rho_B)^{\otimes n}]&=O(2^{-z}/\sqrt n).
 \label{eq:pure-errors}
\end{align}
If $\mathbb E\operatorname{Var}(Z\mid X)=0$, each $|\psi_x\rangle_B$ lies in one eigenspace of $\rho_B$. Use the projector $|\psi_{x^n}\rangle\langle\psi_{x^n}|_{B^n}$ when the corresponding spectral value $Z_n$ exceeds $z$, and use zero otherwise. The type~I error probability then equals $\Pr\{Z_n\le z\}$. The type~II error probability equals $\mathbb E[2^{-Z_n}\boldsymbol 1_{\{Z_n>z\}}]$, so \eqref{eq:spectral-local} gives the second bound in \eqref{eq:pure-errors}.

At $z=nI_B+\sqrt{nV_B}\Phi^{-1}(\varepsilon_B)-c_1$ a sufficiently large constant $c_1$ makes the sum of the type~I error probability and $2^z$ times the type~II error probability in \eqref{eq:pure-type-i}--\eqref{eq:pure-errors} at most $\varepsilon_B$. This controls the binary cost in \eqref{eq:cost} and removes the logarithmic packing loss in \eqref{eq:third-pack} for pure conditional outputs.

The test in \eqref{eq:pure-type-i}--\eqref{eq:pure-errors} gives a lower bound on the testing divergence. For the upper bound take $z=nI_B+\sqrt{nV_B}\Phi^{-1}(\varepsilon_B)+c_2$. Every test with type I error probability at most $\varepsilon_B$ and type II error probability $\eta$ achieves signal acceptance at least $1-\varepsilon_B$. Applying \eqref{eq:uniform-roc} to this test gives
\begin{align}
 2^z\eta&\ge\Phi\!\left(\frac{z-nI_B}{\sqrt{nV_B}}\right)-\varepsilon_B-\frac{C''}{\sqrt n}\\
 &\ge c_3/\sqrt n.\label{eq:pure-converse-slack}
\end{align}
The Gaussian density is positive at $\Phi^{-1}(\varepsilon_B)$, so the second line holds for some $c_3>0$ and all sufficiently large $n$ upon choosing a sufficiently large fixed $c_2$. Minimizing over feasible tests gives $\beta_{\varepsilon_B}(\rho_{XB}^{\otimes n}\Vert(\rho_X\otimes\rho_B)^{\otimes n})\ge c_3 2^{-z}/\sqrt n$. Taking minus its binary logarithm recovers the upper bound of Cheng and Liu~\cite[Proposition~6.7]{CL}. Together the two bounds give
\begin{equation}
 D_H^{\varepsilon_B}\!\left(\rho_{XB}^{\otimes n}\middle\Vert
 (\rho_X\otimes\rho_B)^{\otimes n}\right)
 =nI_B+\sqrt{nV_B}\Phi^{-1}(\varepsilon_B)+\tfrac12\log_2 n+O(1)
 \label{eq:pure-rate}
\end{equation}
for finite pure-output ensembles with $V_B>0$. The classical half-logarithmic term is known~\cite[Theorem~2]{HayashiSemi}. Lizarribar-Carrillo, Vazquez-Vilar, and Koch give a one-shot comparison for a pure null state~\cite[Theorem~3]{LVK}. Here each conditional output is pure, while the joint null $\rho_{XB}$ is generally mixed. The lower bound in \eqref{eq:pure-rate} verifies the term proposed in~\cite[Eq.~(18)]{Cheng} for this class.

\section[An explicit public table]{An explicit public table for a finite source alphabet}\label{sec:explicit-table}
The proof of Corollary~\ref{cor:transcript} bounds the size of a table family using a convex decomposition. Here Alice samples a table conditioned on her source symbol at the same asymptotic communication cost. As in \eqref{eq:transcript-size}, $M$ is the number of key values, $K$ the number of sites per key, and $d$ the number of symbols with positive probability. Partition the unit circle $[0,1)$ into half-open
intervals $I_a$ of lengths $p_X(a)$. Draw two independent uniform
numbers $R,S$ on the circle and define each table entry by
\begin{equation}\label{eq:circle-table}
 T_i(R,S)\coloneqq a\quad\text{if }(R+iS)\bmod1\in I_a,
 \qquad 1\le i\le MK .
\end{equation} For $i\ne j$, the
integer matrix with rows $(1,i)$ and $(1,j)$ has nonzero determinant
$j-i$. It therefore defines a surjective homomorphism of the two-torus,
which sends its uniform probability measure to itself. Thus any two
coordinates are independent and uniform before discretization, and
the table symbols have the desired pairwise marginals.

For Alice to sample conditionally on $T_i=a$, draw $Z$ uniformly on $I_a$ and $S$ independently and uniformly on the circle, and set $R\coloneqq(Z-iS)\bmod1$. This generates the exact conditional distribution of $(R,S)$ required by \eqref{eq:circle-table}.
To describe the table, divide the unit square along lines $R+iS=c+k$ for
interval endpoints $c$ and integers $k$. At most $O(d(MK)^2)$ lines meet
the square, since a fixed $i$ and endpoint require at most $i+2$ integers.
They divide it into $O(d^2(MK)^4)$ regions, since the $\ell$th line adds at
most $\ell$ regions. The table is constant within each region. Boundaries
have probability zero, also given a symbol of positive probability.
Indexing the distinct tables gives a public message of $O(\log_2(dMK))$
bits. This construction uses exact real sampling, and listing its regions
can take exponential time in the blocklength.

\section{Separable quantum systems}\label{sec:separable}
A Hilbert space is separable when it has a finite or countably infinite orthonormal basis; the photon-number basis of an optical mode is one example. The one-shot covering and coding bounds allow such quantum systems, and the second-order results allow them when the classical input alphabet is finite and the variance assumptions of Theorem~\ref{thm:rate} hold. We first justify the measurement constructions, then verify the moment estimates for the asymptotics.

\paragraph{Tests and measurements.}
Let all quantum states be positive trace-class operators of unit trace on separable Hilbert spaces. In \eqref{eq:slice-pairing}, the supremum over normal states $\sigma_B$ equals $\norm{\E_A^\tau(Q_{AB})}$, since vector states approach the top of the spectrum. The effects satisfying the type~I error probability constraint in \eqref{eq:beta} form an ultraweakly compact convex set. The payoff is affine and continuous for the ultraweak topology on effects and the trace norm on normal states. Minimax with the compact set of effects gives \eqref{eq:beta-info}. The supremum of the type~II error probabilities is lower semicontinuous, so the minimum in \eqref{eq:beta} exists. The constraints in \eqref{eq:roc} are ultraweakly closed, so its maximum also exists.

This compactness also proves the limiting statement used after Theorem~\ref{thm:ordinary}. For a sequence $\eta_n\downarrow0$, choose effects attaining $G^{\eta_n}(\rho_{AB}\Vert\tau_A)$. An ultraweak accumulation point is feasible at zero, and continuity of the signal expectation gives
\begin{equation}\label{eq:roc-zero-limit}
 \lim_{\eta\downarrow0}G^\eta(\rho_{AB}\Vert\tau_A)
 =G^0(\rho_{AB}\Vert\tau_A).
\end{equation}
Here monotonicity ensures that the limit exists, and feasibility at zero gives the inequality in one direction, while $G^\eta\ge G^0$ gives the other. For the support condition, write $\tau_A=\sum_j\lambda_j|j\rangle\langle j|_A$ with $\lambda_j>0$ on its support, and let $P_A$ project onto that support. For every effect $Q_{AB}$ and vector $|\psi\rangle_B$, \eqref{eq:slice} gives
\begin{equation}\label{eq:roc-zero-quadratic}
 \langle\psi|\E_A^\tau(Q_{AB})|\psi\rangle_B
 =\sum_j\lambda_j\norm{Q_{AB}^{1/2}(|j\rangle_A\otimes|\psi\rangle_B)}^2 .
\end{equation}
Each term is nonnegative. Thus $\E_A^\tau(Q_{AB})=0$ holds exactly when $Q_{AB}(P_A\otimes I_B)=0$. Among these effects, $I_{AB}-P_A\otimes I_B$ maximizes signal acceptance, giving
\begin{equation}\label{eq:roc-zero-support}
 G^0(\rho_{AB}\Vert\tau_A)=\Tr[(I_A-P_A)\rho_A].
\end{equation}
Here $\rho_A\coloneqq\Tr_B[\rho_{AB}]$ is the signal marginal. Dominated convergence in \eqref{eq:ordinary-curve} now gives
\begin{equation}\label{eq:ordinary-limit}
 \lim_{K\to\infty}\int_0^1G^{1/(1+4Kt^2)}(\rho_{AB}\Vert\tau_A)\,dt
 =\Tr[(I_A-P_A)\rho_A].
\end{equation}
This limit is zero precisely when $\operatorname{supp}\rho_A\subseteq\operatorname{supp}\tau_A$.

The positive spectral projector of the trace-class difference of the mixtures attains \eqref{eq:distance}. The averages \eqref{eq:slice} and \eqref{eq:multi-slice} are normal unital completely positive slice maps. In an eigenbasis of $\tau_A$ they sum the diagonal blocks of a bounded observable with the corresponding probabilities. Hence the ordered products and second-moment inequalities of Lemma~\ref{lem:moment} continue to hold. For the spectral measure $P_{AB}$ of $H_{AB}$, Tonelli's theorem applied to its scalar distribution under $\rho_{AB}$ gives
\begin{align}
 \int_0^1\Tr[P_{AB}((t,\infty))\rho_{AB}]\,dt
 &=\int_{[-1,1]} x_+\,\Tr[P_{AB}(dx)\rho_{AB}]\\
 &=\Tr[(H_{AB})_+\rho_{AB}].
\end{align}
This proves the integral step of Theorem~\ref{thm:ordinary}. Integration up to $\tfrac12\norm{p-q}_1$ proves the corresponding step of Theorem~\ref{thm:cover}.

For the decoder in \eqref{eq:pgm-site}, write $S_B(t)$ as in \eqref{eq:pgm-sum}. Since $0\le\rho_B^{t_{mk}}\le S_B(t)$, the form $S_B(t)^{-1/2}\rho_B^{t_{mk}}S_B(t)^{-1/2}$ on the range of $S_B(t)^{1/2}$ is bounded between zero and the identity and thus extends to an effect on the support. Summing these forms gives the identity there. To check \eqref{eq:pgm-cond}, compress all states to increasing finite-rank spectral subspaces of $S_B(t)$ exhausting its support. Let $P_n$ denote these spectral projectors. They commute with $S_B(t)$, so each compressed decoding effect on $P_n$ is $P_n\Lambda_B^{(m,k),t}P_n$. For every letter used in this table, $P_n\rho_B^{t_{mk}}P_n\to\rho_B^{t_{mk}}$ in trace norm since the letter state is supported on $\operatorname{supp}S_B(t)$. By cyclicity the compressed acceptance probability equals $\Tr[\Lambda_B^{(m,k),t} P_n\rho_B^{t_{mk}}P_n]$, which converges to $\Tr[\Lambda_B^{(m,k),t}\rho_B^{t_{mk}}]$ because the effect is bounded. The binary risk converges by \eqref{eq:cost-norm}, and the trace of the compressed sum tends to $\Tr[S_B(t)]=MK$. All sums over sites, letters, and tables in Theorems~\ref{thm:wire} and~\ref{thm:key} are finite. The disagreement probabilities and the joint trace distances obey the same bounds, including the public communication cost of Corollary~\ref{cor:transcript}.

\paragraph{Second-order rates.}
For $J=B,E$, let $Z$ be the spectral log-likelihood of Appendix~\ref{sec:gaussian-cover} for $\rho_{XJ}$ and $\rho_X\otimes\rho_J$. Throughout, $x\in\operatorname{supp}p_X$. To check the moments, write $r_i^x,u_i^x$ and $s_j,v_j$ for the positive eigenvalues and eigenvectors of $\rho_J^x$ and $\rho_J$, respectively. The triple $(x,i,j)$ has probability $p_X(x)r_i^x|\langle u_i^x|v_j\rangle|^2$, and $Z=\log_2(r_i^x/s_j)$. From $p_X(x)r_i^x |u_i^x\rangle\langle u_i^x|\le\rho_J$, testing on the vector $\sum_{j=1}^m s_j^{-1}\langle v_j|u_i^x\rangle v_j$ and letting $m$ increase gives
\begin{align}
 p_X(x)r_i^x\sum_j |\langle u_i^x|v_j\rangle|^2/s_j&\le1,\\
 \mathbb E[2^{-Z}]&\le 1,\\
 \mathbb E[2^Z]&\le |\operatorname{supp}p_X|.\label{eq:separable-moments}
\end{align}
The bound for $2^{-Z}$ follows by summing the squared overlaps over $i$. For $2^Z$, multiply the first inequality by $r_i^x$ and sum over $i,x$. Markov's inequality applied to $2^Z$ and $2^{-Z}$ gives
\begin{equation}\label{eq:separable-exponential-tail}
 \Pr\{|Z|>u\}\le 2^{-u}\bigl(\mathbb E[2^Z]+\mathbb E[2^{-Z}]\bigr)
 \le (1+|\operatorname{supp}p_X|)2^{-u},\qquad u\ge0.
\end{equation}
This exponential tail bound implies that all absolute polynomial moments exist, in particular the third moment needed for Berry--Esseen. For separable outputs, the spectral definitions corresponding to \eqref{eq:rate-information}--\eqref{eq:rate-moments} are
\begin{align}
 I_J&\coloneqq\mathbb E[Z],\label{eq:separable-rate-information}\\
 V_J&\coloneqq\mathbb E[(Z-I_J)^2].\label{eq:separable-rate-moments}
\end{align} The separable-space testing bound~\cite[Proposition~A.2, Eq.~(A24)]{OMW} gives \eqref{eq:li-quantile}. The nonnegative spectral sums in \eqref{eq:harmonic-test} converge by monotone convergence. The remaining Gaussian estimates in Appendix~\ref{sec:gaussian-cover} depend only on these moments. This proves \eqref{eq:third-net} and the commuting-Bob refinement. For pure conditional outputs, the support of $\rho_B$ has dimension at most $|\mathcal X|$, so Appendix~\ref{sec:pure-pack} applies directly.

For commuting Eve, the common eigenbasis may be countable. In Appendix~\ref{sec:classical-cover-order} let $A$ be finite and $b$ a single output symbol of positive probability. Set $L\coloneqq\max_{a:\tau_A(a)>0}\log_2(1/\tau_A(a))$. Under the conditional signal distribution, $Z=\log_2[p_{A|B}(A|b)/\tau_A(A)]$ satisfies
\begin{align}
 Z&\le L,\\
 \Pr\{Z<-u\mid b\}&\le 2^{-u}\quad(u\ge0).
 \label{eq:separable-classical-tail}
\end{align}
The tail bound follows by summing $p_{A|B}(a|b)<2^{-u}\tau_A(a)$ over the indicated letters. Thus the conditional means, variances, and third absolute moments are bounded uniformly in $b$. The Hoeffding estimate for their sums and the conditional Berry--Esseen estimate \eqref{eq:classical-interval} follow as before. For the tilt used in \eqref{eq:conditional-upper-tail}, the conditional mean is a relative entropy and hence nonnegative. Consequently $\mathbb E[e^{\lambda Z}\mid b]\ge1$ for $0\le\lambda\le1$. The tilted third absolute moments are uniformly bounded by $e^L\sup_b\mathbb E[|Z|^3\mid b]$. The third derivative of each log moment-generating function is the tilted third central moment, so $|\psi_b'''(\lambda)|\le Cn$ for a word $b$ of length $n$. On the typical set $\psi_b''(0)\ge nw/2$, whence $nw/4\le\psi_b''(\lambda)\le C'n$ on a fixed interval $[0,\lambda_0]$ when $w>0$. Taking $\kappa$ sufficiently small gives the tilt in \eqref{eq:conditional-upper-tail}. For \eqref{eq:classical-all-exponent}, the unconditional log moment-generating function is differentiable at zero with derivative $D$. An exponential Markov bound at a small positive tilt therefore bounds $\Pr\{Z_n>n(D+\xi/2)\}$ exponentially. These estimates establish the commuting-Eve refinement for countably many eigenvalues.

\end{document}